\documentclass[11pt]{article}
\usepackage{amssymb}
\usepackage{makeidx}
\usepackage[totalwidth=460truept,totalheight=600truept]{geometry}
\usepackage{latexsym,amssymb,amsmath,graphicx,accents,eucal,slashed,subfigure}
\usepackage{dsfont}
\usepackage[T1]{fontenc}
\usepackage[hidelinks]{hyperref}

\newtheorem{theorem}{Theorem}
\newtheorem{proposition}{Proposition}
\def\theequation{\arabic{section}.\arabic{equation}}

\renewcommand{\theequation}{\thesection.\arabic{equation}}
\global\arraycolsep=1truept

\numberwithin{equation}{section}
\renewcommand{\theequation}{\arabic{section}.\arabic{equation}}

\begin{document}

\bigskip \phantom{C}

\vskip1.4truecm

\begin{center}
{\huge \textbf{Ward Identities And Gauge Independence}}

\vskip.5truecm

{\huge \textbf{In General Chiral Gauge Theories}}

\vskip 1truecm

\textsl{Damiano Anselmi}

\vskip .2truecm

\textit{Dipartimento di Fisica ``Enrico Fermi'', Universit\`{a} di Pisa, }

\textit{and INFN, Sezione di Pisa,}

\textit{Largo B. Pontecorvo 3, I-56127 Pisa, Italy}

\vskip .2truecm

damiano.anselmi@df.unipi.it

\vskip 1.5truecm

\textbf{Abstract}
\end{center}

\medskip

Using the Batalin-Vilkovisky formalism, we study the Ward identities and the
equations of gauge dependence in potentially anomalous general gauge
theories, renormalizable or not. A crucial new term, absent in manifestly
nonanomalous theories, is responsible for interesting effects. We prove that
gauge invariance always implies gauge independence, which in turn ensures
perturbative unitarity. Precisely, we consider potentially anomalous
theories that are actually free of gauge anomalies thanks to the
Adler-Bardeen theorem. We show that when we make a canonical transformation
on the tree-level action, it is always possible to re-renormalize the
divergences and re-fine-tune the finite local counterterms, so that the
renormalized $\Gamma $ functional of the transformed theory is also free of
gauge anomalies, and is related to the renormalized $\Gamma $ functional of
the starting theory by a canonical transformation. An unexpected consequence
of our results is that the beta functions of the couplings may depend on the
gauge-fixing parameters, although the physical quantities remain gauge
independent. We discuss nontrivial checks of high-order calculations based
on gauge independence and determine how powerful they are.

\vfill\eject

\section{Introduction}

\label{s1}

\setcounter{equation}{0}

The Ward-Takahashi \cite{ward,takahashi} and Slavnov-Taylor \cite%
{slavnov,taylor} identities are relations among the correlation functions of
quantum field theory, and follow from gauge and global symmetries. They are
usually studied in theories that are manifestly nonanomalous, that is to say
admit a manifestly gauge invariant regularization technique, for example QED
and nonchiral Yang-Mills theories. Chiral gauge theories, such as the
standard model, are potentially anomalous, because they do not admit a
manifestly gauge invariant regularization technique. The Adler-Bardeen (AB)
theorem \cite{adlerbardeen,ABrenoYMLR,ABnonreno} is the main tool that can
establish whether a potentially anomalous theory is in the end truly
anomalous or nonanomalous. It ensures that, if the gauge anomalies are
trivial at one loop, they can be cancelled to all orders.

The potentially anomalous theories that are actually free of gauge anomalies
thanks to the Adler-Bardeen theorem will be called \textit{AB\ nonanomalous}%
. In this paper, we study the Ward identities of the AB nonanomalous general
gauge theories, including the nonrenormalizable ones, and clarify the
relation between gauge invariance and gauge independence. Our investigation
upgrades the ones available in the literature in several respects.

Gauge invariance and gauge independence are two different concepts, to the
extent that a functional can be gauge invariant and gauge dependent at the
same time. For example, the renormalized action of non-Abelian Yang-Mills
theory contains a term propotional to $Z_{A}\int F_{\mu \nu }^{a}F^{a\mu \nu
}$, where $F_{\mu \nu }^{a}$ is the field strength and $Z_{A}$ is the wave
function renormalization constant of the gauge field. This expression is
gauge invariant, but not gauge independent, because $Z_{A}$ may depend on
the gauge-fixing parameters.

Yet, the two concepts are related to each other, and crucial to prove
perturbative unitarity. Gauge invariance is necessary, because its violation
makes unphysical degrees of freedom, such as the longitudinal photons,
propagate. On the other hand, gauge independence is important, because it
allows us to switch back and forth between gauges that exhibit perturbative
unitarity, but do not have good power-counting behaviors (such as the
Coulomb gauge), and gauges that have good power-counting behaviors, but do
not exhibit unitarity (such as the Lorenz gauge). The Lorenz gauges are very
convenient to make calculations and prove theorems to all orders. They make
renormalizability manifest, when the theory is power-counting
renormalizable. When the theory is nonrenormalizable, they make the locality
of counterterms manifest. However, the Lorenz gauges hide unitarity, because
they introduce unphysical, propagating degrees of freedom, such as the
longitudinal components of the gauge fields and the Fadeev-Popov ghosts.
This is where gauge independence plays a key role, because it ensures that
every physical quantity can be equivalently defined by using the Coulomb
gauge, where the propagators have no unphysical poles and perturbative
unitarity is manifest. The equivalence of the two gauges allows us to
loosely say that \textquotedblleft the unphysical degrees of freedom of the
Lorenz gauges compensate one another and drop out of the physical
quantities\textquotedblright .

Thus, in quantum field theory we need both gauge invariance and gauge
independence. If a theory is AB nonanomalous, it is by definition gauge
invariant. It is not obvious that the Adler-Bardeen theorem also ensures
that\ the physical quantities are ultimately gauge independent. Is it so, or
do we need extra assumptions to ensure that the physics does not depend on
the gauge fixing? Among other things, in this paper we answer this question
by proving that gauge invariance \textit{always} implies gauge independence.

In our approach, the Ward identities of AB nonanomalous general gauge
theories are corrected by a term that is absent in manifestly nonanomalous
theories. The correction is evanescent at the bare level, but can generate
finite corrections at the renormalized level, by simplifying some
divergences. One of the main consequences is that the beta functions of the
couplings can depend on the parameters introduced by means of the gauge
fixing. However, the physical quantities are protected, and remain gauge
independent.

We study how the renormalized $\Gamma $ functional $\Gamma _{R}$ depends on
the parameters introduced by the canonical transformations of fields and
sources. Canonical transformations encode field redefinitions and changes of
the gauge fixing, both of which are expected to have no effect on the
physical quantities. When we speak of \textquotedblleft gauge
dependence\textquotedblright\ we refer to the dependence on all types of
parameters introduced by a canonical transformation, including those
associated with field redefinitions.

We work out how a canonical transformation on the (bare) action $S$ affects
the renormalized $\Gamma $ functional $\Gamma _{R}$. After the
transformation, the theory must be renormalized anew. We show that in this
process of re-renormalization, it is always possible to redefine the
subtraction scheme, by fine-tuning the finite local counterterms, so that
the transformed theory is also AB nonanomalous. Moreover, the gauge
dependence of the transformed $\Gamma_R $ is encoded into a canonical
transformation, up to evanescent corrections.

This result allows us to prove that the physical quantities are gauge
independent. However, quantities that are useful for intermediate purposes,
such as the beta functions of the couplings, are normally gauge dependent.
Their gauge dependence can be absorbed inside finite redefinitions of the
couplings.

In manifestly nonanomalous theories we are, to a large extent, free to use a
preferred subtraction scheme, such as the minimal one, both before and after
the canonical transformation. The physical quantities and the beta functions
of the couplings are unaffected by the transformation (see for example \cite%
{back}). In AB nonanomalous theories, instead, we can use a preferred
subtraction scheme neither before, nor after the transformation. Before the
transformation, we need to choose a specific class of subtraction schemes to
take advantage of the Adler-Bardeen theorem and cancel the gauge anomalies
to all orders. After the transformation, we need to choose (another)
specific class of subtraction schemes, to enforce the cancellation of gauge
anomalies again. In this process, some gauge-fixing parameters move out of
the gauge-fixing sector into another unphysical sector, the one encoded by
the choice of the subtraction scheme. The result is that the beta functions
are gauge dependent, in general. Nevertheless, we can make their gauge
dependences disappear, if we specify the new subtraction scheme even further.

Both gauge invariance and gauge independence can be used to make powerful
checks of high-order calculations. As said, a consequence of our
investigation is that in AB\ nonanomalous theories, including the standard
model, the beta functions of the couplings are not completely gauge
independent. We show that, in spite of this, sufficiently powerful checks of
high-order calculations are still available. The reason is that the gauge
dependence cannot be arbitrary, because it cannot affect the physical
quantities.

To keep track of gauge invariance through renormalization, we use the
Batalin-Vilkovisky (BV) formalism \cite{bata}. The gauge invariant
regularization techniques commonly used for manifestly nonanomalous theories
are also convenient to treat AB nonanomalous theories, because they minimize
the number of terms that are potentially anomalous. In this paper we use the
dimensional regularization \cite{dimreg}, or any regularization technique
that underlies the dimensional one, such as the chiral dimensional (CD)
regularization of ref. \cite{chiraldimreg} and the
(chiral)dimensional/higher-derivative regularization of refs. \cite%
{ABrenoYMLR,ABnonreno,chiraldimreg,lvsm}, obtained by merging the (chiral)
dimensional one with the covariant higher-derivative regularization of ref. 
\cite{higherder}. We recall that the CD regularization is particularly
convenient for studying nonrenormalizable theories, to avoid certain
ambiguities that show up when we extract the divergent parts of the BV
antiparentheses $(X,Y)$ of two functionals $X$ and $Y$, as well as other
nuisances that the ordinary dimensional regularization is responsible for.

We also take the chance to revisit some known issues under our perspective.

Before presenting our results in more detail, we comment on the existing
literature on related subjects, and explain the upgrades we make. Most
studies of gauge dependence have been focused on renormalizable theories 
\cite{gaugedep}, or nonrenormalizable, but nonchiral, theories \cite%
{nonchiral,removal}, where the problem is much simpler (see appendix E). We
want to develop an approach that also applies to nonrenormalizable chiral
theories, to include the standard model coupled to quantum gravity. In our
opinion, it is not necessary to wait for the ultimate theory of quantum
gravity to prove general statements about it. The other investigations of
gauge dependence we are aware of use the so-called algebraic approach to
renormalization \cite{algebraica}. The main feature of the algebraic
approach is that it does not make use of an explicit regularization
technique. Instead, it relies on tools such as the \textquotedblleft quantum
action principle\textquotedblright \cite{qap}.

We think that it is important to develop more standard approaches to the
problem of gauge dependence, like the one of the present paper, which uses
the dimensional regularization or modified versions of it. For example,
anomalies have taught us that working without an explicit regularization may
not be completely safe. Another advantage of using an explicit
regularization is that we can identify convenient subtraction schemes, where
simplifications occur and several properties are easier to deal with, to all
orders in the perturbative expansion. Examples are those provided by refs. 
\cite{ABrenoYMLR,ABnonreno}, where it was shown that in suitable subtraction
schemes the gauge anomalies automatically vanish from two loops onwards, if
they cancel out at one loop. By construction, it is not possible to identify
special subtraction schemes in regularization-independent approaches.

Now we state the main results of our investigation. We study canonical
transformations that are continuously connected with the identity. Their
generating functionals have the form%
\begin{equation}
F(\Phi ,K^{\prime },\theta )=\int \Phi ^{\alpha }K_{\alpha }^{\prime }+%
\mathcal{O}(\theta ),  \label{fx}
\end{equation}%
where $\theta $ denotes the \textquotedblleft gauge
parameters\textquotedblright , which are associated with both changes of
field variables and changes of the gauge fixing. In the first part of our
analysis, we prove the main theorem, which states that if the theory is AB\
nonanomalous at $\theta =0$, after making the canonical transformation (\ref%
{fx}) it is always possible to re-renormalize the divergences and
re-fine-tune the finite local counterterms, continuously in $\theta $, so
that the equations 
\begin{eqnarray}
\text{(ABT)} &&\text{ \qquad }(\Gamma _{R\theta },\Gamma _{R\theta })=%
\mathcal{O}(\varepsilon ),  \label{ABT} \\
\text{(GDE)} &&\text{ \qquad }\frac{\partial \Gamma _{R\theta }}{\partial
\theta }-(\Gamma _{R\theta },\langle \tilde{Q}_{R\theta }\rangle )=\mathcal{O%
}(\varepsilon )  \label{GDE}
\end{eqnarray}%
hold for arbitrary $\theta $, where $\Gamma _{R\theta }$ is the renormalized 
$\Gamma $ functional of the transformed theory and $\tilde{Q}_{R\theta }$ is
a suitable renormalized local functional. The right-hand sides of both
equations are (generically nonlocal) functionals that vanish when the
continued spacetime dimension $D=d-\varepsilon $ tends to the physical
spacetime dimension $d$. We denote such functionals by $\mathcal{O}%
(\varepsilon )$ and call them \textquotedblleft evanescent\textquotedblright
.

Equation (\ref{ABT}) ensures that the theory is AB\ nonanomalous for
arbitrary values of the gauge parameters. Thus, it encodes gauge invariance.
Formula (\ref{GDE}) is the equation of gauge dependence, and follows from
the generalized Ward identities. The equations (GDE) can be integrated to
show that the entire gauge dependence of $\Gamma _{R\theta }$ can be
absorbed inside a (convergent, but generically nonlocal) canonical
transformation, up to $\mathcal{O}(\varepsilon )$. The results encoded in
formulas (ABT) and (GDE) are so general that they do not require any
particular assumption (see section \ref{s8}).

We also derive the equations of gauge dependence at the level of the
renormalized action and show that RG\ invariance is preserved by the
canonical transformation.

A simple, but important application of the theorem is to power-counting
renormalizable chiral gauge theories gauge-fixed by means of a
nonrenormalizable gauge fixing. We show that the theory remains
renormalizable in a nonmanifest form, because the parameters of negative
dimensions introduced by the gauge fixing do not propagate into the physical
sector. Another application of the theorem is a crucial step in the proof of
the Adler-Bardeen theorem for nonrenormalizable theories \cite{ABnonreno}.

In some situations, we can prove formula (ABT) for arbitrary values of a
certain gauge parameter $\theta $ within a given class of subtraction
schemes. Then, it is not necessary to re-renormalize the divergences and the
re-fine-tune the finite local counterterms. Under the assumption that the
theory satisfies a certain cohomological property, which is a generalized
version of the well-known Kluberg-Stern--Zuber conjecture \cite{kluberg}, we
can derive an more specific version of equations (GDE), which reads 
\begin{equation}
\text{(GDE2)}\text{ \qquad }\frac{\partial \Gamma _{R\theta }}{\partial
\theta }-(\Gamma _{R\theta },\langle H_{R\theta }\rangle )-\sum_{i}\rho _{i}%
\frac{\partial \Gamma _{R\theta }}{\partial \lambda _{i}}=\mathcal{O}%
(\varepsilon ),  \label{GDE2}
\end{equation}%
where $\lambda _{i}$ are the independent parameters of the classical action, 
$\rho _{i}$ are constants that depend on $\lambda _{i}$ and the other
parameters of the theory and $H_{R\theta }$ is a renormalized local
functional.

We can write (\ref{GDE2}) in the form (\ref{GDE}) by suitably
\textquotedblleft evolving the parameters $\lambda $ in the $\theta $
direction\textquotedblright . Such redefinitions encode how the beta
functions of the couplings depend on $\theta $.

So far, the Adler-Bardeen theorem has been proved in a variety of cases. The
original proof given by Adler and Bardeen \cite{adlerbardeen} was designed
to work in QED. Most generalizations to renormalizable non-Abelian gauge
theories used\ arguments based on the renormalization group \cite%
{zee,collins,tonin,sorella}, which work well unless the first coefficients
of the beta functions satisfy peculiar conditions \cite{sorella} (for
example, they should not vanish). Then there exist algebraic/geometric
derivations \cite{witten} based on the Wess-Zumino consistency conditions 
\cite{wesszumino} and the quantization of the Wess-Zumino-Witten action.
Another method to prove the Adler-Bardeen theorem in renormalizable theories
is obtained by extending the coupling constants to spacetime-dependent
fields \cite{kraus}. A proof that covers all power-counting renormalizable
gauge theories was given in ref. \cite{ABrenoYMLR}. It was obtained by
elaborating on a previous proof \cite{lvsm} given for quantum field theories
that violate Lorentz symmetry at high energies (in particular, Lorentz
violating extensions of the standard model) and are renormalizable by
weighted power counting \cite{halat}. Recently, the proof of \cite%
{ABrenoYMLR} was further extended in ref. \cite{ABnonreno}, to include a
large class of nonrenormalizable theories, such as the standard model
coupled to quantum gravity. We emphasize that a byproduct of our
investigation is that the standard model, coupled to quantum gravity or not,
is perturbatively unitary, and so are most of its extensions.

The paper is organized as follows. In section \ref{s2} we compare the Ward
identities of chiral and nonchiral gauge theories, and illustrate the
crucial new term that appears when the theory is potentially anomalous. In
section \ref{s8} we prove the main theorem of this paper, by deriving and
integrating the equations of gauge dependence in AB\ nonanomalous theories.
We show that every canonical transformation on the classical action is
mapped into a canonical transformation on the renormalized $\Gamma $
functional, provided that the finite local counterterms are appropriately
re-fine-tuned. We also integrate the equations of gauge dependence. In
section \ref{renogauge} we derive the equations of gauge dependence of the
renormalized action. In section \ref{s24} we prove that the canonical
transformation preserves RG\ invariance and discuss two applications of the
main theorem. In section \ref{s3} we study the gauge dependence of the beta
functions in detail. In section \ref{s4} we explain how to switch off the
ghosts, the antighosts, the Lagrange multipliers for the gauge fixing, and
the sources for the symmetry transformations, and get to the physical
quantities, collected into a \textquotedblleft physical\textquotedblright\ $%
\Gamma $ functional $\Gamma _{\text{ph}}$. We derive the (nonlocal) gauge
symmetry of $\Gamma _{\text{ph}}$, and prove that it closes off shell.
Finally, we prove that $\Gamma _{\text{ph}}$ is gauge independent, up to
field redefinitions, and perturbatively unitary. In section \ref{s5} we
investigate the checks of high-order calculations provided by gauge
independence and estimate how powerful they are. Section \ref{s9} contains
our conclusions. In the appendices we prove some properties used in the
paper, recall earlier results and collect some reference formulas for the
standard model coupled to quantum gravity. Moreover, we revisit the gauge
dependence of manifestly nonanomalous theories in the light of the new
results.

\section{Generalized Ward identities}

\label{s2}

\setcounter{equation}{0}

In this section we fix some notation, recall the main properties of the
Batalin-Vilkovisky formalism for general gauge theories \cite{bata} and
derive the generalized Ward identities.

Let $D=d-\varepsilon $ denote the continued, complex dimension of spacetime,
and $d$ the physical spacetime dimension. The $D$-dimensional spacetime
manifold $\mathbb{R}^{D}$ is split into the product $\mathbb{R}^{d}\times 
\mathbb{R}^{-\varepsilon }$ of the ordinary $d$-dimensional spacetime $%
\mathbb{R}^{d}$ times a residual $(-\varepsilon )$-dimensional evanescent
space, $\mathbb{R}^{-\varepsilon }$. The spacetime indices $\mu ,\nu ,\ldots 
$ of vectors and tensors are split into the bar indices $\bar{\mu},\bar{\nu}%
,\ldots $, which take the values of $0,1,\cdots ,d-1$, and the formal hat
indices $\hat{\mu},\hat{\nu},\ldots $, which denote the $\mathbb{R}%
^{-\varepsilon }$ components. For example, the momenta $p^{\mu }$ are split
into the pairs $p^{\bar{\mu}}$, $p^{\hat{\mu}}$, also written as $\bar{p}%
^{\mu }$, $\hat{p}^{\mu }$, and the coordinates $x^{\mu }$ are split into $%
\bar{x}^{\mu }$, $\hat{x}^{\mu }$. The formal flat-space metric $\eta _{\mu
\nu }$ is split into the usual $d\times d$ flat-space metric $\eta _{\bar{\mu%
}\bar{\nu}}=$diag$(1,-1,\cdots ,-1)$ and the formal evanescent metric $\eta
_{\hat{\mu}\hat{\nu}}=-\delta _{\hat{\mu}\hat{\nu}}$. The off-diagonal
components $\eta _{\bar{\mu}\hat{\nu}}$ vanish. The evanescent components
are contracted among themselves by means of the metric $\eta _{\hat{\mu}\hat{%
\nu}}$, so for example $\hat{p}^{2}=p^{\hat{\mu}}\eta _{\hat{\mu}\hat{\nu}%
}p^{\hat{\nu}}$. Full $SO(1,D-1)$ invariance is lost in most expressions,
replaced by $SO(1,d-1)\times SO(-\varepsilon )$ invariance.

We recall that in the CD\ regularization the fields $\Phi $ have strictly $d$%
-dimensional components. The metric tensor $g_{\mu \nu }$ is block-diagonal:
the diagonal blocks are $g_{\bar{\mu}\bar{\nu}}(x)$ and $\eta _{\hat{\mu}%
\hat{\nu}}$, while $g_{\bar{\mu}\hat{\nu}}=0$. Moreover, the $\gamma $
matrices are strictly $d$ dimensional, and satisfy the usual Dirac algebra $%
\{\gamma ^{\bar{a}},\gamma ^{\bar{b}}\}=2\eta ^{\bar{a}\bar{b}}$, where the
indices $\bar{a},\bar{b},\ldots $ refer to the Lorentz group. If $d=2k$ is
even, the $d$-dimensional generalization of $\gamma _{5}$ is defined as 
\begin{equation*}
\tilde{\gamma}=-i^{k+1}\gamma ^{0}\gamma ^{1}\cdots \gamma ^{2k-1},
\end{equation*}%
and satisfies $\tilde{\gamma}^{\dagger }=\tilde{\gamma}$, $\tilde{\gamma}%
^{2}=1$. The left and right projectors $P_{L}=(1-\tilde{\gamma})/2$, $%
P_{R}=(1+\tilde{\gamma})/2$ are defined as usual. The tensor $\varepsilon ^{%
\bar{a}_{1}\cdots \bar{a}_{d}}$ and the charge-conjugation matrix $\mathcal{C%
}$ also coincide with the usual ones.

The set of fields $\Phi ^{\alpha }=\{\phi ^{i},C,\bar{C},B\}$ contains the
classical fields $\phi $, the Fadeev-Popov ghosts $C$, the antighosts $\bar{C%
}$ and the Lagrange multipliers $B$ for the gauge fixing. An external source 
$K_{\alpha }$ with opposite statistics is associated with each $\Phi
^{\alpha }$, and coupled to the $\Phi ^{\alpha }$ transformations $R^{\alpha
}(\Phi )$. If $X$ and $Y$ are functionals of $\Phi $ and $K$, their \textit{%
antiparentheses} are defined as 
\begin{equation}
(X,Y)\equiv \int \left( \frac{\delta _{r}X}{\delta \Phi ^{\alpha }}\frac{%
\delta _{l}Y}{\delta K_{\alpha }}-\frac{\delta _{r}X}{\delta K_{\alpha }}%
\frac{\delta _{l}Y}{\delta \Phi ^{\alpha }}\right) ,  \label{usa}
\end{equation}%
where the integral is over spacetime points associated with repeated indices
and the subscripts $l$ and $r$ in $\delta _{l}$ and $\delta _{r}$ denote the
left and right functional derivatives, respectively.

The action $S$ should solve the \textit{master equation} $(S,S)=0$ in $D$
dimensions, with the \textquotedblleft boundary condition\textquotedblright\ 
$S(\Phi ,K)=S_{c}(\phi )$ at $C=\bar{C}=B=K=0$, where $S_{c}(\phi )$ is the
classical action.

If the gauge algebra closes off shell, there exists a choice of field/source
variables such that the non-gauge-fixed solution $\bar{S}_{d}(\Phi ,K)$ of
the master equation has the form%
\begin{equation}
\bar{S}_{d}(\Phi ,K)=S_{c}(\phi )+S_{K},\qquad S_{K}(\Phi ,K)=-\int
R^{\alpha }(\Phi )K_{\alpha }.  \label{sbard}
\end{equation}%
In this case, $(\bar{S}_{d},\bar{S}_{d})=0$ splits into the two identities%
\begin{equation*}
\int R^{i}(\phi )\frac{\delta _{l}S_{c}(\phi )}{\delta \phi ^{i}}=0,\qquad
\int R^{\beta }(\Phi )\frac{\delta _{l}R^{\alpha }(\Phi )}{\delta \Phi
^{\beta }}=0,
\end{equation*}%
which express the gauge invariance of the classical action and the closure
of the algebra, respectively. The gauge-fixed solution $S_{d}(\Phi ,K)$ of
the master equation reads%
\begin{equation}
S_{d}(\Phi ,K)=S_{c}(\phi )+(S_{K},\Psi )+S_{K}=\bar{S}_{d}+(S_{K},\Psi ),
\label{sk}
\end{equation}%
where $\Psi (\Phi )$ is the \textit{gauge fermion}, that is to say a local
functional of ghost number $-1$ that encodes the gauge fixing. Reference
formulas for $S_{c}$, $S_{K}$ and $\Psi $ in the case of the standard model
coupled to quantum gravity can be found in appendix D. Typically, $\Psi $
has the form 
\begin{equation}
\Psi (\Phi )=\int \bar{C}\left( G(\phi ,\xi )+\frac{1}{2}P(\phi ,\xi
^{\prime },\partial )B\right) ,  \label{psif}
\end{equation}%
where $G(\phi ,\xi )$ is the gauge-fixing function, $P$ is an operator that
may contain derivatives acting on $B$, and $\xi $, $\xi ^{\prime }$ are
gauge-fixing parameters. For example, $G(\phi )=\partial ^{\mu }A_{\mu }$
for the Lorenz gauge in Yang-Mills theories. Clearly, $S_{d}$ also solves
the master equation $(S_{d},S_{d})=0$ in $D$ dimensions.

If the gauge algebra does not close off shell, $\bar{S}_{d}(\Phi ,K)$ is not
linear in $K$ and $S_{d}$ is obtained from $\bar{S}_{d}$ by applying the
canonical transformation generated by 
\begin{equation}
F(\Phi ,K^{\prime })=\int \Phi ^{\alpha }K_{\alpha }^{\prime }+\Psi (\Phi ).
\label{cang}
\end{equation}

In manifestly nonanomalous theories we can solve $(S,S)=0$ in $D$ dimensions
at the regularized level. Typically, the solution coincides with (\ref{sk}).
In potentially anomalous theories, instead, we cannot achieve this goal.
There, the functional $S_{d}(\Phi ,K)$ does solve $(S_{d},S_{d})=0$ in $D$
dimensions, but is not well regularized. The most common reason is the
presence of chiral fermions. We can deform $S_{d}$ into a well-regularized
action 
\begin{equation}
S(\Phi ,K)=S_{d}+S_{\text{ev}}  \label{sdev}
\end{equation}%
by adding an evanescent part $S_{\text{ev}}$ that collects suitable
regularizing terms \cite{chiraldimreg}. The deformed action $S$ does not
solve $(S,S)=0$ in $D$ dimensions. Instead, it solves the \textit{deformed
master equation} 
\begin{equation}
(S,S)=\mathcal{O}(\varepsilon ),  \label{sleva}
\end{equation}%
where the right-hand side denotes terms that vanish for $D\rightarrow d$.

Given a generic action $S(\Phi ,K)$, the generating functionals $Z$ and $W$
of the (connected) correlation functions are defined by the formulas 
\begin{equation}
Z(J,K)=\int [\mathrm{d}\Phi ]\exp \left( iS(\Phi ,K)+i\int \Phi ^{\alpha
}J_{\alpha }\right) =\exp iW(J,K),  \label{zg}
\end{equation}%
and the generating functional $\Gamma (\Phi ,K)=W(J,K)-\int \Phi ^{\alpha
}J_{\alpha }$ of the one-particle irreducible diagrams is the Legendre
transform of $W(J,K)$ with respect to $J$. The anomaly functional is defined
as 
\begin{equation}
\mathcal{A}=(\Gamma ,\Gamma )=\langle (S,S)\rangle  \label{anom}
\end{equation}%
and collects the set of one-particle irreducible correlation functions that
contain one insertion of $(S,S)$, where $\left\langle \cdots \right\rangle $
denotes the average defined by $S$ at arbitrary $J$ . The last equality of (%
\ref{anom}) can be proved by making the change of variables 
\begin{equation}
\Phi ^{\alpha }\rightarrow \Phi ^{\alpha }+\varpi (S,\Phi ^{\alpha })=\Phi
^{\alpha }-\varpi \frac{\delta _{r}S}{\delta K_{\alpha }},  \label{chv}
\end{equation}%
in the functional integral (\ref{zg}), where $\varpi $ is a constant
anticommuting parameter. For the detailed proof, see for example the
appendices of refs. \cite{ABrenoYMLR,back}. See also appendix A.

Let us explain the meaning of formula (\ref{anom}). The functional $(S,S)$
represents the symmetry violation, so it is basically the integral of the
divergence of the gauge current $J^{\mu }$ multiplied by the ghosts:%
\begin{equation*}
(S,S)\hspace{0.01in}\sim 2\int \mathrm{d}^{D}x\hspace{0.01in}C(x)\hspace{%
0.01in}\partial _{\mu }J^{\mu }(x),
\end{equation*}%
where the sign \textquotedblleft $\sim $\textquotedblright\ means that the
right-hand side is written up to terms proportional to the field equations
and other terms that we can neglect in the present discussion. As said,
formula (\ref{anom}) collects the one-particle irreducible diagrams that
contain one insertion of $(S,S)$ and arbitrary external $\Phi $ and $K$
legs. The key diagram of this type in four dimensions is the one-loop
triangle diagram that is responsible for the well-known ABJ anomaly \cite%
{ABJ}, which arises by considering one $(S,S)$ insertion and two external
gauge field legs. Amputating those legs, we get 
\begin{equation}
\frac{1}{2}\left\langle (S,S)\hspace{0.01in}J_{\mu }(x)\hspace{0.01in}J_{\nu
}(y)\right\rangle \approx \int \mathrm{d}^{D}z\hspace{0.01in}\hspace{0.01in}%
C(z)\left\langle \partial _{\rho }J^{\rho }(z)\hspace{0.01in}J_{\mu }(x)%
\hspace{0.01in}J_{\nu }(y)\right\rangle .  \label{ssjj}
\end{equation}%
The sign \textquotedblleft $\approx $\textquotedblright\ comes from the leg
amputation and the fact that we have taken the ghosts out of the average,
because this is the only way to get nontrivial contributions to anomalies at
one loop. See ref. \cite{chiraldimreg} for the calculation of the one-loop
triangle anomaly in chiral Yang-Mills theories with formula (\ref{anom}) and
the CD regularization technique.

The Adler-Bardeen theorem is the statement that if the gauge anomalies are
trivial at one loop, there exists a class of subtraction schemes where they
vanish to all orders, that is to say%
\begin{equation}
\mathcal{A}_{R}=(\Gamma _{R},\Gamma _{R})=\left\langle
(S_{R},S_{R})\right\rangle =\mathcal{O}(\varepsilon ),  \label{ABth}
\end{equation}%
$S_{R}$ and $\Gamma _{R}$ being the renormalized action and the renormalized 
$\Gamma $ functional, respectively. The right-hand side of (\ref{ABth})
vanishes for $D\rightarrow d$, which ensures that the renormalized $\Gamma $
functional is gauge invariant in the physical limit. The AB nonanomalous
theories are those that admit subtraction schemes where (\ref{ABth}) holds.

While the AB identity (\ref{ABth}) ensures gauge invariance, it does not say
much about gauge independence, which is a different statement, namely the
property that a certain class of correlation functions (that we call
\textquotedblleft physical\textquotedblright ) do not depend on the gauge
fixing.

One way to study the gauge independence is through Ward identities. We begin
by\ recalling how those identities work in manifestly nonanomalous theories,
where the master equation $(S,S)=0$ is satisfied exactly at the regularized
level. Let $\Upsilon (\Phi )$ denote a $K$-independent, but otherwise
completely arbitrary, product of elementary and local composite fields at
distinct points. By making the change of field variables (\ref{chv}) in the
functional integral%
\begin{equation*}
\int [\mathrm{d}\Phi ]\hspace{0.01in}\Upsilon \hspace{0.01in}\mathrm{e}^{iS},
\end{equation*}%
we find%
\begin{equation}
\int [\mathrm{d}\Phi ]\hspace{0.01in}\left( S,\Upsilon \right) \hspace{0.01in%
}\mathrm{e}^{iS}=0.  \label{ncw}
\end{equation}%
We omit details of the derivation, because the proof of this formula is a
particular case of the more general proof given below. We just stress that
it is crucial to use the master equation $(S,S)=0$, which implies that $S$
is invariant under the field redefinition (\ref{chv}).

Equation (\ref{ncw}) is the usual Ward identity. For example, if we take $%
\Upsilon =\bar{C}(x)\hspace{0.01in}\hspace{0.01in}\partial ^{\mu }A_{\mu
}(y) $ and $\Upsilon =\bar{C}(x)\hspace{0.01in}\bar{\psi}(y)\hspace{0.01in}%
\psi (z)$ in QED, we can derive the well-known formula $Z_{e}Z_{A}^{1/2}=1$
that relates the renormalization constants $Z_{e}$ and $Z_{A}$ of the
electric charge and the gauge field \cite{collins2}.

In this paper, the average $\left\langle \cdots \right\rangle $ denotes the
sum of connected diagrams. For example, if $X$ and $Y$ are local
functionals, we have $\langle X\hspace{0.01in}Y\rangle =\langle X\hspace{%
0.01in}Y\rangle _{\text{nc}}-\langle X\rangle \langle Y\rangle $, where $%
\langle X\hspace{0.01in}Y\rangle _{\text{nc}}$ includes disconnected
diagrams. The subscript 0 in $\left\langle \cdots \right\rangle _{0}$ means
that the correlation functions are evaluated at $J=0$. An equivalent form of
the identity (\ref{ncw}) is 
\begin{equation}
\left\langle \left( S,\Upsilon \right) \right\rangle _{0}=0.  \label{ncw2}
\end{equation}

If we repeat the argument leading to (\ref{ncw}) without assuming $(S,S)=0$,
we get the generalized Ward identity that we consider in this paper, which
reads 
\begin{equation}
\left\langle \left( S,\Upsilon \right) \right\rangle _{0}+\frac{i}{2}%
\left\langle (S,S)\hspace{0.01in}\Upsilon \right\rangle _{0}=0.  \label{genw}
\end{equation}%
The extra term on the left-hand side of this formula is going to appear in
many other contexts and is responsible for the new effects anticipated in
the introduction.

To prove (\ref{genw}), express $\Upsilon $ as the product $%
\prod\limits_{i}X_{i}$ of $K$-independent elementary and local composite
fields $X_{i}$. Then, consider the functional integral%
\begin{equation}
\int [\mathrm{d}\Phi ]\hspace{0.01in}\mathrm{e}^{iS+\sum_{i}X_{i}\sigma
_{i}},  \label{int}
\end{equation}%
where $\sigma _{i}$ are arbitrary constants. Under the field redefinition (%
\ref{chv}), the action $S$ and the functionals $X_{i}$ transform as follows:%
\begin{equation*}
S\rightarrow S-\varpi \int \frac{\delta _{r}S}{\delta K_{\alpha }}\frac{%
\delta _{l}S}{\delta \Phi ^{\alpha }}=S+\frac{\varpi }{2}(S,S),\qquad
X_{i}\rightarrow X_{i}-\varpi \int \frac{\delta _{r}S}{\delta K_{\alpha }}%
\frac{\delta _{l}X_{i}}{\delta \Phi ^{\alpha }}=X_{i}+\varpi (S,X_{i}).
\end{equation*}%
In the last step we have used the assumption that $X_{i}$ depends only on
the fields $\Phi $. When we make the change of variables (\ref{chv}) inside (%
\ref{int}) and divide by (\ref{int}), we get 
\begin{equation*}
\frac{\int [\mathrm{d}\Phi ]\left( \sum_{j}\varpi (S,X_{j})\sigma _{j}+\frac{%
i}{2}\varpi (S,S)\right) \mathrm{e}^{iS+\sum_{i}X_{i}\sigma _{i}}}{\int [%
\mathrm{d}\Phi ]\hspace{0.01in}\mathrm{e}^{iS+\sum_{k}X_{k}\sigma _{k}}}=0.
\end{equation*}%
The left-hand side of this formula is a sum of connected diagrams.
Differentiating it once to the right with respect to each $\sigma
_{1},\ldots ,\sigma _{n}$ and setting $\sigma _{i}=0$ at the end, we project
onto the diagrams that have one external $\sigma _{i}$ leg for each $i$. So
doing, we get precisely formula (\ref{genw}).

When the local functionals $X_{i}$ of the product $\Upsilon
=\prod\limits_{i}X_{i}$ depend on both $\Phi $ and $K$, and the sources $J$
are not set to zero, the generalized Ward identities can be worked out from
formula (\ref{anom}), by deforming the action $S$ into $S+\sum_{i}X_{i}%
\sigma _{i}$, where $\sigma _{i}$ are constants, and taking the first order
in all $\sigma _{i}$s.

In particular, if $\Upsilon $ is equal to a local functional $X$, it is easy
to show that when the action $S$ is deformed into $S+X\sigma $, where $%
\sigma $ is a constant, the $\Gamma $ functional deforms into $\Gamma
+\langle X\rangle \sigma +\mathcal{O}(\sigma ^{2})$, while the average $%
\langle Y\rangle $ of a local functional $Y$\ deforms into $\langle Y\rangle
+i\langle \hspace{0.01in}YX\rangle _{\Gamma }\sigma +\mathcal{O}(\sigma
^{2}) $, where $\langle \prod\limits_{i}A_{i}\rangle _{\Gamma }$ denotes the
set of one-particle irreducible diagrams that contain one $A_{i}$ insertion
for each $i$, $A_{i}$ being local functionals (details are given in appendix
A). Expanding $(\Gamma ,\Gamma )=\langle (S,S)\rangle $ in powers of $\sigma 
$ and taking the first order of the expansion, we obtain the identity \cite%
{back} 
\begin{equation}
\left\langle (S,X)\right\rangle +\frac{i}{2}\langle (S,S)\hspace{0.01in}%
X\rangle _{\Gamma }=(\Gamma ,\langle X\rangle ).  \label{genwa}
\end{equation}%
Both sides of (\ref{genwa}) are viewed as functionals of $\Phi $ and $K$
(rather than functionals of $J$ and $K$). Note that, in particular, $\langle
X\rangle =\langle X\rangle _{\Gamma }$.

Repeating the derivation for $\Upsilon =XY$, where $X$ and $Y$ are both
local functionals, we get the identity%
\begin{equation}
\left\langle (S,XY)\right\rangle _{\Gamma }+\frac{i}{2}\langle (S,S)\hspace{%
0.01in}XY\rangle _{\Gamma }=(\Gamma ,\langle XY\rangle _{\Gamma
})-i(-1)^{\varepsilon _{X}}(\langle X\rangle ,\langle Y\rangle
)+i(-1)^{\varepsilon _{X}}\left\langle (X,Y)\right\rangle ,  \label{genwa2}
\end{equation}%
where $\varepsilon _{X}$ denotes the statistics of the functional $X$ (which
is 0 if $X$ is bosonic, 1 if it is fermionic). When $\Upsilon $ is the
product of more local functionals, we can proceed similarly.

An important application of the generalized Ward identities is the
derivation of the \textit{equations of gauge dependence}, which tell us how
the generating functional $\Gamma $ depends on the gauge parameters. We
first recall such equations in manifestly nonanomalous theories and then
switch to AB nonanomalous theories.

In manifestly nonanomalous theories $(S,S)=0$ in $D$ dimensions and $S_{%
\text{ev}}=0$, $S=S_{d}$. The functional $\Gamma $ satisfies the equation%
\begin{equation}
\frac{\partial \Gamma }{\partial \xi }=\left\langle \frac{\partial S}{%
\partial \xi }\right\rangle =\left\langle (S,\Psi _{\xi })\right\rangle
=(\Gamma ,\left\langle \Psi _{\xi }\right\rangle ),  \label{gman}
\end{equation}%
where $\xi $ is any gauge-fixing parameter and $\Psi _{\xi }=\partial \Psi
/\partial \xi $ is the $\xi $-derivative of the gauge fermion $\Psi $. The
first equality is obvious. The second equality follows from formula (\ref{sk}%
). Indeed, recalling that the parameters $\xi $ are contained only in $\Psi $%
, we have $\partial S/\partial \xi =(S_{K},\Psi _{\xi })=(S,\Psi _{\xi })$.
The third equality follows from formula (\ref{genwa}).

More generally, if $\theta $ denotes any gauge parameter, introduced by a
canonical transformation generated by (\ref{fx}), we find 
\begin{equation}
\frac{\partial \Gamma }{\partial \theta }=\left\langle \frac{\partial S}{%
\partial \theta }\right\rangle =\langle (S,\tilde{Q}_{\theta })\rangle
=(\Gamma ,\langle \tilde{Q}_{\theta }\rangle ),  \label{gman1}
\end{equation}%
where $\tilde{Q}_{\theta }$ is the derivative $F(\Phi ,K^{\prime },\theta )$
with respect to $\theta $, reexpressed as a functional of $\Phi $ and $K$.

Equations (\ref{gman}) can be renormalized and integrated (see \cite{back}
and appendix C). The result is that the $\xi $ dependence can be absorbed
into a canonical transformation on $\Gamma $. Therefore, the contributions
due to the right-hand side of (\ref{gman}), which are in general
nonvanishing, do not affect the physical quantities, for example the
S-matrix elements. See subsection \ref{s43} for details.

In AB\ nonanomalous theories the equations of gauge dependence are corrected
by an extra term, which corresponds to the extra term of (\ref{genw}).
Formula (\ref{gman1}) turns into \cite{back} 
\begin{equation}
\frac{\partial \Gamma }{\partial \theta }=\left\langle \frac{\partial S}{%
\partial \theta }\right\rangle =\langle (S,\tilde{Q}_{\theta })\rangle
=(\Gamma ,\langle \tilde{Q}_{\theta }\rangle )-\frac{i}{2}\langle (S,S)%
\tilde{Q}_{\theta }\rangle _{\Gamma }.  \label{gman2}
\end{equation}%
Assuming that the primes denote the $\theta $-independent quantities, the
second equality of (\ref{gman2}) follows from formula (\ref{bu}) recalled in
appendix A, since $\partial S^{\prime }/\partial \theta =0$. The last
equality of (\ref{gman2}) follows from formula (\ref{genwa}).

The identities (\ref{genw}), (\ref{genwa}), (\ref{genwa2}) and (\ref{gman2})
are so general that they also hold in truly anomalous theories. However,
their most interesting applications are to AB nonanomalous theories, which
are the main focus of this paper.

In the next sections we are going to renormalize the equations (\ref{gman2})
and integrate their renormalized versions. The nontrivial part of this task
is to work out the effects of the last term of \ formula (\ref{gman2}). The
result is that the $\theta $ dependence can be absorbed into a canonical
transformation on the renormalized $\Gamma $ functional $\Gamma _{R}$,
provided that the finite local counterterms are appropriately fine-tuned.

We stress again that gauge invariance, which is expressed by formula (\ref%
{ABth}), does not imply gauge independence in an obvious way. However, in
this paper we prove that ultimately it does. Gauge independence allows us to
prove the perturbative unitarity of the theory (see subsection \ref{s44}).

Before concluding this section, we make some remarks to emphasize the role
played by the evanescent terms $\mathcal{O}(\varepsilon )$ in our
discussion. With respect to the limit\ $D\rightarrow d$ we can distinguish
divergent, nonevanescent and evanescent terms. A contribution is called
\textquotedblleft nonevanescent\textquotedblright\ if it has a regular limit
for $D\rightarrow d$ and coincides with the value of that limit. In the
(ordinary, as well as chiral) dimensional regularization the evanescences\
can be of two types: \textit{formal} or \textit{analytic}. Analytically
evanescent terms are those that factorize at least one $\varepsilon $, such
as $\varepsilon F_{\bar{\mu}\bar{\nu}}F^{\bar{\mu}\bar{\nu}}$, $\varepsilon 
\bar{\psi}_{L}ie_{\bar{a}}^{\bar{\mu}}\gamma ^{\bar{a}}D_{\bar{\mu}}\psi
_{L} $, etc., where $\psi _{L}$ is a left-handed fermion. Formally
evanescent terms are those that formally disappear when $D\rightarrow d$,
although they do not factorize powers of $\varepsilon $, such as $\psi
_{L}^{T}\hat{\partial}^{2}\psi _{L}$. The divergences are poles in $%
\varepsilon $, and can multiply either nonevanescent terms or formally
evanescent terms. In the latter case they are called \textit{divergent
evanescences}. An example is $\psi _{L}^{T}\hat{\partial}^{2}\psi
_{L}/\varepsilon $. It is convenient to subtract away the divergent
evanescences like any other divergences.

In most derivations it is necessary to extract the divergent parts of
functionals and antiparentheses of functionals. We have to take some
precautions to ensure that this operation can safely cross the
antiparentheses, so that for example $(S,X)_{\text{div}}=(S,X_{\text{div}})$%
. The first thing to do is define the classical action (\ref{sdev}) so that
it does not contain analytically evanescent terms, but only nonevanescent
and formally evanescent terms, multiplied by $\varepsilon $-independent
coefficients. In this way, $S$ does not contain dangerous $\varepsilon $
factors that could simplify the divergences of $X$ inside $(S,X)$. For the
same reason, it is convenient to use the chiral dimensional regularization
of \cite{chiraldimreg}, instead of the ordinary dimensional regularization.
In particular, we must use the CD regularization when the theory in not
power-counting renormalizable. So doing, we avoid a number of ambiguities
that would complicate our operations. For details on this subject, see refs. 
\cite{ABrenoYMLR,chiraldimreg}.

\section{The theorem of gauge dependence}

\label{s8}

\setcounter{equation}{0}

Consider a general gauge theory with action $S(\Phi ,K,\omega )$, where $%
\omega $ denotes its parameters. Let $S_{R}$ denote the renormalized action
and $\Gamma _{R}$ the renormalized $\Gamma $ functional. Assume that the
theory is AB nonanomalous, i.e.%
\begin{equation}
(\Gamma _{R},\Gamma _{R})=\mathcal{O}(\varepsilon ).  \label{abba}
\end{equation}

For the purposes of this section, we do not need to make other assumptions.
The gauge algebra may be irreducible or reducible, and close off shell or on
shell. The theory may be renormalizable or nonrenormalizable, perturbatively
unitary or not. In particular, it may contain higher-derivative fields. The
action $S$ does not need to satisfy special cohomological properties. We can
also include local composite fields $\mathcal{O}^{I}(x)$, in renormalizable
and nonrenormalizable theories, by coupling them to external sources $%
L_{I}(x)$ and appropriately extending the actions $S_{c}$, $\bar{S}_{d}$, $%
S_{d}$ and $S$. In the arguments that follow, the dependence on such types
of external sources is not made explicit. However, we understand that it may
be there, whenever necessary.

Consider a canonical transformation $\Phi ,K\rightarrow \Phi ^{\prime
},K^{\prime }$ with generating functional 
\begin{equation}
F(\Phi ,K^{\prime },\theta )=\int \Phi ^{\alpha }K_{\alpha }^{\prime
}+Q(\Phi ,K^{\prime },\theta ),  \label{trasfa}
\end{equation}%
where $Q=\mathcal{O}(\theta )$ is a local functional. Let $S_{\theta }$
denote the action obtained by applying (\ref{trasfa}) to $S$, $S_{R\theta }$
the renormalized version of $S_{\theta }$ and $\Gamma _{R\theta }$ the
renormalized $\Gamma $ functional associated with $S_{R\theta }$. We assume,
for simplicity, that $Q$ does not contain analytically evanescent
contributions.

We work out how $\Gamma _{R}$ and the identity (\ref{abba}) change when we
make the transformation (\ref{trasfa}) on $S$. To reach $S_{\theta }$ from $%
S $, it is useful to embed the theory into a more general theory, by
considering the extended action 
\begin{equation}
\Sigma (\Phi ,K,\omega ,\hbar \tau )\equiv S(\Phi ,K,\omega )+\sum_{i}\hbar
\tau _{i}\mathcal{H}_{i}(\Phi ,K),  \label{estensi}
\end{equation}%
where $\tau _{i}$ are arbitrary parameters and $\{\mathcal{H}_{i}\}$ is a
basis of local functionals of $\Phi $ and $K$. Specifically, the $\mathcal{H}%
_{i}$ are integrals of local monomials constructed with the fields, the
sources and their derivatives. They can be restricted by demanding that they
be invariant under the nonanomalous symmetries of the theory. However, they
are not restricted by gauge invariance, or power counting. To simplify a
number of formulas, we include duplicates of the terms that are already
present in $\Sigma $, multiplied by new independent parameters $\hbar \tau
_{i}$. The difference $\Sigma -S$ is made of $\mathcal{O}(\hbar )$-terms and
is also assumed to contain evanescent terms (including those that are
already present in $S$). Basically, $\Sigma -S$ parametrizes the
arbitrariness of the subtraction scheme.\ We denote the $\Gamma $ functional
calculated with the action $\Sigma $ by $\Omega (\Phi ,K,\omega ,\hbar \tau
) $.

Now, we renormalize $\Sigma $. We denote its renormalized action by $\Sigma
_{R}$ and the $\Gamma $ functional associated with $\Sigma _{R}$ by $\Omega
_{R}$. We can imagine, for a moment, that we replace each $\hbar \tau _{i}$
with an ordinary parameter $\rho _{i}$ of order zero in $\hbar $. In that
case, the construction of $\Sigma _{R}$ is straightforward, since every
divergence can be subtracted by means of $\rho _{i}$ redefinitions. At a
second stage, we raise the order of the parameters $\rho _{i}$ by restoring $%
\hbar \tau _{i}$ in their places. The consistency of this operation is
justified by the arguments that follow.

We organize the renormalization of $\Sigma $ so that $\Sigma _{R}$ coincides
with $S_{R}$ when the parameters $\tau _{i}$ are equal to suitable finite
functions $\tau _{i}^{\ast }(\omega )$, which identify the subtraction
scheme where formula (\ref{abba}) holds: 
\begin{equation}
\Sigma _{R}(\Phi ,K,\omega ,\hbar \tau ^{\ast })=S_{R}(\Phi ,K,\omega
),\qquad \Omega _{R}(\Phi ,K,\omega ,\hbar \tau ^{\ast })=\Gamma _{R}(\Phi
,K,\omega ).  \label{billo}
\end{equation}%
At arbitrary $\tau $, the action $\Sigma _{R}$ can be viewed as an extended
renormalization of $S$, which includes the most general subtraction scheme.
We say that $\Sigma _{R}$ is the\textit{\ arbitrary renormalization} of $S$.
When we set $\tau _{i}=\tau _{i}^{\ast }$ we specialize the subtraction
scheme to the one used for $S_{R}$, which, by assumption (\ref{abba}),
preserves gauge invariance to all orders.

Since it is consistent to set $\tau _{i}\equiv \tau _{i}^{\ast }$, it is
also consistent to set $\tau _{i}=\tau _{i}^{\ast }+\hbar ^{n}\tilde{\nu}%
_{n+1\hspace{0.01in}i}$, $n\geqslant 0$, for arbitrary new parameters $%
\tilde{\nu}_{n+1\hspace{0.01in}i}$. By this we mean that the renormalization
of each $\tilde{\nu}_{n+1\hspace{0.01in}i}$ remains analytic in $\hbar $. We
can better explain this fact by noting that the renormalizations of the
differences $\delta _{i}\equiv \tau _{i}-\tau _{i}^{\ast }$ vanish at $%
\delta _{j}=0$, so they must be proportional to $\delta _{j}$. Thus, if we
replace $\delta _{i}$ by $\hbar ^{n}\tilde{\nu}_{n+1\hspace{0.01in}i}$, $n>1$%
, the renormalizations of $\tilde{\nu}_{n+1\hspace{0.01in}i}$ remain
analytic in $\hbar $. These remarks illustrate a trick that we use in the
recursive proof given below. Precisely, at each step we raise the $\hbar $
order of certain residual parameters by one unit, till we make those
parameters disappear, and show that we can do this while preserving the
analyticity in $\hbar $.

The definition (\ref{estensi}) understands that the difference $\Sigma -S$
starts from $\mathcal{O}(\hbar )$. Indeed, we do not want to modify the
classical action, but just parametrize the arbitrariness of the subtraction
scheme. The reason why we move to the more general theory $\Sigma $ is that
if we want to cancel the anomalies after the canonical transformation, we
generically need to re-fine-tune all sorts of finite, local terms, including
the gauge noninvariant ones.

As said, $S_{\theta }(\Phi ,K,\omega ,\theta )$ denotes the action obtained
by applying (\ref{trasfa})\ to $S(\Phi ,K,\omega )$. Let $\Sigma _{\theta
}(\Phi ,K,\omega ,\hbar \tau ,\theta )$ denote the action obtained by
applying (\ref{trasfa})\ to $\Sigma $. We obviously have $\Sigma _{\theta
}=S_{\theta }+\mathcal{O}(\hbar )$. We denote the renormalized version of $%
\Sigma _{\theta }$ by $\Sigma _{R\theta }(\Phi ,K,\omega ,\hbar \tau ,\theta
)$. Since $\Sigma _{R\theta }=\Sigma _{\theta }+\mathcal{O}(\hbar
)=S_{\theta }+\mathcal{O}(\hbar )$, $\Sigma _{R\theta }$ can be viewed as
the arbitrary renormalization of $S_{\theta }$. Note that $\Sigma _{\theta }$
is not gauge invariant, so its renormalization is not subject to particular
restrictions, aside from the continuity condition $\Sigma _{R\theta }(\Phi
,K,\omega ,\hbar \tau ,\theta )=\Sigma _{R}(\Phi ,K,\omega ,\hbar \tau )+%
\mathcal{O}(\theta )$. We denote the $\Gamma $ functional associated with $%
\Sigma _{R\theta }$ by $\Omega _{R\theta }$.

Finally, consider the local functional $Q(\Phi ,K^{\prime })$ defined by the
canonical transformation (\ref{trasfa}), and define $Q_{\theta }(\Phi
,K^{\prime })=\partial Q(\Phi ,K^{\prime })/\partial \theta $ and $\tilde{Q}%
_{\theta }(\Phi ,K)=Q_{\theta }(\Phi ,K^{\prime }(\Phi ,K))$. Let $\mathcal{%
\tilde{Q}}_{R\theta }$ denote the renormalized version of $\tilde{Q}_{\theta
}(\Phi ,K)$ at generic $\tau $.

We prove that

\begin{theorem}
there exist finite functions $\tau _{j}^{\ast \prime }(\omega ,\theta )=%
\mathcal{O}(\theta )$, such that, defining $\tilde{\tau}_{j}^{\ast }(\omega
,\theta )=\tau _{j}^{\ast }(\omega )+\tau _{j}^{\prime \ast }(\omega ,\theta
)$, the action 
\begin{equation}
S_{R\theta }(\Phi ,K,\omega ,\theta )\equiv \Sigma _{R\theta }(\Phi
,K,\omega ,\hbar \tilde{\tau}_{j}^{\ast }(\omega ,\theta ),\theta )
\label{th1}
\end{equation}%
gives a $\Gamma $ functional $\Gamma _{R\theta }$ that satisfies the
identities%
\begin{eqnarray}
(\Gamma _{R\theta },\Gamma _{R\theta }) &=&\mathcal{O}(\varepsilon ),
\label{resi1} \\
\frac{\partial \Gamma _{R\theta }}{\partial \theta }-(\Gamma _{R\theta
},\langle \tilde{Q}_{R\theta }\rangle ) &=&\mathcal{O}(\varepsilon ),
\label{resi2}
\end{eqnarray}%
for arbitrary $\theta $, where $\tilde{Q}_{R\theta }$ denotes the functional 
$\mathcal{\tilde{Q}}_{R\theta }$ calculated at $\hbar \tau _{i}=$ $\hbar 
\tilde{\tau}_{i}^{\ast }$.
\end{theorem}

Note that formula (\ref{th1}) ensures that $S_{R\theta }$ also satisfies the
continuity condition $S_{R\theta }(\Phi ,K,\omega ,\theta )=S_{R}(\Phi
,K,\omega )+\mathcal{O}(\theta )$. In fact, all the operations we make
preserve the continuity in $\theta $.

For clarity, it is useful to summarize the definitions given so far in a
table:%
\begin{equation*}
\begin{tabular}{lllllll}
&  &  &  & $S_{R}$ & $\overset{\Gamma }{\longrightarrow }$ & $\Gamma _{R}$
\\ 
&  &  &  & $\uparrow _{\hbar \tau ^{\ast }}$ &  & $\uparrow _{\hbar \tau
^{\ast }}$ \\ 
$S$ & $\overset{\hbar \tau }{\longrightarrow }$ & $\Sigma $ & $\overset{R}{%
\longrightarrow }$ & $\Sigma _{R}$ & $\overset{\Gamma }{\longrightarrow }$ & 
$\Omega _{R}$ \\ 
$\downarrow ^{\theta }$ &  & $\downarrow ^{\theta }$ &  & $\downarrow
^{\theta }$ &  & $\downarrow ^{\theta }$ \\ 
$S_{\theta }$ & $\overset{\hbar \tau }{\longrightarrow }$ & $\Sigma _{\theta
}$ & $\overset{R}{\longrightarrow }$ & $\Sigma _{R\theta }$ & $\overset{%
\Gamma }{\longrightarrow }$ & $\Omega _{R\theta }$ \\ 
&  &  &  & $\downarrow ^{\hbar \tilde{\tau}^{\ast }}$ &  & $\downarrow
^{\hbar \tilde{\tau}^{\ast }}$ \\ 
&  &  &  & $S_{R\theta }$ & $\overset{\Gamma }{\longrightarrow }$ & $\Gamma
_{R\theta }$%
\end{tabular}%
\end{equation*}

\subsection{The equations of gauge dependence}

\label{s81}

If we apply the identity (\ref{gx}) of appendix A to the renormalized action 
$\Sigma _{R\theta }$ and the renormalized $\Gamma $ functional $\Omega
_{R\theta }$, with $X=\mathcal{\tilde{Q}}_{R\theta }$, we obtain 
\begin{equation}
\frac{\partial \Omega _{R\theta }}{\partial \theta }-(\Omega _{R\theta
},\langle \mathcal{\tilde{Q}}_{R\theta }\rangle )=\left\langle \frac{%
\partial \Sigma _{R\theta }}{\partial \theta }-(\Sigma _{R\theta },\mathcal{%
\tilde{Q}}_{R\theta })-\frac{i}{2}(\Sigma _{R\theta },\Sigma _{R\theta })%
\mathcal{\tilde{Q}}_{R\theta }\right\rangle _{\Gamma }.  \label{proveg}
\end{equation}%
It is convenient to organize this formula in the form 
\begin{equation}
\frac{\partial \Omega _{R\theta }}{\partial \theta }=(\Omega _{R\theta
},\langle \mathcal{\tilde{Q}}_{R\theta }\rangle )+\left\langle \mathcal{Y}%
_{R\theta }\right\rangle _{\Gamma },  \label{pri}
\end{equation}%
where 
\begin{equation}
\mathcal{Y}_{R\theta }\equiv -\frac{i}{2}(\Sigma _{R\theta },\Sigma
_{R\theta })\mathcal{\tilde{Q}}_{R\theta }+\frac{\partial \Sigma _{R\theta }%
}{\partial \theta }-(\Sigma _{R\theta },\mathcal{\tilde{Q}}_{R\theta }).
\label{prin}
\end{equation}

If the right-hand side of formula (\ref{pri}) contained no $\left\langle 
\mathcal{Y}_{R\theta }\right\rangle _{\Gamma }$ (which happens, for example,
in manifestly nonanomalous theories) or we knew that $\left\langle \mathcal{Y%
}_{R\theta }\right\rangle _{\Gamma }$ is for some reason equal to $\mathcal{%
O(}\varepsilon )$, the solution of our problem would be straightforward.
Formula (\ref{pri}) would turn into a much simpler equation, which is
integrated in ref. \cite{back} and in appendix C. The result would be that
the entire $\theta $ dependence of $\Omega _{R\theta }$ can be absorbed into
a convergent canonical transformation acting on $\Omega _{R}$, up to $%
\mathcal{O(}\varepsilon )$. Moreover, there would be no reason to keep $\tau 
$ generic. More simply, we could just work with $\tau =\tau ^{\ast }$ from
the start. Then, formula (\ref{pri}) would give (\ref{resi2}). Integrating (%
\ref{resi2}) with the procedure of appendix C, we would find a convergent
canonical transformation that turns $\Gamma _{R}$ into $\Gamma _{R\theta }$,
again up to $\mathcal{O(}\varepsilon )$. That canonical transformation would
also turn formula (\ref{abba}) directly into (\ref{resi1}), since the
right-hand side would remain evanescent.

Unfortunately, $\left\langle \mathcal{Y}_{R\theta }\right\rangle _{\Gamma }$
is there, because the theory we are considering is potentially anomalous, so
we must study the effects of such an extra term. To achieve this goal, a few
facts need to be noticed.

($i$) By construction, $\Omega _{R\theta }$ and $\langle \mathcal{\tilde{Q}}%
_{R\theta }\rangle $ are convergent.

($ii$) The local functional $(\Sigma _{R\theta },\Sigma _{R\theta })$ is
already renormalized. Indeed, formula (\ref{anom}) tells us that $\langle
(\Sigma _{R\theta },\Sigma _{R\theta })\rangle =(\Omega _{R\theta },\Omega
_{R\theta })$, which is convergent. Since $\Sigma _{R\theta }=S_{\theta }+%
\mathcal{O}(\hbar )$, we can say that $(\Sigma _{R\theta },\Sigma _{R\theta
})$ is the arbitrary renormalization of $(S_{\theta },S_{\theta })$.

($iii$) By points ($i$)\ and ($ii$), all the subdiagrams of the diagrams
that contribute to the average $\langle (\Sigma _{R\theta },\Sigma _{R\theta
})\hspace{0.01in}\mathcal{\tilde{Q}}_{R\theta }\rangle _{\Gamma }$ are
already renormalized, except those that contain \textit{both} insertions of $%
(\Sigma _{R\theta },\Sigma _{R\theta })$ and $\mathcal{\tilde{Q}}_{R\theta }$%
.

($iv$) The object $\mathcal{Y}_{R\theta }$ is a bit peculiar, because at the
tree level it is equal to 
\begin{equation}
Y_{\theta }\equiv -\frac{i}{2}(S_{\theta },S_{\theta })\hspace{0.01in}%
\hspace{0.01in}\tilde{Q}_{\theta }.  \label{yx}
\end{equation}%
The reason why the last two terms of (\ref{prin}) do not contribute at $%
\hbar =0$ is that 
\begin{equation}
\frac{\partial S_{\theta }}{\partial \theta }-(S_{\theta },\tilde{Q}_{\theta
})=\frac{\partial S}{\partial \theta }=0,  \label{steq}
\end{equation}%
which follows from formula (\ref{bu}), if we understand that the primes
denote the fields and the sources before the transformation, i.e. write $%
S=S(\Phi ^{\prime },K^{\prime })$ and $S_{\theta }=S_{\theta }(\Phi ,K)$. We
see that $Y_{\theta }$ is the product of two local functionals. We call $%
Y_{\theta }$ a local \textit{bifunctional}. We extend the definition of
local bifunctional to any expression of the form%
\begin{equation}
\mathcal{B}=\sum_{i}A_{i}B_{i}+C  \label{locbi}
\end{equation}%
where $A_{i}$, $B_{i}$ and $C$ are local functionals. An evanescent local
bifunctional is a local bifunctional (\ref{locbi}) where $C$ and $A_{i}$ (or 
$B_{i}$) are evanescent.

Now, $(S,S)$ is an evanescent local functional, by formula (\ref{sleva}),
and $S_{\theta }$ is obtained from $S$ by means of a finite canonical
transformation, which preserves the antiparentheses and maps $\mathcal{O}%
(\varepsilon )$ into $\mathcal{O}(\varepsilon )$. Thus, $(S_{\theta
},S_{\theta })$ is also evanescent, and $\mathcal{Y}_{R\theta }$ is an
evanescent local bifunctional. Actually,

($v$) $\mathcal{Y}_{R\theta }$ is a renormalized evanescent local
bifunctional, since formula (\ref{pri}) implies that $\left\langle \mathcal{Y%
}_{R\theta }\right\rangle _{\Gamma }$ is convergent.

The procedure to renormalize a local bifunctional is explained in appendix
B. There, it is also shown how to renormalize an evanescent local
bifunctional $\mathcal{E}$ in such a way that $\langle \mathcal{E}_{R}%
\mathcal{\rangle }_{\Gamma }=\mathcal{O}(\varepsilon )$. To describe what
happens order by order in the perturbative expansion, consider for
simplicity an evanescent local bifunctional of the form $\mathcal{E}=EB+F$
where $E$ and $F$ are evanescent local functionals. Let $E_{n}$ and $B_{n}$
denote the functionals $E$ and $B$ renormalized up to and including $n$
loops, and inductively assume that $E_{n}$ satisfies $\langle E_{n}\rangle =%
\mathcal{O}(\varepsilon )+\mathcal{O}(\hbar ^{n+1})$. Also assume that $%
F_{n} $ is a local functional such that $\langle \mathcal{E}_{n}\mathcal{%
\rangle }_{\Gamma }=\mathcal{O}(\varepsilon )+\mathcal{O}(\hbar ^{n+1})$,
where $\mathcal{E}_{n}=E_{n}B_{n}+F_{n}$. Then, the $\mathcal{O}(\hbar
^{n+1})$ contributions to $\langle \mathcal{E}_{n}\mathcal{\rangle }_{\Gamma
}$ are the sum of a local divergent part, a local nonevanescent part and a
generically nonlocal evanescent part. If $B_{n+1}$ is the functional $B$
renormalized up to and including $n+1$ loops, there exist local functionals $%
E_{n+1}$ and $F_{n+1}$ such that $\langle E_{n+1}\rangle =\mathcal{O}%
(\varepsilon )+\mathcal{O}(\hbar ^{n+2})$ and $\langle \mathcal{E}_{n+1}%
\mathcal{\rangle }_{\Gamma }=\mathcal{O}(\varepsilon )+\mathcal{O}(\hbar
^{n+2})$, where $\mathcal{E}_{n+1}=E_{n+1}B_{n+1}+F_{n+1}$. The subtraction
can be iterated in $n$ to obtain $\langle E_{R}\rangle =\mathcal{O}%
(\varepsilon )$ and $\langle \mathcal{E}_{R}\mathcal{\rangle }_{\Gamma }=%
\mathcal{O}(\varepsilon )$, where $E_{R}=E_{\infty }$ and $\mathcal{E}_{R}=%
\mathcal{E}_{\infty }$.

Although $\mathcal{Y}_{R\theta }$ is renormalized, it does not satisfy $%
\left\langle \mathcal{Y}_{R\theta }\right\rangle _{\Gamma }=\mathcal{O}%
(\varepsilon )$, as far as we know. However, we will obtain $\left\langle 
\mathcal{Y}_{R\theta }\right\rangle _{\Gamma }=\mathcal{O}(\varepsilon )$ by
identifying the functions $\tau _{j}^{\ast \prime }(\omega ,\theta )$ and
setting $\tau _{i}=\tau _{i}^{\ast }+\tau _{i}^{\prime \ast }$.

To prove (\ref{th1}), (\ref{resi1}) and (\ref{resi2}), we proceed by
induction. Let $\nu _{nj}$ denote free parameters of order $\hbar ^{n}$. The
first inductive assumption is that

($a_{n}$) there exist finite functions $\mu _{nj}(\omega ,\nu _{n+1\hspace{%
0.01in}k},\theta )=\mathcal{O}(\theta )\mathcal{O}(\hbar )$, such that the
action 
\begin{equation}
\Sigma _{n}(\Phi ,K,\omega ,\nu _{n+1\hspace{0.01in}j},\theta )\equiv \Sigma
_{R\theta }(\Phi ,K,\omega ,\hbar \tau _{j}^{\ast }+\mu _{nj}+\nu _{n+1%
\hspace{0.01in}j},\theta )  \label{sign}
\end{equation}%
gives a $\Gamma $ functional $\Omega _{n}$ that satisfies 
\begin{equation}
(\Omega _{n},\Omega _{n})=\left\langle (\Sigma _{n},\Sigma _{n})\hspace{%
0.01in}\right\rangle _{n}=\mathcal{O}(\varepsilon )+\mathcal{O}(\hbar
^{n+1}),  \label{sign2}
\end{equation}%
where $\left\langle \cdots \right\rangle _{n}$ denotes the average
calculated with the action $\Sigma _{n}$.

Now, define%
\begin{eqnarray}
\tilde{Q}_{n} &\equiv &\mathcal{\tilde{Q}}_{R\theta }(\Phi ,K,\omega ,\hbar
\tau _{j}^{\ast }+\mu _{nj}+\nu _{n+1\hspace{0.01in}j},\theta ),  \notag \\
Y_{n} &\equiv &-\frac{i}{2}(\Sigma _{n},\Sigma _{n})\hspace{0.01in}\tilde{Q}%
_{n}+\frac{\partial \Sigma _{n}}{\partial \theta }-(\Sigma _{n},\tilde{Q}%
_{n}).  \label{yn1}
\end{eqnarray}%
Applying formula (\ref{gx}) to the action $\Sigma _{n}$ and its $\Gamma $
functional $\Omega _{n}$, with $X=\tilde{Q}_{n}$, we obtain%
\begin{equation}
\frac{\partial \Omega _{n}}{\partial \theta }=(\Omega _{n},\langle \tilde{Q}%
_{n}\rangle _{n})+\left\langle Y_{n}\right\rangle _{n\Gamma },  \label{io}
\end{equation}%
where $\left\langle \cdots \right\rangle _{n\Gamma }$ denotes the
one-particle irreducible diagrams of the average $\left\langle \cdots
\right\rangle _{n}$. The second inductive assumption is that

($b_{n}$)%
\begin{equation}
\left\langle Y_{n}\right\rangle _{n\Gamma }=\mathcal{O}(\varepsilon )+%
\mathcal{O}(\hbar ^{n+1}).  \label{ion}
\end{equation}

Statement ($a_{0}$) is true with $\mu _{0j}=0$, because $\Sigma
_{0}=S_{\theta }+\mathcal{O}(\hbar )$ and $(S_{\theta },S_{\theta })$ is
evanescent, so $\left\langle (\Sigma _{0},\Sigma _{0})\hspace{0.01in}%
\right\rangle _{0}=\mathcal{O}(\varepsilon )+\mathcal{O}(\hbar )$. Statement
($b_{0}$) is also true, because $\tilde{Q}_{0}=\tilde{Q}_{\theta }+\mathcal{O%
}(\hbar )$, so $Y_{0}=Y_{\theta }+\mathcal{O}(\hbar )$.

\subsection{Inductive proof}

\label{indproof}

Assume that ($a_{n}$) and ($b_{n}$) hold. Then, the averages $\left\langle
(\Sigma _{n},\Sigma _{n})\hspace{0.01in}\right\rangle _{n}$ and $%
\left\langle Y_{n}\right\rangle _{n\Gamma }$ are evanescent up to and
including $n$ loops. The arguments of appendix B ensure that the $(n+1)$%
-loop contributions $Y_{n}^{(n+1)}$ to $\left\langle Y_{n}\right\rangle
_{n\Gamma }$, which are convergent by formula (\ref{io}), are the sum of a
local nonevanescent part $Y_{n\text{nonev}}^{(n+1)}$ plus a generically
nonlocal evanescent part. We have 
\begin{equation}
\left\langle Y_{n}\right\rangle _{n\Gamma }=\mathcal{O}(\varepsilon )+Y_{n%
\text{nonev}}^{(n+1)}+\mathcal{O}(\hbar ^{n+2}).  \label{ynn}
\end{equation}

We can write an explicit expression for $Y_{n\text{nonev}}^{(n+1)}$. Recall,
from formula (\ref{estensi}), that the derivatives $\partial \Sigma
/\partial (\hbar \tau _{j})$ form a basis for the local functionals of $\Phi 
$ and $K$. Obviously, so do the derivatives $\partial \Sigma _{\theta
}/\partial (\hbar \tau _{j})\equiv \mathcal{H}_{j\theta }$. Up to higher
orders in $\hbar $, the derivatives $\partial \Sigma _{R\theta }/\partial
(\hbar \tau _{j})=\mathcal{H}_{j\theta }+\mathcal{O}(\hbar )$ are also a
basis, as well as the derivatives $\partial \Sigma _{n}/\partial \nu _{n+1%
\hspace{0.01in}j}$. Thus, there exist finite order-$\hbar ^{n+1}$ functions $%
\sigma _{j}^{(n)}$, which depend analytically on $\omega $, $\nu _{n+1%
\hspace{0.01in}k}$ and $\theta $, such that 
\begin{equation}
Y_{n\text{nonev}}^{(n+1)}=\sum_{j}\sigma _{j}^{(n)}\frac{\partial \Sigma _{n}%
}{\partial \nu _{n+1\hspace{0.01in}j}}+\mathcal{O}(\hbar ^{n+2}).
\label{bu2}
\end{equation}

Now, define 
\begin{equation}
\mathcal{Y}_{n+1}=Y_{n}-\sum_{j}\sigma _{j}^{(n)}\frac{\partial \Sigma _{n}}{%
\partial \nu _{n+1\hspace{0.01in}j}}.  \label{ynp}
\end{equation}%
Taking the average of both sides, and using (\ref{basic1}), we get 
\begin{equation}
\left\langle \mathcal{Y}_{n+1}\right\rangle _{n\Gamma }=\left\langle
Y_{n}\right\rangle _{n\Gamma }-\sum_{j}\sigma _{j}^{(n)}\frac{\partial
\Omega _{n}}{\partial \nu _{n+1\hspace{0.01in}j}}.  \label{ij}
\end{equation}%
Using (\ref{ynn}) and (\ref{bu2}), we obtain 
\begin{equation}
\left\langle \mathcal{Y}_{n+1}\right\rangle _{n\Gamma }=Y_{n\text{nonev}%
}^{(n+1)}-Y_{n\text{nonev}}^{(n+1)}+\mathcal{O}(\varepsilon )+\mathcal{O}%
(\hbar ^{n+2})=\mathcal{O}(\varepsilon )+\mathcal{O}(\hbar ^{n+2}).
\label{yn2}
\end{equation}%
Using (\ref{ij}) inside (\ref{io}), we also find 
\begin{equation}
\frac{\partial \Omega _{n}}{\partial \theta }=(\Omega _{n},\langle \tilde{Q}%
_{n}\rangle _{n})+\sum_{j}\sigma _{j}^{(n)}\frac{\partial \Omega _{n}}{%
\partial \nu _{n+1\hspace{0.01in}j}}+\left\langle \mathcal{Y}%
_{n+1}\right\rangle _{n\Gamma }.  \label{geneqgda}
\end{equation}

Define finite functions $\nu _{n+1\hspace{0.01in}j}(\omega ,\bar{\nu}_{n+1%
\hspace{0.01in}k},\theta )$ as the solutions of the evolution equations 
\begin{equation}
\frac{\partial \nu _{n+1\hspace{0.01in}j}}{\partial \theta }=-\sigma
_{j}^{(n)}(\omega ,\nu _{n+1\hspace{0.01in}k},\theta ),  \label{rgg}
\end{equation}%
with the initial conditions $\nu _{n+1\hspace{0.01in}j}(\omega ,\bar{\nu}%
_{n+1\hspace{0.01in}k},0)=\bar{\nu}_{n+1\hspace{0.01in}j}$. Clearly, $\nu
_{n+1\hspace{0.01in}j}=\bar{\nu}_{n+1\hspace{0.01in}j}+\mathcal{O}(\theta )$%
. Given a functional $X(\Phi ,K,\omega ,\nu _{n+1\hspace{0.01in}j},\theta )$%
, define 
\begin{equation}
\bar{X}(\Phi ,K,\omega ,\bar{\nu}_{n+1\hspace{0.01in}j},\theta )=X(\Phi
,K,\omega ,\nu _{n+1\hspace{0.01in}j}(\omega ,\bar{\nu}_{n+1\hspace{0.01in}%
k},\theta ),\theta ).  \label{defo}
\end{equation}%
Then, 
\begin{equation}
\frac{\partial \bar{X}}{\partial \theta }=\overline{\frac{\partial X}{%
\partial \theta }}-\sum_{i}\bar{\sigma}_{j}^{(n)}\overline{\frac{\partial X}{%
\partial \nu _{n+1\hspace{0.01in}j}}},  \label{defi}
\end{equation}%
where $\bar{\sigma}_{j}^{(n)}$ are the functions obtained by applying the
redefinitions $\nu _{n+1\hspace{0.01in}j}(\omega ,\bar{\nu}_{n+1\hspace{%
0.01in}k},\theta )$ to $\sigma _{j}^{(n)}$. Choosing $X=\Omega _{n}$, we can
turn equation (\ref{geneqgda})\ into%
\begin{equation}
\frac{\partial \bar{\Omega}_{n}}{\partial \theta }=(\bar{\Omega}_{n},%
\overline{\langle \tilde{Q}_{n}\rangle _{n}})+\overline{\left\langle 
\mathcal{Y}_{n+1}\right\rangle _{n\Gamma }},  \label{gn2}
\end{equation}

Applying the redefinitions $\nu _{n+1\hspace{0.01in}j}(\omega ,\bar{\nu}_{n+1%
\hspace{0.01in}k},\theta )$ to the functions $\mu _{nj}(\omega ,\nu _{n+1%
\hspace{0.01in}k},\theta )$ of assumption ($a_{n}$), and including the
contributions coming from $\nu _{n+1\hspace{0.01in}j}-\bar{\nu}_{n+1\hspace{%
0.01in}j}$, which are proportional to $\theta $, we can define new $\mathcal{%
O}(\theta )\mathcal{O}(\hbar )$ functions $\bar{\mu}_{nj}(\omega ,\bar{\nu}%
_{n+1\hspace{0.01in}k},\theta )$ by the formula%
\begin{equation*}
\bar{\mu}_{nj}(\omega ,\bar{\nu}_{n+1\hspace{0.01in}k},\theta )\equiv \mu
_{nj}(\omega ,\nu _{n+1\hspace{0.01in}k}(\omega ,\bar{\nu}_{n+1\hspace{0.01in%
}l},\theta ),\theta )+\nu _{n+1\hspace{0.01in}j}(\omega ,\bar{\nu}_{n+1%
\hspace{0.01in}k},\theta )-\bar{\nu}_{n+1\hspace{0.01in}j}.
\end{equation*}%
Then, using (\ref{defo})\ and (\ref{sign}), we have 
\begin{equation*}
\bar{\Sigma}_{n}(\Phi ,K,\omega ,\bar{\nu}_{n+1\hspace{0.01in}j},\theta
)=\Sigma _{R\theta }(\Phi ,K,\omega ,\hbar \tau _{j}^{\ast }+\bar{\mu}_{nj}+%
\bar{\nu}_{n+1\hspace{0.01in}j},\theta ).
\end{equation*}

At this point, the independent parameters are $\omega $, $\bar{\nu}_{n+1%
\hspace{0.01in}j}$ and $\theta $. The formulas we have written so far hold
for every value of $\bar{\nu}_{n+1\hspace{0.01in}j}$, as long as it is $%
\mathcal{O}(\hbar ^{n+1})$. Now we want to raise the $\hbar $ order of $\bar{%
\nu}_{n+1\hspace{0.01in}j}$ by one unit. The validity of this choice will be
self-evident. By this we mean that it allows us to iterate all the arguments
of the proof without difficulties till the very end and preserve the
analyticity in $\hbar $.

Define \ 
\begin{equation}
\bar{\nu}_{n+1\hspace{0.01in}j}=\nu _{n+2\hspace{0.01in}j},\qquad \mu _{n+1%
\hspace{0.01in}j}(\omega ,\nu _{n+2\hspace{0.01in}k},\theta )=\left. \bar{\mu%
}_{nj}(\omega ,\bar{\nu}_{n+1\hspace{0.01in}k},\theta )\right\vert _{\bar{\nu%
}_{n+1\hspace{0.01in}k}\rightarrow \nu _{n+2\hspace{0.01in}k}}.  \label{ride}
\end{equation}%
So doing, we obtain the action $\Sigma _{n+1}$, given by formula (\ref{sign}%
) with the replacement $n\rightarrow n+1$: 
\begin{equation}
\Sigma _{n+1}(\Phi ,K,\omega ,\nu _{n+2\hspace{0.01in}j},\theta )=\Sigma
_{R\theta }(\Phi ,K,\omega ,\hbar \tau _{j}^{\ast }+\mu _{n+1\hspace{0.01in}%
j}+\nu _{n+2\hspace{0.01in}j},\theta )=\bar{\Sigma}_{n}(\Phi ,K,\omega ,\nu
_{n+2\hspace{0.01in}j},\theta ).  \label{tella}
\end{equation}%
Recalling that $\Sigma _{R\theta }=\Sigma _{R}+\mathcal{O}(\theta )$ and $%
\mu _{n+1\hspace{0.01in}j}=\mathcal{O}(\theta )\mathcal{O}(\hbar )$, formula
(\ref{tella}) tells us that, at $\theta =0$,%
\begin{equation*}
\left. \Sigma _{n+1}\right\vert _{\theta =0}=\Sigma _{R}(\Phi ,K,\omega
,\hbar \tau _{j}^{\ast }+\nu _{n+2\hspace{0.01in}j})=\Sigma _{R}(\Phi
,K,\omega ,\hbar \tau _{j}^{\ast })+\mathcal{O}(\hbar ^{n+2})=S_{R}(\Phi
,K,\omega )+\mathcal{O}(\hbar ^{n+2}),
\end{equation*}%
where the last equality follows from the first equation of (\ref{billo}).
Finally, the second equation of (\ref{billo}) and formula (\ref{abba}) give 
\begin{equation}
\left. (\Omega _{n+1},\Omega _{n+1})\right\vert _{\theta =0}=(\Gamma
_{R},\Gamma _{R})+\mathcal{O}(\hbar ^{n+2})=\mathcal{O}(\varepsilon )+%
\mathcal{O}(\hbar ^{n+2}).  \label{t0}
\end{equation}%
This is a check that the new action $\Sigma _{n+1}$ is AB\ nonanomalous at $%
\theta =0$, up to $\mathcal{O}(\varepsilon )$ and $\mathcal{O}(\hbar ^{n+2})$%
. Now we show that $\Sigma _{n+1}$ satisfies the same property for every $%
\theta $.

Using formula (\ref{gn2}), we get%
\begin{equation}
\frac{\partial \Omega _{n+1}}{\partial \theta }=(\Omega _{n+1},\langle 
\tilde{Q}_{n+1}\rangle _{n+1})+\left\langle Y_{n+1}\right\rangle _{n+1%
\hspace{0.01in}\Gamma },  \label{gy}
\end{equation}%
where the functionals $\tilde{Q}_{n+1}$ and $Y_{n+1}$ are obtained from $%
\tilde{Q}_{n}$ and $\mathcal{Y}_{n+1}$ by applying the redefinitions $\nu
_{n+1\hspace{0.01in}j}(\omega ,\bar{\nu}_{n+1\hspace{0.01in}k},\theta )$ and
(\ref{ride}). Using (\ref{defo}), (\ref{defi}) and (\ref{ynp}), it is easy
to see that formulas (\ref{yn1}) hold with $n\rightarrow n+1$.

Moreover, formula (\ref{yn2}) ensures that 
\begin{equation*}
\left\langle Y_{n+1}\right\rangle _{n+1\hspace{0.01in}\Gamma }=\mathcal{O}%
(\varepsilon )+\mathcal{O}(\hbar ^{n+2}),
\end{equation*}%
that is to say formulas (\ref{io}) and (\ref{ion}) hold with $n\rightarrow
n+1$.

Taking the antiparentheses of (\ref{gy}) with $\Omega _{n+1}$ and using the
Jacobi identity, we also find%
\begin{equation}
\frac{\partial }{\partial \theta }(\Omega _{n+1},\Omega _{n+1})=((\Omega
_{n+1},\Omega _{n+1}),\langle \tilde{Q}_{n+1}\rangle _{n+1})+2(\Omega
_{n+1},\left\langle Y_{n+1}\right\rangle _{n+1\hspace{0.01in}\Gamma }).
\label{teta}
\end{equation}%
The last term of the right-hand side is $\mathcal{O}(\varepsilon )+\mathcal{O%
}(\hbar ^{n+2})$. In appendix C we show how to integrate equation (\ref{teta}%
) and prove that the $\theta $ dependence of $(\Omega _{n+1},\Omega _{n+1})$
is encoded into a canonical transformation, up to $\mathcal{O}(\varepsilon )$
and $\mathcal{O}(\hbar ^{n+2})$. By formula (\ref{t0}), the value of $%
(\Omega _{n+1},\Omega _{n+1})$ at $\theta =0$ is also of such orders.
Moreover, the canonical transformation is convergent, because it is uniquely
determined by $\langle \tilde{Q}_{n+1}\rangle _{n+1}$, which is convergent.
Therefore, we find%
\begin{equation*}
(\Omega _{n+1},\Omega _{n+1})=\mathcal{O}(\varepsilon )+\mathcal{O}(\hbar
^{n+2})
\end{equation*}%
for arbitrary $\theta $, which is formula (\ref{sign2}) with $n\rightarrow
n+1$. As promised, the action $\Sigma _{n+1}$ is AB nonanomalous for
arbitrary $\theta $, up to $\mathcal{O}(\varepsilon )$ and $\mathcal{O}%
(\hbar ^{n+2})$. We have thus proved statements ($a_{n+1}$) and ($b_{n+1}$).

Finally, formulas (\ref{resi1}) and (\ref{resi2}) follow by taking $n$ to
infinity, with $\nu _{\infty j}=0$ and $\hbar \tau _{j}^{\ast \prime
}(\omega ,\theta )=\mu _{\infty j}(\omega ,0,\theta )$. Indeed, because of (%
\ref{sign}), if we define $S_{R\theta }$ according to (\ref{th1}), we have $%
\Sigma _{\infty }=S_{R\theta }$, so $\Omega _{\infty }=\Gamma _{R\theta }$.
Then, formula (\ref{sign2}) becomes (\ref{resi1}) at $n=\infty $. By (\ref%
{ion}), formula (\ref{io}) turns into (\ref{resi2}) at $n=\infty $, with $%
\tilde{Q}_{R\theta }=\tilde{Q}_{\infty }$.

\subsection{Integrating the equations of gauge dependence}

\label{s34}

Equation (\ref{resi2}) can be integrated with the method of appendix C (see
also \cite{back}). There, it is shown that we can consistently ignore the
terms $\mathcal{O}(\varepsilon )$ appearing on the right-hand side, in the
sense that the solution we find by ignoring those terms is correct up to $%
\mathcal{O}(\varepsilon )$. The basic reason is that the equations involve
only convergent functionals. Alternatively, we can just remove the cutoff by
taking the physical limit $\varepsilon \rightarrow 0$ in (\ref{resi2}) and
then work in the physical dimension $d$. The result is that every $\theta $
dependence of $\Gamma _{R\theta }$ can be absorbed into a convergent
canonical transformation, up to $\mathcal{O}(\varepsilon )$.

More precisely, the theorem of appendix C ensures that there exists a
canonical transformation $\Phi ,K\rightarrow \Phi ^{\prime },K^{\prime }$
such that the $\Gamma $ functional $\Gamma _{R}^{\prime }$ defined by 
\begin{equation}
\Gamma _{R}^{\prime }(\Phi ^{\prime },K^{\prime },\omega )=\Gamma _{R\theta
}(\Phi (\Phi ^{\prime },K^{\prime },\omega ,\theta ),K(\Phi ^{\prime
},K^{\prime },\omega ,\theta ),\omega ,\theta )  \label{panp0}
\end{equation}%
is $\theta $ independent, up to $\mathcal{O}(\varepsilon )$. Setting $\theta
=0$, we find $\Gamma _{R}^{\prime }=\Gamma _{R}$, since%
\begin{equation*}
\Gamma _{R}^{\prime }(\Phi ^{\prime },K^{\prime },\omega )=\Gamma _{R\theta
}(\Phi ^{\prime },K^{\prime },\omega ,0)=\Gamma _{R}(\Phi ^{\prime
},K^{\prime },\omega ).
\end{equation*}%
Finally, inverting the transformations, we get 
\begin{equation}
\Gamma _{R\theta }(\Phi ,K,\omega ,\theta )=\Gamma _{R}(\Phi ^{\prime }(\Phi
,K,\omega ,\theta ),K^{\prime }(\Phi ,K,\omega ,\theta ),\omega ).
\label{panp}
\end{equation}%
As promised, the dependence of $\Gamma _{R\theta }$ on the gauge parameter $%
\theta $ can be fully absorbed inside a canonical transformation.

We recall that the canonical transformations we are talking about, which are
convergent, nonlocal and act on the renormalized $\Gamma $ functional,
originate from a local canonical transformation of the form (\ref{trasfa})
that acts on the tree-level action. The connection between the two is a
procedure of re-renormalization and a re-fine-tuning of the finite local
counterterms. We call such canonical transformations on $\Gamma _{R}$ 
\textit{special}. Clearly, the composition of special canonical
transformations is a special canonical transformation. If we repeat the
argument of this subsection for any other gauge parameter $\theta $ that
satisfies (\ref{resi2}), taking one at a time, we can prove that the entire
dependence of the $\Gamma $ functional on the gauge parameters can be
absorbed into a special canonical transformation.

In subsection \ref{s43} the equations of gauge dependence are used to prove
that the physical quantities are gauge independent.

\section{Gauge dependence of the renormalized action}

\label{renogauge}

\setcounter{equation}{0}

In this section we study the counterparts of equations (\ref{resi1}) and (%
\ref{resi2}) at the level of the renormalized action. Using the identity (%
\ref{anom}), formula (\ref{resi1}) gives $(\Gamma _{R\theta },\Gamma
_{R\theta })=\langle (S_{R\theta },S_{R\theta })\rangle =\mathcal{O}%
(\varepsilon )$, which implies that $(S_{R\theta },S_{R\theta })$ is a
\textquotedblleft truly evanescent\textquotedblright\ local functional, i.e.
a local functional such that its average is evanescent. We use the symbol $%
\mathcal{E}$ to denote such type of functionals. Thus, we have the formula%
\begin{equation}
(S_{R\theta },S_{R\theta })=\mathcal{E},  \label{abthsr}
\end{equation}%
where $\langle \mathcal{E}\rangle =$ $\mathcal{O}(\varepsilon )$. Equation (%
\ref{abthsr}) expresses the cancellation of the gauge anomalies to all
orders at the level of the renormalized action.

Next, if we apply formulas (\ref{basic1}) and (\ref{basic2}) to (\ref{resi2}%
), we obtain\ 
\begin{equation}
\left\langle \frac{\partial S_{R\theta }}{\partial \theta }-(S_{R\theta },%
\tilde{Q}_{R\theta })\right\rangle -\frac{i}{2}\langle (S_{R\theta
},S_{R\theta })\hspace{0.01in}\tilde{Q}_{R\theta }\rangle _{\Gamma }=%
\mathcal{O}(\varepsilon ).  \label{esso}
\end{equation}%
By formula (\ref{abthsr}), $(S_{R\theta },S_{R\theta })$ is a renormalized
local functional such that its average is evanescent. In appendix B we prove
that there exists an $\mathcal{O}(\hbar )$ local functional $F_{R}$, such
that the local bifunctional $Y_{R}\equiv -(i/2)(S_{R\theta },S_{R\theta })%
\hspace{0.01in}\tilde{Q}_{R\theta }+F_{R}$ is renormalized and the average $%
\left\langle Y_{R}\right\rangle _{\Gamma }$ is evanescent to all orders. We
denote such $F_{R}$ by the symbolic expression $(S_{R\theta },S_{R\theta };%
\tilde{Q}_{R\theta })$. Thus, formula (\ref{esso}) gives 
\begin{equation*}
\left\langle \frac{\partial S_{R\theta }}{\partial \theta }-(S_{R\theta },%
\tilde{Q}_{R\theta })-(S_{R\theta },S_{R\theta };\tilde{Q}_{R\theta
})\right\rangle =\mathcal{O}(\varepsilon ).
\end{equation*}%
In turn, this equation implies%
\begin{equation}
\frac{\partial S_{R\theta }}{\partial \theta }=(S_{R\theta },\tilde{Q}%
_{R\theta })+(S_{R\theta },S_{R\theta };\tilde{Q}_{R\theta })+\mathcal{E}.
\label{gegdsr}
\end{equation}%
Formula (\ref{gegdsr}) is the equation of gauge dependence for the
renormalized action $S_{R\theta }$. Note that $(S_{R\theta },S_{R\theta };%
\tilde{Q}_{R\theta })$ encodes the re-fine-tuning of the finite local
counterterms.

Equation (\ref{gegdsr}) can be integrated with the method explained in
appendix C. Although the term $(S_{R\theta },S_{R\theta };\tilde{Q}_{R\theta
})$ depends on $S_{R\theta }$, a recursive procedure allows us to treat it
as a known functional at every step.

\section{RG invariance and other applications}

\label{s24}

\setcounter{equation}{0}

In this section we give a few applications of the theorem proved in section %
\ref{s8}. The first application is the proof that RG invariance is preserved
by the canonical transformation. The second application is the proof that
renormalizable chiral gauge theories gauge-fixed by means of a
nonrenormalizable gauge fixing remain renormalizable, although in a
nonmanifest way. The third application is a step of the proof of the
Adler-Bardeen theorem in nonrenormalizable theories \cite{ABnonreno}.

RG\ invariance is expressed by the Callan-Symanzik equation (which is
derived at the end of this section) 
\begin{equation}
\mu \frac{\partial \Gamma _{R}}{\partial \mu }+\hat{\beta}^{i}\frac{\partial
\Gamma _{R}}{\partial \omega _{i}}-(\Gamma _{R},\langle U_{R}\rangle )=%
\mathcal{O}(\varepsilon ),  \label{rg1}
\end{equation}%
where $\hat{\beta}^{i}$ are the $\omega _{i}$ beta functions (at $%
\varepsilon \neq 0$) and $U_{R}$ is a local functional. At the level of the
renormalized action $S_{R}$, the Callan-Symanzik equation reads%
\begin{equation}
\mu \frac{\partial S_{R}}{\partial \mu }+\hat{\beta}^{i}\frac{\partial S_{R}%
}{\partial \omega _{i}}-(S_{R},U_{R})-(S_{R},S_{R};U_{R})=\mathcal{E}.
\label{rgg1}
\end{equation}

Let $\tilde{\Gamma}_{R}$ denote the renormalized $\Gamma $ functional where
the parameters $\omega _{i}$ are written in terms of their running versions $%
\tilde{\omega}_{i}(\mu )$ and $\mu $, where $\tilde{\omega}_{i}$ are the
solutions of $\mu {\mathrm{d}}\tilde{\omega}_{i}/{\mathrm{d}}\mu =-\hat{\beta%
}^{i}(\tilde{\omega})$ with initial conditions $\omega _{i}$. We have%
\begin{equation*}
\tilde{\Gamma}_{R}(\Phi ,K,\tilde{\omega},\mu )=\Gamma _{R}(\Phi ,K,\omega
,\mu ).
\end{equation*}%
Let $\tilde{U}_{R}$ denote the functional $\langle U_{R}\rangle $
reparametrized in a similar way. Then the Callan-Symanzik equation becomes 
\begin{equation*}
\mu \frac{\tilde{\partial}\tilde{\Gamma}_{R}}{\tilde{\partial}\mu }-(\tilde{%
\Gamma}_{R},\tilde{U}_{R})=\mathcal{O}(\varepsilon ),
\end{equation*}%
where $\tilde{\partial}$ denotes the derivative at fixed $\tilde{\lambda}$.
The new equation has the same form as (\ref{resi2}), so it is solved by
making a canonical transformation. From formula (\ref{panp0}), we learn that
there exists a canonical transformation that takes us to new fields and
sources $\tilde{\Phi}$, $\tilde{K}$ and a reference value of $\mu $, which
we denote by $\bar{\mu}$ and leave implicit, such that%
\begin{equation*}
\tilde{\Gamma}_{R}(\Phi ,K,\tilde{\omega},\mu )=\bar{\Gamma}_{R}(\tilde{\Phi}%
(\Phi ,K,\tilde{\omega},\mu ),\tilde{K}(\Phi ,K,\tilde{\omega},\mu ),\tilde{%
\omega}),
\end{equation*}%
for a certain other functional $\bar{\Gamma}_{R}$.

Now, if we make the canonical transformation (\ref{trasfa}) on the
tree-level action, we get, by formula (\ref{panp}),%
\begin{eqnarray*}
\tilde{\Gamma}_{R\theta }(\Phi ,K,\tilde{\omega},\mu ,\theta ) &=&\tilde{%
\Gamma}_{R}(\Phi ^{\prime }(\Phi ,K,\tilde{\omega},\mu ,\theta ),K^{\prime
}(\Phi ,K,\tilde{\omega},\mu ,\theta ),\tilde{\omega},\mu )= \\
&=&\bar{\Gamma}_{R}(\tilde{\Phi}(\Phi ^{\prime },K^{\prime },\tilde{\omega}%
,\mu ),\tilde{K}(\Phi ^{\prime },K^{\prime },\tilde{\omega},\mu ),\tilde{%
\omega}).
\end{eqnarray*}%
Going back to the parameters $\omega $, we also have%
\begin{equation*}
\Gamma _{R\theta }(\Phi ,K,\omega ,\mu ,\theta )\equiv \tilde{\Gamma}%
_{R\theta }(\Phi ,K,\tilde{\omega},\mu ,\theta )=\bar{\Gamma}_{R}(\bar{\Phi}%
^{\prime }(\Phi ,K,\omega ,\mu ,\theta ),\bar{K}^{\prime }(\Phi ,K,\omega
,\mu ,\theta ),\tilde{\omega}(\omega ,\mu )).
\end{equation*}%
having defined $\tilde{\Phi}(\Phi ^{\prime },K^{\prime },\tilde{\omega},\mu
)=\bar{\Phi}^{\prime }(\Phi ,K,\omega ,\mu ,\theta )$ and similarly for $%
\tilde{K}$. Differentiating with respect to $\ln \mu $, and recalling that
our canonical transformations are special, we get%
\begin{equation}
\mu \frac{\partial \Gamma _{R\theta }}{\partial \mu }+\hat{\beta}^{i}\frac{%
\partial \Gamma _{R\theta }}{\partial \omega _{i}}-(\Gamma _{R\theta
},\langle U_{R\theta }\rangle )=\mathcal{O}(\varepsilon ),  \label{rg2}
\end{equation}%
for some new local functional $U_{R\theta }$. Formula (\ref{rg2}) is the
transformed RG\ equation. Note that the beta functions do not depend on $%
\theta $ in this approach.

Another application that we mention is to power-counting renormalizable
chiral gauge theories gauge-fixed by means of a nonrenormalizable gauge
fixing. If a renormalizable theory is nonchiral, it is rather
straightforward to prove that it remains renormalizable when a
nonrenormalizable gauge fixing is used. When the theory is chiral, on the
other hand, the matter is more complicated. In principle, the
simplifications between divergences and evanescences can make the parameters
of negative dimensions, introduced by the gauge fixing, propagate into the
physical sector and turn the theory into a truly nonrenormalizable one. The
theorem of section \ref{s8}, combined with RG invariance, ensures that this
cannot happen.

Consider for example the standard model in flat space and gauge fix a
non-Abelian gauge symmetry by means of a gauge-fixing function such as 
\begin{equation*}
\bar{G}^{a}(\phi )=\partial ^{\mu }A_{\mu }^{a}+\kappa A_{\mu }^{b}A_{\nu
}^{a}F^{b\mu \nu }.
\end{equation*}%
Since the constant $\kappa $ has dimension $-2$ in units of mass, power
counting alone is not sufficient to classify the counterterms in a
convenient way at $\kappa \neq 0$. However, the change of gauge fixing that
turns $G^{a}(\phi )=\partial ^{\mu }A_{\mu }^{a}$ into $\bar{G}^{a}(\phi )$
is a canonical transformation, so we can apply the theorem of section \ref%
{s8}. Formula (\ref{estensi}) teaches us that infinitely many terms $%
\mathcal{H}_{i}$ of arbitrary dimensions are switched on, including the
gauge noninvariant ones. Nevertheless, the theorem ensures that once we have
done that, it is possible to express the coefficients of all of those terms\
as functions of the other parameters of the theory, and fine-tune those
functions to enforce again the cancellation of gauge anomalies to all
orders. Moreover, the argument given above ensures that RG\ invariance is
preserved. We conclude that no new independent parameters are necessary to
subtract the divergences and cancel the gauge anomalies in a RG invariant
way:\ the physical couplings are still finitely many. Thus, when a power
counting renormalizable chiral gauge theory, such as the standard model in
flat space, is gauge-fixed by means of a nonrenormalizable gauge fixing, it
remains a renormalizable theory, although its renormalizability is not
manifest anymore. A\ similar conclusion holds when the theory is
renormalizable by weighted power counting \cite{halat} or any other
criterion.

The third application we mention is the proof of the Adler-Bardeen theorem
in nonrenormalizable theories, recently obtained in ref. \cite{ABnonreno} by
upgrading the arguments of \cite{ABrenoYMLR}. It applies to the theories
whose gauge symmetries are general covariance, local Lorentz symmetry and
Abelian and non-Abelian Yang-Mills symmetries, and satisfy a variant of the
Kluberg-Stern--Zuber conjecture. Quantum gravity coupled to the standard
model satisfies all the assumptions and so is free of gauge anomalies to all
orders. In the approach of \cite{ABnonreno}, the CD regularization is
combined with a higher-derivative regularization. If the scale $\Lambda $
associated with the higher-derivative terms is kept fixed, we obtain a
super-renormalizable higher-derivative (HD) theory, which satisfies the
Adler-Bardeen theorem by simple power-counting arguments. When the scale $%
\Lambda $ is sent to infinity, the $\Lambda $ divergences are renormalized
inductively. At each step, the theorem of section \ref{s8} allows us to
resubtract the divergences in $\varepsilon $ and re-fine-tune the finite
local terms, in order to enforce the cancellation of gauge anomalies to all
orders at $\Lambda $ fixed. In the end, thanks to this, the cancellation of
gauge anomalies survives the renormalization of both types of divergences.
Moreover, the approach of ref. \cite{ABnonreno} identifies a special
subtraction scheme where the cancellation of gauge anomalies is manifest
from two loops onwards, within any given truncation. We stress that it is
not possible to achieve a similar goal by means of
regularization-independent methods.

The Callan-Symanzik equation (\ref{rg1}) can be proved from the results of
ref. \cite{ABnonreno} as follows, under the assumptions specified there. At $%
\Lambda $ fixed the HD theory is renormalized by redefinitions of
parameters, while the trivially anomalous terms are canceled by adding a
finite local functional $-\chi /2$ to the action. The renormalized action
coincides with its bare version apart from $\chi $ itself, which satisfies $%
\chi =\mu ^{-\varepsilon }\chi _{\mathrm{B}}$, where $\chi _{\mathrm{B}}$ is
RG invariant. Then, the HD theory satisfies formulas (\ref{rg1}) and (\ref%
{rgg1}) with $U_{R}=0$, the right-hand side of (\ref{rgg1}) being equal $%
\varepsilon \chi /2$. The average $\langle \chi \rangle $ is convergent in
the HD theory (since its divergences, which would start from two loops, are
excluded by the arguments of \cite{ABnonreno}), so the product $\varepsilon
\langle \chi \rangle /2$ is truly evanescent at $\Lambda $ fixed. At a
second stage, the renormalization is completed by removing the $\Lambda $
divergences. This is done by means of special canonical transformations and
redefinitions of parameters. In this section we have proved that those
operations preserve the Callan-Symanzik equation, although they can affect
the beta functions and the functional $U_{R}$. In the end, we obtain
equations of the forms (\ref{rg1}) and (\ref{rgg1}).

\section{Gauge dependence of the beta functions}

\label{s3}

\setcounter{equation}{0}

Often, we can prove that a theory is AB\ nonanomalous in a family of gauges,
parametrized by certain gauge-fixing parameters $\xi $. In various common
situations we can achieve this goal by applying the results of ref. \cite%
{ABrenoYMLR}, where the Adler-Bardeen theorem was proved for arbitrary
values of the gauge-fixing parameter $\xi $ of the Lorenz gauge, in power
counting renormalizable gauge theories that have unitary free-field limits.
More generally, if the theory is coupled to quantum gravity, we can apply
the results of \cite{ABnonreno}. Then, when we study the dependence of the
correlation functions on $\xi $, we can proceed more straightforwardly than
in section \ref{s8}, since we already know that $(\Gamma _{R},\Gamma _{R})=%
\mathcal{O}(\varepsilon )$ for arbitrary $\xi $. It is worth recalling that
in section \ref{s8} we had to derive this result from just knowing that $%
(\Gamma _{R},\Gamma _{R})$ was $\mathcal{O}(\varepsilon )$ for $\xi $ equal
to some initial value $\xi ^{\ast }$.

In this section we study the equations of gauge dependence in theories that
are AB\ nonanomalous for arbitrary values of some gauge parameter $\theta $
and satisfy some additional assumptions. Those assumptions are not very
restrictive, since they are fulfilled quite commonly. When $\theta $ varies,
we do not need to readjust the subtraction scheme by fine-tuning the finite
local counterterms. Then, however, the beta functions of the couplings are
in general gauge dependent. Their gauge dependence can be removed by
redefining the couplings themselves.

We begin by listing the assumptions we need.

(I)\ We assume that the gauge algebra is irreducible and closes off shell.
This assumption is satisfied by the theories whose gauge symmetries are
general covariance, local Lorentz symmetry and Abelian and non-Abelian
Yang-Mills symmetries, such as the standard model coupled to quantum
gravity. It allows us to make a number of simplifications. For example, we
can choose the fields $\Phi $ and the sources $K$ so that the gauge-fixed
tree-level solution $S_{d}$ of the $D$-dimensional master equation $%
(S_{d},S_{d})=0$ is linear in $K$ and has the very simple structure (\ref{sk}%
).

We have already remarked that in various cases, for example when the theory
is chiral or parity violating, the action $S_{d}$, embedded in $D$
dimensions using the standard rules of the dimensional regularization
technique, is in general not well regularized, due to the key role played by
the $d$-dimensional analogue $\tilde{\gamma}$ of the matrix $\gamma _{5}$,
or the tensor $\varepsilon ^{\bar{a}_{1}\cdots \bar{a}_{d}}$. Using the
chiral dimensional regularization, a well-regularized classical action $%
S(\Phi ,K)$ is obtained by adding a number of evanescent corrections $S_{%
\text{ev}}$ to $S_{d}$ \cite{chiraldimreg}, as shown in formula (\ref{sdev}%
). We denote the parameters contained in $S_{\text{ev}}$ by $\eta _{I}$. For
convenience, we assume that $S_{\text{ev}}$ depends linearly on the
parameters $\eta $, and vanishes for $\eta =0$.

Let $\{\mathcal{G}_{i}(\phi )\}$ denote a basis of local gauge invariant
functionals of the classical fields $\phi $. Expand the classical action as%
\begin{equation}
S_{c}(\phi )=\sum_{i}\lambda _{i}\mathcal{G}_{i}(\phi ),  \label{sc}
\end{equation}%
where $\lambda _{i}$ are independent parameters. We call the constants $%
\lambda _{i}$ \textquotedblleft physical parameters\textquotedblright ,
since they contain or are related to the gauge coupling constants, the
masses, the Yukawa couplings, etc. In our notation some parameters $\lambda
_{i}$ may be actually redundant. Nevertheless, to simplify some derivations
we prefer to keep an independent $\lambda _{i}$ for every $\mathcal{G}_{i}$.
For example, it is often useful to restrict $S_{c}$ by dropping the terms
that are proportional to the $S_{c}$ field equations, because those terms
can be renormalized by means of canonical transformations, rather than $%
\lambda _{i}$ redefinitions. We do not implement this restriction right now,
to make some arguments of the derivations that follow more transparent. We
can always remove that class of redundant terms at the end by means of a
convergent canonical transformation, by applying either the procedure of
section \ref{s8}, which is more general, or the one of this section, which
holds under specific assumptions. Both procedures preserve the cancellation
of gauge anomalies and the equations of gauge dependence.

In total, we have physical parameters $\lambda $, gauge-fixing parameters $%
\xi $, contained in $\Psi $, and regularizing parameters $\eta $. The
classical action is written as $S(\Phi ,K,\lambda ,\xi ,\eta )$.

The action $S_{c}$ may contain \textit{accidental} symmetries, which are the
global symmetries unrelated to the gauge transformations. Some accidental
symmetries are dynamically lost, because they are anomalous, others are
nonanomalous. Let $G_{\text{nas}}$ denote the group of nonanomalous
accidental symmetries, or the identity group, depending on whether the gauge
group contains $U(1)$ factors or not. By definition, the set $\{\mathcal{G}%
_{i}(\phi )\}$ includes the invariants that explicitly break the anomalous
accidental symmetries, but excludes the invariants, denoted by $\mathcal{%
\check{G}}_{i}(\phi )$, that explicitly break $G_{\text{nas}}$. Then the
actions$\ S_{c}$ and $S_{d}$ do not contain the invariants $\mathcal{\check{G%
}}_{i}$, so we define extended actions $\check{S}_{c}$ and $\check{S}_{d}=%
\check{S}_{c}+(S_{K},\Psi )+S_{K}$ that do include them, multiplied by
independent parameters $\check{\lambda}_{i}$. Both choices of including and
excluding the invariants $\mathcal{\check{G}}_{i}$, are consistent, from the
point of view of renormalization.

We say that the action $S_{d}$ satisfies the Kluberg-Stern--Zuber assumption 
\cite{kluberg}, if every nonevanescent local functional $X$ of ghost number
zero that solves the equation $(S_{d},X)=0$ has the form%
\begin{equation*}
\qquad X=\sum_{i}a_{i}\mathcal{G}_{i}+(S_{d},Y),
\end{equation*}%
where $a_{i}$ are constants depending on the parameters of the theory, and $%
Y $ is a local functional of ghost number $-1$.

We say that the action $S_{d}$ is \textit{cohomologically complete} if its
extension $\check{S}_{d}$ satisfies the extended Kluberg-Stern--Zuber
assumption, that is to say every nonevanescent local functional $X$ of ghost
number zero that solves $(\check{S}_{d},X)=0$ has the form%
\begin{equation}
\qquad X=\sum_{i}a_{i}\mathcal{G}_{i}+\sum_{i}b_{i}\mathcal{\check{G}}_{i}+(%
\check{S}_{d},Y),  \label{cohopr}
\end{equation}%
where $b_{i}$ are other constants, and $Y$ is a local functional.

(II)\ We assume that the action $S_{d}$ of (\ref{sk}) is cohomologically
complete and the group $G_{\text{nas}}$ is compact.

The Kluberg-Stern--Zuber assumption is satisfied when the Yang-Mills gauge
group is semisimple and the action $S_{d}$ satisfies generic properties \cite%
{coho2}. It is not satisfied when the gauge group has $U(1)$ factors and
accidental symmetries are present. In particular, it is not satisfied by the
standard model. However, it can be proved, using the Ward identities that
hold in the Lorenz gauge, that the standard model is cohomologically
complete \cite{ABrenoYMLR}. So are the Lorentz violating extensions of the
standard model of refs. \cite{lvsm,lvsm2}, which are renormalizable by
weighted power counting \cite{halat}. Starting from the cohomological
theorems proved in ref. \cite{coho2}, it can be proved that the standard
model coupled to quantum gravity is also cohomologically complete \cite%
{ABnonreno}, and so are most of its extensions.

The condition $(S_{d},X)=0$ is the one typically satisfied by the
counterterms. In this section we show that the contributions of the extra
term contained in the generalized Ward identity (\ref{genw}) satisfy the
same condition. Thus, assumption (II) will give us control on the effects of
the new term.

We can imagine that $\theta $ is one of the parameters $\xi $, or another
parameter introduced by a field redefinition. We keep it distinct from the
other parameters $\lambda ,\xi ,\eta $ contained in the action $S$ and
assume that $S(\Phi ,K,\lambda ,\xi ,\eta )$ denotes the action at some
specific value $\theta ^{\ast }$ of $\theta $. With no loss of generality,
we take $\theta ^{\ast }=0$. By definition of gauge parameter, when we vary $%
\theta $, we make a canonical transformation generated by a functional of
the form (\ref{trasfa}) on the action $S$, and this operation gives the
action $S_{\theta }$. As before, let $S_{R\theta }$ denote the renormalized
action and $\Gamma _{R\theta }$ the $\Gamma $ functional associated with it.

(III)\ We assume that the theory is AB nonanomalous for arbitrary values of
some gauge parameter $\theta $. Precisely, we assume that there exists a
class of subtraction schemes where the renormalized $\Gamma $ functional $%
\Gamma _{R\theta }$ satisfies the identity%
\begin{equation}
(\Gamma _{R\theta },\Gamma _{R\theta })=\mathcal{O}(\varepsilon ),
\label{ahid}
\end{equation}%
where $\theta $ takes values in some continuous range that includes $\theta
=0$. From now on we understand that we work in that class of subtraction
schemes.

Assumption (III) has been proved, for common families of gauge conditions,
in the power-counting renormalizable gauge theories that have unitary
free-field limits \cite{ABrenoYMLR}, in the Lorentz violating extensions of
the standard model that are renormalizable by weighted power counting \cite%
{lvsm,lvsm2}, in the standard model coupled to quantum gravity and a large
class of other nonrenormalizable theories \cite{ABnonreno}.

We prove that there exist finite functions $\rho _{j}$ of $\lambda $, $\xi $%
, $\eta $ and $\theta $, which start from $\mathcal{O}(\hbar )$, and a
renormalized local functional $H_{R\theta }(\Phi ,K)=\tilde{Q}_{\theta
}(\Phi ,K)+\mathcal{O}(\hbar )$, where $\tilde{Q}_{\theta }(\Phi
,K)=Q_{\theta }(\Phi ,K^{\prime }(\Phi ,K))$ and $Q_{\theta }=\partial
Q/\partial \theta $, such that $\Gamma _{R\theta }$ satisfies the equation%
\begin{equation}
\frac{\partial \Gamma _{R\theta }}{\partial \theta }=\sum_{j}\rho _{j}\frac{%
\partial \Gamma _{R\theta }}{\partial \lambda _{j}}+(\Gamma _{R\theta
},\langle H_{R\theta }\rangle )+\mathcal{O}(\varepsilon ).  \label{geneqgd2}
\end{equation}

The first term on the right-hand side of (\ref{geneqgd2}) can be absorbed by
means of finite redefinitions of the parameters $\lambda $ (which correspond
to the re-fine-tuning of the previous section). The second term is the one
that can be absorbed into a canonical transformation.

In the rest of this section we derive the equations of gauge dependence (\ref%
{geneqgd2}) under the assumptions listed above, and integrate them. Before
beginning the derivation, a few preliminary remarks are in order. If we
differentiate (\ref{ahid}) with respect to any parameter $\zeta $, we find 
\begin{equation}
\left( \Gamma _{R\theta },\frac{\partial \Gamma _{R\theta }}{\partial \zeta }%
\right) =\mathcal{O}(\varepsilon ).  \label{dgr}
\end{equation}%
Now we take the antiparentheses of both sides of formula (\ref{resi2}) or (%
\ref{geneqgd2}) with $\Gamma _{R\theta }$, and use (\ref{dgr}) for $\zeta
=\theta $ and $\zeta =\lambda _{j}$, the Jacobi identity satisfied by the
antiparentheses and formula (\ref{ahid}) again. At the end, we find a
consistent relation of the form $\mathcal{O}(\varepsilon )=\mathcal{O}%
(\varepsilon )$. Thus, we can view formulas (\ref{resi2}) and (\ref{geneqgd2}%
) as the solutions to the condition (\ref{dgr}) for $\zeta =\theta $.

To explain this issue more clearly, let us define an operator $\delta
_{\Gamma }$ that acts on a (generically nonlocal) functional $Y$ by taking
its antiparentheses with $\Gamma _{R\theta }$: $\delta _{\Gamma }Y=(\Gamma
_{R\theta },Y)$. Formula (\ref{ahid}) ensures that $\delta _{\Gamma }$ is
nilpotent up to $\mathcal{O}(\varepsilon )$, because the Jacobi identity
gives 
\begin{equation}
\delta _{\Gamma }^{2}Y=(\Gamma _{R\theta },(\Gamma _{R\theta },Y))=\frac{1}{2%
}((\Gamma _{R\theta },\Gamma _{R\theta }),Y)=\mathcal{O}(\varepsilon ).
\label{d2x}
\end{equation}%
Therefore, it is meaningful to study the cohomology of $\delta _{\Gamma }$.
Consider the problem $\delta _{\Gamma }Y=0$, of which the $\varepsilon
\rightarrow 0$ limit of (\ref{dgr}) is an example. It is a nonlocal upgrade
of the more standard cohomological problem $(S_{d},X)=0$, where $X$ is
local. Formula (\ref{dgr}) tells us that $\partial \Gamma _{R\theta
}/\partial \theta $ is closed, in the sense of the $\delta _{\Gamma }$
cohomology, up to $\mathcal{O}(\varepsilon )$. On the other hand, formula (%
\ref{geneqgd2}) ensures that there exist finite linear combinations of $%
\partial \Gamma _{R\theta }/\partial \zeta $ that are $\delta _{\Gamma }$%
-exact, up to $\mathcal{O}(\varepsilon )$.

However, nonlocal cohomological problems are difficult to solve and must be
treated with care, because if we do not specify which nonlocalities are
allowed and which are not, any closed functional can in principle be exact.
In other words, we cannot derive (\ref{geneqgd2}) immediately from (\ref{dgr}%
), which is why gauge dependence deserves a separate investigation.

\subsection{The equations of gauge dependence}

We apply formula (\ref{gx}) of appendix A to the renormalized action $%
S_{R\theta }$ and the renormalized $\Gamma $ functional $\Gamma _{R\theta }$%
, with $X=\tilde{Q}_{R\theta }$, where $\tilde{Q}_{R\theta }$ denotes the
renormalized version of the functional $\tilde{Q}_{\theta }(\Phi ,K)$. We
obtain 
\begin{equation}
\frac{\partial \Gamma _{R\theta }}{\partial \theta }=(\Gamma _{R\theta
},\langle \tilde{Q}_{R\theta }\rangle )+\left\langle U_{R\theta
}\right\rangle _{\Gamma },  \label{pril}
\end{equation}%
where 
\begin{equation}
U_{R\theta }=-\frac{i}{2}(S_{R\theta },S_{R\theta })\tilde{Q}_{R\theta }+%
\frac{\partial S_{R\theta }}{\partial \theta }-(S_{R\theta },\tilde{Q}%
_{R\theta }).  \label{trt}
\end{equation}%
Taking the antiparentheses of both sides of (\ref{pril}) with $\Gamma
_{R\theta }$ and using (\ref{dgr}) and (\ref{d2x}), we obtain%
\begin{equation}
(\Gamma _{R\theta },\left\langle U_{R\theta }\right\rangle _{\Gamma })=%
\mathcal{O}(\varepsilon ).  \label{gY}
\end{equation}%
Differently from (\ref{dgr}), this nonlocal cohomological problem can be
reduced to a local one, and solved. The reason is that $U_{R\theta }$ is
originated by an evanescent local bifunctional. We prove that there exist 
\textit{finite} functions $\rho _{j}=$ $\mathcal{O}(\hbar )$ of $\lambda $, $%
\xi $, $\eta $ and $\theta $, and a renormalized local functional $%
W_{R\theta }=\mathcal{O}(\hbar )$, such that 
\begin{equation}
\left\langle U_{R\theta }\right\rangle _{\Gamma }=\sum_{j}\rho _{j}\frac{%
\partial \Gamma _{R\theta }}{\partial \lambda _{j}}+(\Gamma _{R\theta
},\left\langle W_{R\theta }\right\rangle )+\mathcal{O}(\varepsilon ).
\label{finiY}
\end{equation}

We proceed by induction. Assume that there exist finite functions $\rho _{n%
\hspace{0.01in}j}=\mathcal{O}(\hbar )$ of $\lambda $, $\xi $, $\eta $ and $%
\theta $, and a renormalized local functional $W_{n}=\mathcal{O}(\hbar )$,
such that the partially subtracted functional 
\begin{equation}
U_{n}\equiv U_{R\theta }-\sum_{j}\rho _{n\hspace{0.01in}j}\frac{\partial
S_{R\theta }}{\partial \lambda _{j}}-(S_{R\theta },W_{n})-\frac{i}{2}%
(S_{R\theta },S_{R\theta })\hspace{0.01in}W_{n},  \label{iondu}
\end{equation}%
satisfies%
\begin{equation}
\left\langle U_{n}\right\rangle _{\Gamma }=\mathcal{O}(\varepsilon )+%
\mathcal{O}(\hbar ^{n+1}).  \label{iu}
\end{equation}%
This assumption is clearly satisfied at the zeroth order, where $\rho _{0%
\hspace{0.01in}j}=0$ and $W_{0}=0$, because by formula (\ref{steq}) we have $%
U_{R\theta }=Y_{\theta }+\mathcal{O}(\hbar )$, where $Y_{\theta }$ is
evanescent and given by (\ref{yx}).

Using formulas (\ref{basic1}) and (\ref{basic2}), we obtain the average 
\begin{equation}
\left\langle U_{n}\right\rangle _{\Gamma }=\left\langle U_{R\theta
}\right\rangle _{\Gamma }-\sum_{j}\rho _{n\hspace{0.01in}j}\frac{\partial
\Gamma _{R\theta }}{\partial \lambda _{j}}-(\Gamma _{R\theta },\left\langle
W_{n}\right\rangle ),  \label{yn}
\end{equation}%
which is clearly convergent. Consider the $(n+1)$-loop contributions $%
U_{n}^{(n+1)}$ to $\left\langle U_{n}\right\rangle _{\Gamma }$. They are
convergent, because so is the right-hand side of (\ref{yn}). Moreover, the
inductive assumption (\ref{iu}) states that the average $\left\langle
U_{n}\right\rangle _{\Gamma }$ is evanescent up to and including $n$ loops,
while (\ref{ahid}) ensures that $\langle (S_{R\theta },S_{R\theta })\rangle
=(\Gamma _{R\theta },\Gamma _{R\theta })$ is evanescent to all orders. The
arguments of appendix B ensure that the functional $U_{n}^{(n+1)}$ is the
sum of a local nonevanescent part $U_{n\text{nonev}}^{(n+1)}$ plus a
generically nonlocal evanescent part:%
\begin{equation}
\left\langle U_{n}\right\rangle _{\Gamma }=U_{n\text{nonev}}^{(n+1)}+%
\mathcal{O}(\varepsilon )+\mathcal{O}(\hbar ^{n+2}).  \label{iu2}
\end{equation}

Thus, using (\ref{yn}) and (\ref{iu2}), we have%
\begin{equation}
\left\langle U_{R\theta }\right\rangle _{\Gamma }=\sum_{j}\rho _{n\hspace{%
0.01in}j}\frac{\partial \Gamma _{R\theta }}{\partial \lambda _{j}}+(\Gamma
_{R\theta },\left\langle W_{n}\right\rangle )+U_{n\text{nonev}}^{(n+1)}+%
\mathcal{O}(\varepsilon )+\mathcal{O}(\hbar ^{n+2}).  \label{finiYn}
\end{equation}%
Inserting this expression inside (\ref{gY}) and using (\ref{dgr}), (\ref{d2x}%
) and (\ref{ahid}), we obtain%
\begin{equation*}
(\Gamma _{R\theta },U_{n\text{nonev}}^{(n+1)})=\mathcal{O}(\varepsilon )+%
\mathcal{O}(\hbar ^{n+2}).
\end{equation*}%
Taking the $(n+1)$-loop nonevanescent contributions to this formula, we find%
\begin{equation}
(S_{d\theta },U_{n\text{nonev}}^{(n+1)})=0,  \label{tr}
\end{equation}%
where $S_{d\theta }$ is the action obtained by applying the canonical
transformation (\ref{trasfa}) to $S_{d}$. In deriving the result (\ref{tr}),
it is important to recall that the tree-level action (\ref{sdev}) and the
canonical transformation (\ref{trasfa}) do not contain analytically
evanescent terms. In turn, $S_{\theta }$ and $S_{d\theta }$ satisfy the same
property, and $S_{d\theta }$ is the full nonevanescent part of $S_{\theta }$.

Applying the inverse of the transformation (\ref{trasfa}) to equation (\ref%
{tr}) and letting $\tilde{U}_{n\text{nonev}}^{(n+1)}$denote the functional
obtained from $U_{n\text{nonev}}^{(n+1)}$, we get%
\begin{equation}
(S_{d},\tilde{U}_{n\text{nonev}}^{(n+1)})=0.  \label{sdx}
\end{equation}

At this point, we apply assumption (II). Let us imagine that instead of
working with the classical action $S_{c}$ we work with its extension $\check{%
S}_{c}$, which includes the invariants $\mathcal{\check{G}}_{i}$ that break
the nonanomalous accidental symmetries belonging to the group $G_{\text{nas}%
} $. Similarly, we extend $S_{d}$ to $\check{S}_{d}$, $S_{\text{ev}} $ to $%
\check{S}_{\text{ev}}$ and $S=S_{d}+S_{\text{ev}}$ to $\check{S}$. Every
extended functional reduces to the nonextended one when we set $\check{%
\lambda}=\check{\eta}=0$, where $\check{\lambda}$ and $\check{\eta}$ are the
extra parameters of $\check{S}_{c}$ and $\check{S}_{\text{ev}}$,
respectively. If we repeat the operations that lead to (\ref{sdx}), we
obtain an extended, nonevanescent local functional $\check{U}_{n\text{nonev}%
}^{(n+1)}$ that satisfies $(\check{S}_{d},\check{U}_{n\text{nonev}%
}^{(n+1)})=0$. By assumption (II), the action $\check{S}_{d}$\ satisfies the
extended Kluberg-Stern--Zuber assumption. Therefore, there exist finite
order-$\hbar ^{n+1}$ constants $\check{\sigma}_{i}^{(n+1)}$, $\check{\tau}%
_{i}^{(n+1)}$, depending on the parameters, and a finite nonevanescent local
functional $\check{V}_{\hspace{0.01in}\theta }^{(n+1)}$ of order $\hbar
^{n+1}$ such that%
\begin{equation*}
\check{U}_{n\text{nonev}}^{(n+1)}=\sum_{i}\check{\sigma}_{i}^{(n+1)}\mathcal{%
G}_{i}+\sum_{i}\check{\tau}_{i}^{(n+1)}\mathcal{\check{G}}_{i}+(\check{S}%
_{d},\check{V}_{\hspace{0.01in}\theta }^{(n+1)}).
\end{equation*}%
If we set $\check{\lambda}=\check{\eta}=0$ in this equation, we obtain%
\begin{equation}
\tilde{U}_{n\text{nonev}}^{(n+1)}=\sum_{i}\bar{\sigma}_{i}^{(n+1)}\mathcal{G}%
_{i}+\sum_{i}\bar{\tau}_{i}^{(n+1)}\mathcal{\check{G}}_{i}+(S_{d},\bar{V}_{%
\hspace{0.01in}\theta }^{(n+1)}),  \label{inn}
\end{equation}%
where $\bar{\sigma}_{i}^{(n+1)}$, $\bar{\tau}_{i}^{(n+1)}$ and $\bar{V}_{%
\hspace{0.01in}\theta }^{(n+1)}$ are equal to $\check{\sigma}_{i}^{(n+1)}$, $%
\check{\tau}_{i}^{(n+1)}$ and $\check{V}_{\hspace{0.01in}\theta }^{(n+1)}$
at $\check{\lambda}=\check{\eta}=0$. However, $\tilde{U}_{n\text{nonev}%
}^{(n+1)}$ and $S_{d}$ are invariant under $G_{\text{nas}}$, while the
functionals $\mathcal{\check{G}}_{i}$ are not. If we average on $G_{\text{nas%
}}$ (which we can do, since $G_{\text{nas}}$ is assumed to be compact), the $%
\mathcal{\check{G}}_{i}$ disappear or give linear combinations of the
invariants $\mathcal{G}_{i}$, and $\bar{V}_{\theta }^{(n+1)}$ turns into
some $\tilde{V}_{\hspace{0.01in}\theta }^{(n+1)}$. We obtain\footnote{%
If the terms proportional to the $S_{c}$ field equations are dropped from $%
S_{c}$, the average on $G_{\text{nas}}$ may generate them back. In the case
of general covariance, local Lorentz symmetry and Yang-Mills symmetries, the
average of $\mathcal{\check{G}}_{i}$ may also affect $\tilde{V}_{\hspace{%
0.01in}\theta }^{(n+1)}$, besides the coefficients $\sigma _{i}^{(n+1)}$,
but the final result is still of the form (\ref{uttila}).}%
\begin{equation}
\tilde{U}_{n\text{nonev}}^{(n+1)}=\sum_{j}\sigma _{j}^{(n+1)}\frac{\partial
S_{d}}{\partial \lambda _{j}}+(S_{d},\tilde{V}_{\hspace{0.01in}\theta
}^{(n+1)}),  \label{uttila}
\end{equation}%
for some new constants $\sigma _{j}^{(n+1)}$. We have used $\mathcal{G}%
_{i}=\partial S_{d}/\partial \lambda _{j}$. At this point, we apply\ the
canonical transformation (\ref{trasfa}) again, and note that, by formula (%
\ref{bu}) the difference between the transformed $\partial S_{d}/\partial
\lambda _{j}$ and $\partial S_{d\theta }/\partial \lambda _{j}$ is equal to $%
(S_{d\theta },X_{\theta })$ for some local functional $X_{\theta }$. In the
end, we get%
\begin{equation}
U_{n\text{nonev}}^{(n+1)}=\sum_{j}\sigma _{j}^{(n+1)}\frac{\partial
S_{d\theta }}{\partial \lambda _{j}}+(S_{d\theta },V_{\hspace{0.01in}\theta
}^{(n+1)})  \label{unev}
\end{equation}%
for some new local functional $V_{\hspace{0.01in}\theta }^{(n+1)}$ of order $%
\hbar ^{n+1}$. Now, define%
\begin{eqnarray*}
U_{n+1} &=&U_{R\theta }-\sum_{j}\rho _{n+1\hspace{0.01in}j}\frac{\partial
S_{R\theta }}{\partial \lambda _{j}}-(S_{R\theta },W_{n+1})-\frac{i}{2}%
(S_{R\theta },S_{R\theta })\hspace{0.01in}W_{n+1}, \\
\rho _{n+1\hspace{0.01in}j} &=&\rho _{n\hspace{0.01in}j}+\sigma
_{j}^{(n+1)},\qquad W_{n+1}=W_{n\hspace{0.01in}}+V_{R\theta }^{(n+1)},\qquad
\end{eqnarray*}%
where $V_{R\theta }^{(n+1)}$ are the renormalized versions of the
functionals $V_{\theta }^{(n+1)}$. Using (\ref{iondu}), we also have%
\begin{equation}
U_{n+1}=U_{n}-\sum_{j}\sigma _{j}^{(n+1)}\frac{\partial S_{R\theta }}{%
\partial \lambda _{j}}-(S_{R\theta },V_{R\theta }^{(n+1)})-\frac{i}{2}%
(S_{R\theta },S_{R\theta })\hspace{0.01in}V_{R\theta }^{(n+1)}.  \label{both}
\end{equation}%
Recall that $S=S_{d}+S_{\text{ev}}$, which implies $S_{\theta }=S_{d\theta }+%
\mathcal{O}(\varepsilon )$ and $S_{R\theta }=S_{d\theta }+\mathcal{O}%
(\varepsilon )+\mathcal{O}(\hbar )$. Taking the average of both sides of (%
\ref{both}), and using (\ref{basic1}), (\ref{basic2}), (\ref{unev}) and then
(\ref{iu2}), we find 
\begin{equation*}
\left\langle U_{n+1}\right\rangle _{\Gamma }=\left\langle U_{n}\right\rangle
_{\Gamma }-U_{n\text{nonev}}^{(n+1)}+\mathcal{O}(\varepsilon )+\mathcal{O}%
(\hbar ^{n+2})=\mathcal{O}(\varepsilon )+\mathcal{O}(\hbar ^{n+2}),
\end{equation*}%
which extends the inductive assumption (\ref{iu}) to the order $n+1$.
Formula (\ref{finiY}) follows from formula (\ref{finiYn}) for $n=\infty $,
with $\rho _{j}=\rho _{\infty \hspace{0.01in}j}$ and $W_{R\theta }=W_{\infty
}=\mathcal{O}(\hbar )$.

Finally, using (\ref{finiY}) inside (\ref{pril}), we get%
\begin{equation*}
\frac{\partial \Gamma _{R\theta }}{\partial \theta }=\sum_{j}\rho _{j}\frac{%
\partial \Gamma _{R\theta }}{\partial \lambda _{j}}+(\Gamma _{R\theta
},\langle \tilde{Q}_{R\theta }+W_{R\theta }\rangle )+\mathcal{O}(\varepsilon
).
\end{equation*}%
This formula is equivalent to (\ref{geneqgd2}) with the identification $%
H_{R\theta }=\tilde{Q}_{R\theta }+W_{R\theta }$. Observe that $H_{R\theta }$
is another renormalized version of the functional $\tilde{Q}_{\theta }(\Phi
,K)$, and just differs from $\tilde{Q}_{R\theta }$ by a choice of
subtraction scheme.

\subsection{Integrating the new equations and RG\ invariance}

\label{s33}

Now we integrate the equations (\ref{geneqgd2}). We can easily absorb away
the first term on the right-hand side by making finite redefinitions $%
\lambda (\lambda ^{\prime },\theta )$ of the parameters $\lambda $. We
choose functions $\lambda _{i}(\lambda ^{\prime },\xi ,\eta ,\theta )$ that
solve the evolution equations 
\begin{equation}
\frac{\partial \lambda _{i}}{\partial \theta }=-\rho _{i}(\lambda ,\xi ,\eta
,\theta ),  \label{rgg2}
\end{equation}%
with the initial conditions $\lambda _{i}(\lambda ^{\prime },\xi ,\eta
,0)=\lambda _{i}^{\prime }$. Using formulas (\ref{geneqgd2}) and (\ref{defi}%
), we obtain 
\begin{equation}
\frac{\partial \bar{\Gamma}_{R\theta }}{\partial \theta }=(\bar{\Gamma}%
_{R\theta },\overline{\langle H_{R\theta }\rangle })+\mathcal{O}(\varepsilon
),  \label{blu}
\end{equation}%
where $\bar{\Gamma}_{R\theta }$ is related to $\Gamma _{R\theta }$ according
to the definition $\lambda (\lambda ^{\prime },\xi ,\eta ,\theta )$ [see (%
\ref{defo}) and the arguments given right after that formula].

Observe that equation (\ref{blu}) is equivalent to formula (\ref{resi2}) of
section \ref{s8}. This means the redefinitions $\lambda _{i}(\lambda
^{\prime },\xi ,\eta ,\theta )$ perform the re-fine-tuning of finite local
counterterms (automatically incorporated in the approach of section \ref{s8}%
) that was missing so far in the approach of the present section. As in
subsection \ref{s34}, equation (\ref{blu}) can be integrated with the method
of appendix C. We find that there exists a canonical transformation $\Phi
,K\rightarrow \Phi ^{\prime },K^{\prime }$ such that the $\Gamma $
functional $\Gamma _{R}^{\prime }$ defined by 
\begin{equation}
\Gamma _{R}^{\prime }(\Phi ^{\prime },K^{\prime },\lambda ^{\prime },\xi
,\eta )=\Gamma _{R\theta }(\Phi (\Phi ^{\prime },K^{\prime },\lambda
^{\prime },\xi ,\eta ,\theta ),K(\Phi ^{\prime },K^{\prime },\lambda
^{\prime },\xi ,\eta ,\theta ),\lambda (\lambda ^{\prime },\xi ,\eta ,\theta
),\xi ,\eta ,\theta )  \label{grp0}
\end{equation}%
is $\theta $ independent, up to $\mathcal{O}(\varepsilon )$. Since $\theta
=0 $ gives $\Gamma _{R}^{\prime }=\Gamma _{R}$, we also have%
\begin{equation*}
\Gamma _{R}^{\prime }(\Phi ^{\prime },K^{\prime },\lambda ^{\prime },\xi
,\eta )=\Gamma _{R\theta }(\Phi ^{\prime },K^{\prime },\lambda ^{\prime
},\xi ,\eta ,0)=\Gamma _{R}(\Phi ^{\prime },K^{\prime },\lambda ^{\prime
},\xi ,\eta ).
\end{equation*}%
Inverting the transformations, we obtain the formula 
\begin{equation}
\Gamma _{R\theta }(\Phi ,K,\lambda ,\xi ,\eta ,\theta )=\Gamma _{R}(\Phi
^{\prime }(\Phi ,K,\lambda ,\xi ,\eta ,\theta ),K^{\prime }(\Phi ,K,\lambda
,\xi ,\eta ,\theta ),\lambda ^{\prime }(\lambda ,\xi ,\eta ,\theta ),\xi
,\eta ),  \label{gru}
\end{equation}%
which shows that the dependence of $\Gamma _{R\theta }$ on the gauge
parameter $\theta $ can be fully absorbed inside a finite redefinition of
the parameters $\lambda $ and a canonical transformation.

According to formulas (\ref{grp0}) and (\ref{gru}), the beta functions $%
\beta _{\lambda ^{\prime }}^{\prime }$ of the parameters $\lambda ^{\prime }$
(in the framework where the fields and the sources have primes) are $\theta $
independent. That means, however, that the beta functions $\beta _{\lambda }$
of the couplings $\lambda $ do depend on $\theta $.\ However, their $\theta $
dependence is not arbitrary, because it disappears by making the
redefinitions $\lambda (\lambda ^{\prime },\xi ,\eta ,\theta )$.

We can repeat the argument for any other gauge parameter $\theta $ for which
formula (\ref{ahid}) is known to hold, taking one at a time. Since the
composition of special canonical transformations and redefinitions of
parameters is a special canonical transformation combined with a
redefinition of parameters, we reach the conclusion that the entire
dependence on the gauge parameters can be absorbed into such operations,
which do not affect the physical quantities (see subsection \ref{s43}).%
\newline

We can also repeat the arguments of section \ref{s24} and prove that RG
invariance is preserved. The difference is that now instead of (\ref{rg2})
we get a transformed Callan-Symanzik equation that contains $\theta $%
-dependent beta functions.

\section{Gauge independence and unitarity}

\label{s4}

In general gauge theories we need to introduce extra fields, such as the
Fedeev-Popov ghosts $C$, the antighosts $\bar{C}$ and the Lagrange
multipliers $B$, and choose gauge-fixing conditions to make the functional
integral perturbatively well defined. In addition, to implement the
renormalization of divergences to all orders, study the gauge dependence and
prove the Adler-Bardeen theorem, it is also convenient to introduce the
sources $K$ and use the Batalin-Vilkovisky formalism. The extra fields and
the sources must be switched off at some point. In this section we explain
how to define the physical quantities and show that they are gauge
independent, under the sole assumption that the theory is AB nonanomalous, as in section \ref{s8}. We work with convergent functionals, so we can set $\varepsilon
=0$. We denote the $\varepsilon \rightarrow 0$ limits of $\Gamma _{R}$ and
the other functionals involved in our arguments by the same symbols used so
far, since no confusion is expected to arise.

First, we need to \textquotedblleft un-gauge-fix\textquotedblright\ the
theory, by switching off $\bar{C}$, $B$ and their sources $K_{\bar{C}}$, $%
K_{B}$. This operation is regular inside the $\Gamma $ functionals, once
Feynman diagrams have been evaluated, but not inside the actions $S$ and $%
S_{R}$, in the sense that if we un-gauge-fix the action, Feynman diagrams
obviously become ill defined. For this reason, some gauge dependence
survives the un-gauge-fixing procedure. Besides un-gauge-fixing, we must
switch off the sources $K$. The combined switch-off procedure allows us to
define a physical $\Gamma $ functional, identify its gauge symmetries, check
that they close on-shell, and prove that no gauge dependence affects the
physical quantities.

Since the gauge fixing is introduced by means of a canonical transformation,
such as (\ref{cang}), when we vary the gauge-fixing parameters $\theta =\xi $
we make a canonical transformation. Therefore, the equations (\ref{resi2})
and (\ref{geneqgd2}) can be used to study the dependence of the physical
quantities on the parameters $\xi $.

The information gathered so far is encoded in the key formulas%
\begin{eqnarray}
(\Gamma _{R\theta },\Gamma _{R\theta }) &=&0,  \label{equa1} \\
\frac{\partial \Gamma _{R\theta }}{\partial \theta }-\sum_{j}\rho _{j}\frac{%
\partial \Gamma _{R\theta }}{\partial \lambda _{j}}-(\Gamma _{R\theta
},\langle H_{R\theta }\rangle ) &=&0,  \label{equa2}
\end{eqnarray}%
and is sufficient to achieve the goals of this section. We work on the $%
\varepsilon \rightarrow 0$ limit of (\ref{geneqgd2}), rather than the one of
(\ref{resi2}), because everything we say starting from the former can be
easily generalized to the other case.

\subsection{Quantum gauge algebra}

\label{s41}

Formula (\ref{equa1}) gives 
\begin{equation}
0=-\int \frac{\delta _{r}\Gamma _{R\theta }}{\delta K_{\alpha }}\frac{\delta
_{l}\Gamma _{R\theta }}{\delta \Phi ^{\alpha }}=-\int \left\langle \frac{%
\delta _{r}S_{R\theta }}{\delta K_{\alpha }}\right\rangle \frac{\delta
_{l}\Gamma _{R\theta }}{\delta \Phi ^{\alpha }}=\int \langle (S_{R\theta
},\Phi ^{\alpha })\rangle \frac{\delta _{l}\Gamma _{R\theta }}{\delta \Phi
^{\alpha }},  \label{gin}
\end{equation}%
and tells us that $\Gamma _{R\theta }$ is invariant under the infinitesimal
(nonlocal) transformations 
\begin{equation*}
\Phi ^{\alpha }\rightarrow \Phi ^{\alpha }+\delta \Phi ^{\alpha },\qquad
\delta \Phi ^{\alpha }\equiv \varpi \left\langle (S_{R\theta },\Phi ^{\alpha
})\right\rangle =-\varpi \frac{\delta _{r}\Gamma _{R\theta }}{\delta
K_{\alpha }}.
\end{equation*}%
Here and below $\varpi $, $\varpi ^{\prime }$, etc., denote constant
anticommuting parameters. Write $\Phi ^{\alpha }=\{\phi ^{i},C^{a},\bar{C}%
^{a},B^{a}\}$ and $K_{\alpha }=\{K_{\phi }^{i},K_{C}^{a},K_{\bar{C}%
}^{a},K_{B}^{a}\}$, to separate the classical fields $\phi ^{i}$ and their
sources $K_{\phi }^{i}$ from the extra fields and their sources.

Observe that $S$ is independent of $K_{B}$ and contains $K_{\bar{C}}$ only
through the term $-\int B^{a}K_{\bar{C}}^{a}$. This is also true after the
canonical transformation (\ref{trasfa}), if we assume, for simplicity, that
the functional $Q(\Phi ,K^{\prime })$ appearing in (\ref{trasfa}) is
independent of $K_{\bar{C}}$ and $K_{B}$. Then $S_{\theta }$ also satisfies $%
(S_{\theta },\bar{C})=B$ and $(S_{\theta },B)=0$. Moreover, the sources $K_{%
\bar{C}}$ and $K_{B}$ cannot contribute to any nontrivial one-particle
irreducible diagrams. Thus, after renormalization we still have $(S_{R\theta
},\bar{C})=B$ and $(S_{R\theta },B)=0$, i.e. $\delta \bar{C}^{a}=\varpi
B^{a} $ and $\delta B^{a}=0$.

Define%
\begin{equation*}
\hat{\Gamma}_{R}(\phi )\equiv \left. \Gamma _{R\theta }(\Phi ,K)\right\vert
_{\bar{C}=B=K=0},\qquad \hat{\delta}\Phi ^{\alpha }=\left. \delta \Phi
^{\alpha }\right\vert _{\bar{C}=B=K=0}.
\end{equation*}%
Observe that $\hat{\Gamma}_{R}(\phi )$ is independent of the ghosts $C$,
because it has ghost number zero and after suppressing $\bar{C}$ and $K$ no
fields and/or sources of negative ghost numbers survive. For the same
reason, $\hat{\delta}\phi ^{i}$, which has ghost number equal to one, is
linear in $C$. Clearly, $\hat{\delta}\bar{C}=\hat{\delta}B=0$. Thus, when $%
\bar{C}$, $B$ and $K$ are switched off, formula (\ref{gin}) turns into 
\begin{equation}
0=\int \hat{\delta}\phi ^{i}\frac{\delta _{l}\hat{\Gamma}_{R}(\phi )}{\delta
\phi ^{i}}.  \label{more}
\end{equation}%
The terms proportional to $\delta _{l}\Gamma _{R\theta }/\delta C$ do not
contribute to (\ref{more}) because $\hat{\Gamma}_{R}(\phi )$ is $C$
independent. The terms proportional to $\delta _{l}\Gamma _{R\theta }/\delta 
\bar{C}$ and $\delta _{l}\Gamma _{R\theta }/\delta B$ disappear, because
they multiply $\hat{\delta}\bar{C}$ and $\hat{\delta}B$, respectively.

We call $\hat{\Gamma}_{R}(\phi )$ the \textquotedblleft
physical\textquotedblright\ $\Gamma $ functional. The transformations $\hat{%
\delta}\phi ^{i}$ encode the gauge symmetry of $\hat{\Gamma}_{R}$. Indeed,
recall that $\hat{\delta}\phi ^{i}$ is linear in $C$ and of course $\varpi $%
. Replacing each ghost $C$ with $\varpi ^{\prime }\Lambda $, where $\Lambda
(x)$ is a function having statistics opposite to the one of $C$, and
dropping the products\ $\varpi \varpi ^{\prime }$ after moving them to the
left, we can define a symmetry transformation $\delta _{\Lambda }\phi ^{i}$
by the formula 
\begin{equation*}
\varpi \varpi ^{\prime }\delta _{\Lambda }\phi ^{i}=\left. \hat{\delta}\phi
^{i}\right\vert _{C\rightarrow \varpi ^{\prime }\Lambda }
\end{equation*}%
and prove, using equation (\ref{more}), that $\hat{\Gamma}_{R}(\phi )$ is
invariant under this symmetry:%
\begin{equation*}
\delta _{\Lambda }\hat{\Gamma}_{R}(\phi )=\int \delta _{\Lambda }\phi ^{i}%
\frac{\delta _{l}\hat{\Gamma}_{R}(\phi )}{\delta \phi ^{i}}=0.
\end{equation*}%
We call $\delta _{\Lambda }\phi ^{i}$ the quantum gauge transformations. To
the lowest order in $\hbar $ they coincide with the starting gauge
transformations, but at higher orders they are in general nonlocal
functionals. We call the algebra of the transformations $\delta _{\Lambda }$ 
\textit{quantum gauge algebra}.

\subsection{Closure of the quantum gauge algebra}

\label{s42}

Now we study the closure of the quantum gauge algebra. If we differentiate (%
\ref{equa1}) with respect to $K$, we obtain%
\begin{equation*}
(\Gamma _{R\theta },\delta \Phi ^{\alpha })=0.
\end{equation*}%
Consider this equation in the case $\delta \Phi ^{\alpha }\rightarrow \delta
\phi ^{i}$, then switch off $\bar{C}$ and $B$, and set $K=0$ at the end.
Recalling that $\delta \bar{C}=\varpi B$ and $\delta B=0$, and observing
that $\delta \phi ^{i}$ does not depend on $K_{\bar{C}}$ and $K_{B}$, we
obtain%
\begin{equation}
\int \hat{\delta}^{\prime }\phi ^{j}\frac{\delta _{l}(\hat{\delta}\phi ^{i})%
}{\delta \phi ^{j}}+\int \hat{\delta}^{\prime }C^{a}\frac{\delta _{l}(\hat{%
\delta}\phi ^{i})}{\delta C^{a}}=\int \left. \frac{\delta _{r}(\delta \phi
^{i}\varpi ^{\prime })}{\delta K_{\phi }^{j}}\right\vert _{\bar{C}=B=K=0}%
\frac{\delta _{l}\hat{\Gamma}_{R}(\phi )}{\delta \phi ^{j}},  \label{closure}
\end{equation}%
having multiplied to the left by $\varpi ^{\prime }$ and having defined $%
\hat{\delta}^{\prime }\Phi ^{\alpha }=\left. \hat{\delta}\Phi ^{\alpha
}\right\vert _{\varpi \rightarrow \varpi ^{\prime }}$. The right-hand side
of (\ref{closure}) is proportional to the $\phi ^{j}$ \textquotedblleft $%
\Gamma $ field equations\textquotedblright , which means that closure is
achieved on shell. The left-hand side of (\ref{closure}) can be handled as
follows. Since $\hat{\delta}\phi ^{i}$ and $\hat{\delta}C^{a}$ are linearly
and quadratically proportional to the ghosts, respectively, we can write
them in the form 
\begin{equation*}
\hat{\delta}\phi ^{i}=\varpi \int C^{\bar{a}}T_{\bar{a}}^{i}(\phi ),\qquad 
\hat{\delta}C^{a}=-\frac{1}{2}\int C^{\bar{b}}\varpi C^{\bar{c}}T_{\bar{b}%
\bar{c}}^{a}(\phi ),
\end{equation*}%
where $T_{\bar{a}}^{i}$ and $T_{\bar{b}\bar{c}}^{a}$ are nonlocal
functionals. Here the bar indices include the spacetime points where the
corresponding fields are located and the summation over repeated bar indices
understands the integration over those spacetime points. Now, take formula (%
\ref{closure}) and replace $C^{a}$ with $\varpi ^{\prime \prime }\Lambda
^{a}+\varpi ^{\prime \prime \prime }\Sigma ^{a}$, $\Lambda ^{a}$ and $\Sigma
^{a}$ being functions of the coordinates. The left-hand side of (\ref%
{closure}) is turned into $\varpi \varpi ^{\prime }\varpi ^{\prime \prime
}\varpi ^{\prime \prime \prime }$ times%
\begin{equation*}
\int \delta _{\Lambda }\phi ^{j}\frac{\delta _{l}(\delta _{\Sigma }\phi ^{i})%
}{\delta \phi ^{j}}-\int \delta _{\Sigma }\phi ^{j}\frac{\delta _{l}(\delta
_{\Lambda }\phi ^{i})}{\delta \phi ^{j}}-\int \Lambda ^{\bar{a}}\Sigma ^{%
\bar{b}}T_{\bar{a}\bar{b}}^{\bar{c}}(\phi )T_{\bar{c}}^{i}(\phi ).
\end{equation*}%
Finally, the whole formula (\ref{closure}) is equivalent%
\begin{equation}
\lbrack \delta _{\Lambda },\delta _{\Sigma }]\phi ^{i}=\delta _{\lbrack
\Lambda ,\Sigma ]}\phi ^{i}+\int v^{ij}(\phi ,\Lambda ,\Sigma )\frac{\delta
_{l}\hat{\Gamma}_{R}(\phi )}{\delta \phi ^{j}},  \label{clos}
\end{equation}%
where%
\begin{equation*}
\lbrack \Lambda ,\Sigma ]^{a}=\int \Lambda ^{\bar{b}}\Sigma ^{\bar{c}}T_{%
\bar{b}\bar{c}}^{a}(\phi )
\end{equation*}%
and $v^{ij}(\phi ,\Lambda ,\Sigma )$ are suitable functions. Formula (\ref%
{clos}) expresses the on shell closure of the quantum gauge algebra.

The field transformations and the closure relations become clearer if we
switch to a more explicit notation, where they read 
\begin{equation*}
\delta _{\Lambda }\phi ^{i}(x)=\int \mathrm{d}^{d}y\hspace{0.01in}\Lambda
^{a}(y)T_{a}^{i}[\phi ](x,y),\qquad \lbrack \Lambda ,\Sigma ]^{a}(x)=\int 
\mathrm{d}^{d}y\mathrm{d}^{d}z\hspace{0.01in}\Lambda ^{b}(y)\Sigma
^{c}(z)T_{bc}^{a}[\phi ](x,y,z),
\end{equation*}%
$T_{a}^{i}[\phi ]$ and $T_{bc}^{a}[\phi ]$ being (nonlocal) functionals that
depend on two and three spacetime points, respectively.

\subsection{\texorpdfstring{Gauge
dependence
of
the
physical
$%
\Gamma
$
functional}{}}

\label{s43}

The last goal is to study the gauge dependence of $\hat{\Gamma}_{R}(\phi )$.
Observe that the functional $\langle H_{R\theta }\rangle $ that appears in
formula (\ref{equa2}) has ghost number equal to $-1$. Therefore, it must be
proportional to the antighosts $\bar{C}$ and/or some sources $K$. This fact
implies that the derivatives $\delta _{l}\langle H_{R\theta }\rangle /\delta
\phi ^{i}$ and $\delta _{l}\langle H_{R\theta }\rangle /\delta C^{a}$ are
zero at $\bar{C}=K=0$. Moreover, $\langle H_{R\theta }\rangle $ does not
depend on $K_{\bar{C}}$ and $K_{B}$, if the functional $Q(\Phi ,K^{\prime })$
of (\ref{trasfa}) satisfies the same property, as we are assuming here.
Setting $\bar{C}=B=K=0$ in (\ref{equa2}) we obtain 
\begin{equation}
\frac{\partial \hat{\Gamma}_{R}(\phi )}{\partial \theta }=\sum_{j}\rho _{j}%
\frac{\partial \hat{\Gamma}_{R}(\phi )}{\partial \lambda _{j}}+\int
u^{i}(\phi )\frac{\delta _{l}\hat{\Gamma}_{R}(\phi )}{\delta \phi ^{i}},
\label{equa3}
\end{equation}%
where 
\begin{equation*}
u^{i}(\phi )=\left. \frac{\delta _{r}\langle H_{R\theta }\rangle }{\delta
K_{\phi }^{i}}\right\vert _{_{\bar{C}=B=K=0}}.
\end{equation*}%
Formula (\ref{equa3}) is the equation of gauge dependence satisfied by the
physical functional $\hat{\Gamma}_{R}(\phi )$. We can integrate it with the
procedure described in subsection \ref{s33}. The first term on the
right-hand side of (\ref{equa3}) can be absorbed into redefinitions of the
parameters $\lambda $, while the second term can be absorbed into a change
of field variables. We can do this for each gauge parameter $\theta $,
taking one at a time. We obtain that there exists redefinitions $\lambda
(\lambda ^{\prime },\theta )$ and a change of field variables $\phi (\phi
^{\prime },\lambda ^{\prime },\theta )$ such that the transformed physical
functional 
\begin{equation*}
\hat{\Gamma}_{R}^{\prime }(\phi ^{\prime },\lambda ^{\prime })=\hat{\Gamma}%
_{R}(\phi (\phi ^{\prime },\lambda ^{\prime },\theta ),\lambda (\lambda
^{\prime },\theta ),\theta )
\end{equation*}%
is $\theta $ independent. Setting $\theta =0$ we get $\hat{\Gamma}%
_{R}^{\prime }(\phi ^{\prime },\lambda ^{\prime })=\hat{\Gamma}_{R}(\phi
^{\prime },\lambda ^{\prime },0)$, which in the end allows us to write%
\begin{equation*}
\hat{\Gamma}_{R}(\phi ,\lambda ,\theta )=\hat{\Gamma}_{R}(\phi ^{\prime
}(\phi ,\lambda ,\theta ),\lambda ^{\prime }(\lambda ,\theta ),0).
\end{equation*}

Since the entire gauge dependence is encoded into changes of field variables
and redefinitions of parameters, it cannot affect the physical quantities
contained in $\hat{\Gamma}_{R}(\phi )$.

\subsection{Unitarity}

\label{s44}

In this subsection we prove (perturbative) unitarity, to emphasize why gauge
independence is so crucial. For definiteness, we illustrate our arguments in
Yang-Mills theories, but everything we say can be applied to quantum
gravity, as well as any general gauge theory. We recall that perturbative
unitarity is the statement that the identity $SS^{\dag }=1$ holds
diagrammatically, order by order in the perturbative expansion \cite%
{thooftveltman}. A necessary condition is that the free-field theory we
perturb around propagates only physical degrees of freedom. A necessary and
sufficient condition is that when the identity $SS^{\dag }=1$ is written as
a cutting equation no unphysical degrees of freedom contribute to the cut
propagators.

There exists no gauge-fixing conditions where both unitarity and the
locality of counterterms are manifest. If we want manifest unitarity,
propagators must have only physical poles. This happens when we choose
gauge-fixing functions of the Coulomb type, such as $G(\phi )=\partial
_{i}A_{i}$, where $i,j,\ldots $ are space indices, inside the gauge fermion $%
\Psi (\Phi )$\ of (\ref{psif}). However, the locality of counterterms is not
manifest in that gauge, since the Coulomb propagators contain denominators
whose dominant terms (those that determine their ultraviolet behavior) do
not depend on the energy (or do not depend on it in the correct way). Then,
when we differentiate a Feynman diagram with respect to the energies of its
external legs, the overall degree of divergence is not guaranteed to
decrease, so we cannot prove the locality of counterterms in this way.
Besides having a bad power-counting behavior at high energies, the
propagators of the Coulomb gauge generate spurious divergences that are
difficult to handle.

To have a good power-counting behavior we need to equip the propagators with
extra poles, some of which are unphysical. This is achieved for example by
choosing the Lorenz gauge-fixing function $G(\phi )=\partial ^{\mu }A_{\mu }$
in (\ref{psif}). The Fadeev-Popov ghosts then also have poles. The locality
of counterterms is manifest, but unitarity is not.

The extra poles must cancel somehow, but their mutual compensation is not
evident. The best way to prove this compensation is to use the gauge
independence of the physical amplitudes, which allows us to switch back and
forth between gauge-fixing conditions of the Lorentz type and gauge-fixing
conditions of the Coulomb type. The former make the locality of counterterms
manifest and hide unitarity, while the latter make unitarity manifest and
hide the locality of counterterms.

For example, choose the gauge fermion%
\begin{equation}
\Psi (\Phi )=\int \bar{C}^{a}\left( \zeta \partial _{0}A_{0}^{a}-\partial
_{i}A_{i}^{a}+\frac{\xi }{2}B^{a}\right) ,  \label{lc}
\end{equation}%
which contains two gauge-fixing parameters, $\xi $ and $\zeta $. This
functional interpolates between the Lorenz gauge ($\zeta =1$) and the
Coulomb gauge ($\zeta =0$). After integrating $B$ out, the propagators of
the gauge fields are 
\begin{eqnarray}
\left\langle A_{0}(k)\hspace{0.01in}A_{0}(-k)\right\rangle _{0} &=&-\frac{%
i\xi ^{2}}{P(k)}(\xi E^{2}-\bar{k}^{2}),\qquad \left\langle A_{i}(k)\hspace{%
0.01in}A_{0}(-k)\right\rangle _{0}=\frac{i\xi ^{2}}{P(k)}(\zeta -\xi )Ek_{i},
\notag \\
\left\langle A_{i}(k)\hspace{0.01in}A_{j}(-k)\right\rangle _{0} &=&\frac{i}{%
E^{2}-\bar{k}^{2}}\left( \delta _{ij}-\frac{k_{i}k_{j}}{\bar{k}^{2}}\right)
+i(\zeta ^{2}E^{2}-\xi ^{2}\bar{k}^{2})\frac{\xi k^{i}k^{j}}{\bar{k}^{2}P(k)}%
,  \label{bisa1}
\end{eqnarray}%
where $\bar{k}^{2}=k_{i}k_{i}$ and%
\begin{equation*}
P(k)=\xi (\zeta E^{2}-\xi \bar{k}^{2})^{2}-\bar{k}^{2}(1-\xi )(\zeta
^{2}E^{2}-\xi ^{2}\bar{k}^{2}),
\end{equation*}%
while the ghost propagator is 
\begin{equation}
\left\langle C(k)\bar{C}(-k)\right\rangle =\frac{i}{\zeta E^{2}-\bar{k}^{2}}.
\label{bisa2}
\end{equation}%
We see that the propagators are well behaved, from the point of view of
power counting, whenever $\zeta \neq 0$. They are not well behaved for $%
\zeta =0$, which is the Coulomb limit. The parameter $\zeta $ is a sort of
cutoff that regulates the spurious divergences of the Coulomb gauge.
Moreover, at $\zeta =0$ $P(k)$ is equal to $\xi ^{2}(\bar{k}^{2})^{2}$ and
only the physical poles survive. Instead, unphysical poles are present
whenever $\zeta \neq 0$.

In the previous sections we have proved that the physical quantities are
gauge independent. In particular, they are independent of $\xi $ and $\zeta $%
. Thus, they are also unitary, and obey the locality of counterterms. We see
that they are unitary by taking $\zeta =0$. We see that they obey the
locality of counterterms by taking $\zeta \neq 0$.

In the case of the standard model in flat space, we can easily generalize
the proof of the Adler-Bardeen theorem given in ref. \cite{ABrenoYMLR} to
the family of gauge fermions (\ref{lc}), because they are all
renormalizable. Then, the remarks of this subsection allow us to infer that
the standard model in flat space is perturbatively unitary.

In ref. \cite{ABnonreno} a more general proof of the Adler-Bardeen theorem
was given. It holds in a large class of nonrenormalizable theories, which
includes the standard model coupled to quantum gravity. Combining the
results of \cite{ABnonreno} with those of section \ref{s8}, we can extend
the validity of the Adler-Bardeen theorem to the most general local gauge
fermions. In particular, using an analogue of (\ref{lc}), to switch between
the Lorenz and Coulomb gauges of diffeomorphisms and Yang-Mills symmetries,
we infer that the standard model coupled to quantum gravity is unitary as a
perturbative quantum field theory. So are its extensions, as long as they
satisfy the assumptions we have made.

We stress again that gauge independence is crucial to reach these
conclusions, since the Adler-Bardeen theorem \textit{per se} ensures gauge
invariance, but not gauge independence.

\section{Checks of high-order calculations based on gauge independence}

\label{s5}

\setcounter{equation}{0}

In this section we discuss how to use the results of this paper to check
high-order calculations, under the assumptions of section \ref{s3}. We have
proved that

\begin{proposition}
\textit{The beta functions of the physical parameters }$\lambda $\textit{\
may depend on the gauge parameters }$\xi $\textit{, but that dependence can
always be reabsorbed into finite }$\lambda $\textit{\ redefinitions.}\label%
{prop1}
\end{proposition}

This proposition also reminds us that there exists a class of subtraction
schemes where the beta functions are gauge independent, in agreement with
the general theorem proved in section \ref{s8}. If we are extremely lucky,
the framework we choose to simplify high-order calculations might belong to
that class. In ordinary situations, we may expect to be lucky only to the
lowest orders, which may mean till three or four loops, or for special
choices of the gauge fixing. However, we may not be able to identify the
right framework in advance. Therefore, contrary to the usual lore, in
general we cannot make checks of high-order calculations based on the
assumption that $\lambda $ beta functions are completely gauge independent.

Nevertheless, the beta functions cannot be gauge dependent in an arbitrary
way, precisely because their gauge dependence must disappear in a suitable
class of subtraction schemes. Thanks to this, a criterion to make checks of
high-order calculations, based on gauge independence, still exists. It
amounts to verify that every $\xi $ dependence contained in the $\lambda $
beta functions can be cancelled by means of finite $\lambda $ redefinitions.
In this section we show that the correct criterion, although less powerful
than expected, is nontrivial and powerful enough.

For definiteness, consider the standard model in flat space, and let $%
\lambda _{i}$ collect the $\varphi ^{4}$ coupling, the squared gauge
couplings, and the squared Yukawa couplings. The most general $\lambda $
beta functions have the form 
\begin{equation}
\beta _{i}=\sum_{n=2}^{\infty }\hbar ^{n-1}\chi _{i_{1}\cdots i_{n}i}\lambda
_{i_{1}}\cdots \lambda _{i_{n}},  \label{bl}
\end{equation}%
where $\chi _{ii_{1}\cdots i_{n}}$ are constants and the powers of $\hbar $
are inserted to emphasize the order of the loop expansion. The most general\
perturbative $\lambda $ redefinitions can be parametrized as%
\begin{equation}
\lambda _{i}^{\prime }=\lambda _{i}+\sum_{n=2}^{\infty }\hbar
^{n-1}\vartheta _{ii_{1}\cdots i_{n}}\lambda _{i_{1}}\cdots \lambda _{i_{n}},
\label{mod}
\end{equation}%
where $\vartheta _{ii_{1}\cdots i_{n}}$ are other constants. We have%
\begin{eqnarray}
\beta _{i}^{\prime } &=&\sum_{n=2}^{\infty }\hbar ^{n-1}\chi _{i_{1}\cdots
i_{n}i}\lambda _{i_{1}}\cdots \lambda _{i_{n}}+\sum_{n=2}^{\infty }n\hbar
^{n-1}\vartheta _{ii_{1}\cdots i_{n}}\lambda _{i_{1}}\cdots \lambda
_{i_{n-1}}\sum_{m=2}^{\infty }\hbar ^{m-1}\chi _{k_{1}\cdots
k_{m}i_{n}}\lambda _{k_{1}}\cdots \lambda _{k_{m}}  \notag \\
&\equiv &\sum_{n=2}^{\infty }\hbar ^{n-1}\chi _{i_{1}\cdots i_{n}i}^{\prime
}\lambda _{i_{1}}^{\prime }\cdots \lambda _{i_{n}}^{\prime }.  \label{bl2}
\end{eqnarray}%
Proposition \ref{prop1} ensures that the gauge dependence contained in the
beta functions $\beta _{i}$ can be absorbed inside the redefinitions (\ref%
{mod}), that is to say there exist constants $\vartheta _{ii_{1}\cdots
i_{n}} $ such that the couplings $\lambda _{i}^{\prime }$ have gauge
independent beta functions $\beta _{i}^{\prime }$. Using this piece of
information, we can determine which nontrivial checks of high-order
calculations are available.

The one-loop coefficients $\chi _{i_{1}i_{2}i}$ cannot be changed, because
they are scheme independent ($\chi _{i_{1}i_{2}i}^{\prime }=\chi
_{i_{1}i_{2}i}$). Therefore, they are also gauge independent. Comparing (\ref%
{bl}) and (\ref{bl2}), we find that the other coefficients are related by
the formula%
\begin{equation}
\chi _{i_{1}\cdots i_{n}i}^{\prime }=\chi _{i_{1}\cdots
i_{n}i}+(n-1)\vartheta _{ij\{i_{1}\cdots i_{n-2}}\chi
_{i_{n-1}i_{n}\}j}-2\vartheta _{j\{i_{1}\cdots i_{n-1}}\chi
_{i_{n}\}ji}+\cdots ,  \label{chip}
\end{equation}%
where the dots stand for contributions involving $\vartheta _{i_{1}\cdots
i_{k}}$ with $k<n$. We can define an iterative procedure to determine $%
\vartheta _{i_{1}\cdots i_{n}}$ by assuming that the constants $\vartheta
_{i_{1}\cdots i_{k}}$ with $k<n$ are known, and requiring that $\chi
_{i_{1}\cdots i_{n}i}^{\prime }$ be gauge independent.

Now, if the number of couplings $\lambda $ is $N$, the tensors $\chi
_{i_{1}\cdots i_{\ell +1}i}$ have $c_{N,\ell }\equiv N\binom{N+\ell }{\ell +1%
}$ independent components \cite{schouten}, while the tensors $\vartheta
_{i_{1}\cdots i_{\ell }j}$ have $c_{N,\ell -1}$ components, where $\ell $ is
the number of loops. For $N=1$ (that is to say a single coupling $\lambda $)
and $\ell >2$\ it is always possible to absorb the gauge dependence into $%
\lambda $ redefinitions (as long as the one-loop coefficient $\chi $ of the
beta function does not vanish), because $c_{1,\ell }=c_{1,\ell -1}=1$. For $%
\ell =2$\ it is not possible, because the second and third terms on the
right-hand side of formula (\ref{chip}) cancel each other. Thus, two
nontrivial checks are available for $N=1$, due to the gauge independence of
the one-loop and two-loop coefficients of the beta function.

For $N>1$ more nontrivial checks of high-order calculations based on gauge
independence are available, because $c_{N,\ell }>c_{N,\ell -1}$. Proposition %
\ref{prop1} implies that the number of $\xi $-independent components of the
tensors $\chi _{i_{1}\cdots i_{\ell +1}i}$ is obtained by modding out the
redefinitions (\ref{mod}). Generically, this operation leaves%
\begin{equation*}
c_{N,\ell }-c_{N,\ell -1}=N\binom{N+\ell -1}{\ell +1}
\end{equation*}%
independent checks at $\ell $ loops. This number is $(\ell +N)/(N-1)$ times
less than the number we would obtain if the beta functions were completely
gauge independent. Indeed, in that case we would have $c_{N,\ell }$
independent checks at $\ell $ loops, which is equal to the number of
constants $\chi _{i_{1}\cdots i_{\ell +1}i}$.

So far, the beta functions of the standard model have been calculated to
three loops \cite{betas} and the results are fully independent of the
gauge-fixing parameters. Presumably, the convenient gauge-fixing functions
and the clever treatments of the matrix $\gamma _{5}$ used in refs. \cite%
{betas} project onto the class of subtraction schemes where the beta
functions are already gauge independent, at least to the lowest orders.
However, we may expect that this coincidence will stop, sooner or later.
When that happens, we must be aware of the facts pointed out in this
section. Moreover, we stress that in the proofs of properties to all orders,
such as the proof of the Adler-Bardeen theorem in nonrenormalizable theories 
\cite{ABnonreno}, it is often more convenient to use subtraction schemes
that are less practical from the calculational point of view, but more
convenient from the theoretical side. There, it is also important to keep in
mind that the beta functions do not need to be gauge independent.

\section{Conclusions}

\label{s9}

\setcounter{equation}{0}

In this paper we have derived generalized Ward identities for potentially
anomalous theories, and used them to study the problem of gauge
independence. The new equations contain an extra term that is responsible
for a number of interesting effects. We have renormalized the equations of
gauge dependence and integrated them. The result is that every gauge
dependence can be absorbed into a canonical transformation acting on the
renormalized $\Gamma $ functional, provided that the finite local
counterterms are appropriately fine-tuned. RG\ invariance is preserved and,
as expected, the physical quantities are gauge independent. Nevertheless,
the beta functions of the couplings may in general depend on the gauge
choice. Gauge independence is useful to switch back and forth between gauge
conditions that exhibit perturbative unitarity and gauge conditions that
exhibit a correct power-counting behavior and the locality of counterterms.

In several cases, the Adler-Bardeen theorem ensures that the gauge anomalies
cancel to all orders, when they are trivial at one loop. However, it is not
sufficient, \textit{per se}, to ensure that the physical quantities are
independent of the gauge fixing. In this paper we have proved that, in the
end, gauge invariance does imply the gauge independence of the physical
quantities. Precisely, we have shown that it is possible to renormalize\ the
theory and fine-tune its finite local counterterms so that the cancellation
of gauge anomalies ensured by the Adler-Bardeen theorem is preserved for
arbitrary values of the gauge parameters.

Said differently, assume that the gauge anomalies vanish for some specific
choices of the gauge parameters. Varying or turning on a gauge parameter is
equivalent to making a canonical transformation. After the canonical
transformation, it is always possible to re-renormalize the theory and
re-fine-tune its finite local counterterms to enforce the cancellation of
gauge anomalies again. Moreover, the gauge dependence of the renormalized $%
\Gamma $ functional is encoded into a convergent canonical transformation.
The theorem proved in section \ref{s8} is very general, to the extent that
we did not need to make particular assumptions about the gauge algebra or
the properties of the theory under renormalization. In particular, it holds
for renormalizable and nonrenormalizable, chiral and nonchiral, theories and
for arbitrary composite fields. Once we know that the cancellation of gauge
anomalies holds in the framework we prefer, we know that it holds in every
other framework.

One application of the theorem is to power-counting renormalizable chiral
gauge theories gauge-fixed by means of a nonrenormalizable gauge fixing. It
allows us to show that the parameters of negative dimensions introduced by
the gauge fixing do not propagate into the physical quantities. In other
words, the theory remains renormalizable, although in a nonmanifest form. A
second application is a crucial step in the proof of the Adler-Bardeen
theorem for nonrenormalizable theories elaborated in ref. \cite{ABnonreno}.

It is often possible to prove the cancellation of gauge anomalies in a
family of gauges. In that case, if the assumptions listed in section \ref{s3}
hold, we do not need a new fine-tuning to enforce the cancellation of gauge
anomalies after the variation of a gauge parameter. Then, the gauge
dependence of the theory is encoded into a convergent canonical
transformation on the renormalized $\Gamma $ functional, combined with a
finite redefinition of the parameters. This fact makes it apparent that in
general the beta functions of the couplings may depend on the gauge fixing.
We expect that high-order calculations of the beta functions in the standard
model will exhibit, sooner or later, dependences of the type mentioned here.

The gauge dependences of the beta functions can be eliminated by redefining
the couplings in \textit{ad hoc} ways. Thanks to this fact, gauge
independence can still be used to make nontrivial checks of the calculations.

\section*{Appendices}

\renewcommand{\thesection}{A}

\section{Useful formulas}

\label{appA}

\setcounter{equation}{0}\renewcommand{\theequation}{\thesection.%
\arabic{equation}}

In this appendix we collect a few identities that are used in the paper.
First, we recall that 
\begin{equation}
(\Gamma ,\Gamma )=\left\langle (S,S)\right\rangle ,  \label{anoa}
\end{equation}%
where $S$ is any action (renormalized or not), $\Gamma $ denotes the $\Gamma 
$ functional associated with $S$ and 
\begin{equation}
\left\langle X\right\rangle =\frac{1}{Z(J,K)}\int [\mathrm{d}\Phi ]\hspace{%
0.01in}X\exp \left( iS(\Phi ,K)+i\int \Phi ^{\alpha }J_{\alpha }\right)
\label{inte}
\end{equation}%
is the average defined by $S$, $X$ being a local functional. Formula (\ref%
{anoa}) can be proved by making the change of field variables (\ref{chv}) in
the functional integral (\ref{zg}), and recalling that in any dimensional
regularization the local perturbative changes of field variables have
Jacobian determinants identically equal to one. For details on the
derivation, see the appendices of refs. \cite{ABrenoYMLR,back}.

If $\zeta $ is any parameter, we also have the formulas%
\begin{eqnarray}
\frac{\partial \Gamma }{\partial \zeta } &=&\left\langle \frac{\partial S}{%
\partial \zeta }\right\rangle ,  \label{basic1} \\
(\Gamma ,\langle X\rangle ) &=&\left\langle (S,X)\right\rangle +\frac{i}{2}%
\langle (S,S)\hspace{0.01in}X\rangle _{\Gamma },  \label{basic2}
\end{eqnarray}%
where $X$ is an arbitrary local functional and $\left\langle X\hspace{0.01in}%
Y\right\rangle _{\Gamma }$ denotes the set of one-particle irreducible
diagrams that have one $X$ insertion, one $Y$ insertion, and arbitrary $\Phi 
$ and $K$ external legs, $Y$ being another local functional. Formula (\ref%
{basic1}) follows from the definition of $\Gamma $ as the Legendre transform
of $W$. Formula (\ref{basic2}) can be proved by making the change of field
variables (\ref{chv}) in the average (\ref{inte}), and expressing the final
result in terms of $\Phi $ and $K$. For details on this method, see the
appendix of ref. \cite{back}\footnote{%
Note that we have switched from the Euclidean notation used in \cite{back}
to the Minkowskian notation used here.}.

A simpler method to derive formula (\ref{basic2}) is to deform the action $S$
into $S+X\sigma $, where $\sigma $ is a constant, consider the deformed
version of formula (\ref{anoa}) and take the first order of its expansion in
powers of $\sigma $. By (\ref{basic1}), $\Gamma $ is deformed into $\Gamma
+\langle X\rangle \sigma +\mathcal{O}(\sigma ^{2})$. Instead, the average $%
\langle Y\rangle $ of a local functional $Y$ is deformed into $\langle
Y\rangle +i\langle \hspace{0.01in}YX\rangle _{\Gamma }\sigma +\mathcal{O}%
(\sigma ^{2})$. Indeed, the factor $\mathrm{e}^{iS}$ appearing in the
integrands of $Z(J,K)$ and $Z(J,K)\hspace{0.01in}\langle Y\rangle $ [check (%
\ref{zg}) and (\ref{inte})] is deformed into $\mathrm{e}^{iS}(1+iX\sigma +%
\mathcal{O}(\sigma ^{2}))$. Moreover, the deformed average, considered as a
functional of $\Phi $ and $K$, is still a collection of one-particle
irreducible diagrams. Thus, the first correction to $\langle Y\rangle $ is
precisely $i\langle \hspace{0.01in}YX\rangle _{\Gamma }\sigma $. Taking $%
Y=(S,S)$, we obtain $\left\langle (S,S)\right\rangle \rightarrow
\left\langle (S,S)\right\rangle +i\langle \hspace{0.01in}(S,S)X\rangle
_{\Gamma }\sigma +\mathcal{O}(\sigma ^{2})$, wherefrom (\ref{basic2})
follows.

If we subtract the equations (\ref{basic1}) and (\ref{basic2}) we also get 
\begin{equation}
\frac{\partial \Gamma }{\partial \zeta }-(\Gamma ,\langle X\rangle
)=\left\langle \frac{\partial S}{\partial \zeta }-(S,X)-\frac{i}{2}(S,S)%
\hspace{0.01in}X\right\rangle _{\Gamma },  \label{gx}
\end{equation}%
which is the starting point to derive the equations of gauge dependence.

Another useful identity tells us that \cite{removal,back}, if $\Phi
,K\rightarrow \Phi ^{\prime },K^{\prime }$ is a canonical transformation
with generating functional $F(\Phi ,K^{\prime })$, and $Y(\Phi ,K)$ is a
functional behaving as a scalar, i.e. such that $Y^{\prime }(\Phi ^{\prime
},K^{\prime })=Y(\Phi ,K)$, then 
\begin{equation}
\frac{\partial Y^{\prime }}{\partial \zeta }=\frac{\partial Y}{\partial
\zeta }-(Y,\tilde{F}_{\zeta }),  \label{bu}
\end{equation}%
where $\tilde{F}_{\zeta }(\Phi ,K)=F_{\zeta }(\Phi ,K^{\prime }(\Phi ,K))$
and $F_{\zeta }(\Phi ,K^{\prime })=\partial F/\partial \zeta $. The field
and source variables that are kept constant in the $\zeta $ derivative of a
functional are the natural field and source variables of that functional
(that is to say $\Phi ^{\prime }$ and $K^{\prime }$ for $Y^{\prime }$, $\Phi 
$ and $K$ for $Y$, $\Phi $ and $K^{\prime }$ for $F$).

\renewcommand{\thesection}{B}

\section{Renormalization of local bifunctionals}

\label{appB}

\setcounter{equation}{0}\renewcommand{\theequation}{\thesection.%
\arabic{equation}}

In this appendix we show how to renormalize a generic local bifunctional,
and then specialize to evanescent local bifunctionals.\ Given a theory with
action $S$, assume that a local bifunctional $\mathcal{F}$ has the form $AB$%
, where $A$ and $B$ are local functionals. Couple $A$ and $B$ to external
(constant) sources $h_{A}$ and $h_{B}$, by deforming the action $S$ into $%
\breve{S}=S-ih_{A}A-ih_{B}B$. Then, renormalize the extended action $\breve{S%
}$. The renormalized version of $\breve{S}$ has the form 
\begin{equation*}
\breve{S}_{R}=S_{R}-i\breve{h}_{A}A_{R}-i\breve{h}_{B}B_{R}-i\breve{h}_{B}%
\breve{h}_{A}C_{R}+\mathcal{O}(\breve{h}_{A}^{2})+\mathcal{O}(\breve{h}%
_{B}^{2}),
\end{equation*}%
where $A_{R}$ and $B_{R}$ are the renormalized functionals $A$ and $B$,
respectively, $\breve{h}_{A}$, $\breve{h}_{B}$ are the renormalized sources,
and $C_{R}$ is a local functional. Consider the $\Gamma $ functional $\breve{%
\Gamma}_{R}$ associated with $\breve{S}_{R}$. Differentiating it from the
left-hand side with respect to $\breve{h}_{B}$ and then $\breve{h}_{A}$, and
later setting $\breve{h}_{A}=$ $\breve{h}_{B}=0$, we find that the
renormalized $\mathcal{F}$ is equal to $\mathcal{F}_{R}=A_{R}B_{R}+C_{R}$.

It is a known fact (see for example \cite{collins}, chapter 13, or \cite%
{ABrenoYMLR}, section 6) that an evanescent local functional $E$ can be
renormalized so that its renormalized version $E_{R}$ satisfies $%
\left\langle E_{R}\right\rangle =\mathcal{O}(\varepsilon )$. This property
extends to\ evanescent local bifunctionals in a straightforward way.
However, we have to pay attention to some details.

By writing $\hat{\partial}^{\mu }=\hat{\eta}^{\mu \nu }\partial _{\nu }$ and 
$\hat{p}^{\mu }=\hat{\eta}^{\mu \nu }p_{\nu }$ everywhere inside $E$, we can
express each vertex of $E$ in a factorized form $\mathcal{T}_{k}\hat{\delta}%
_{k}$, where $\hat{\delta}_{k}$ denotes the evanescent part, made of tensors 
$\eta ^{\hat{\mu}\hat{\nu}}$, possibly $\varepsilon $ factors and other
structures that stay outside of the diagrams, while $\mathcal{T}_{k}$ is a
nonevanescent local functional and collects all the momenta. We then have $%
E=\sum_{k}\mathcal{T}_{k}\hat{\delta}_{k}$. Instead of considering the
average $\left\langle E\right\rangle $, consider first the diagrams $%
\left\langle \mathcal{T}_{k}\right\rangle $ that contain one insertion of $%
\mathcal{T}_{k}$. Iterating in $n=0,1,\ldots ,$ let $\mathcal{T}_{k\text{div}%
}^{(n+1)}$ denote the $(n+1)$-loop divergent part of $\langle \mathcal{T}%
_{nk}\rangle $, where%
\begin{equation*}
\mathcal{T}_{nk}=\mathcal{T}_{k}-\sum_{p=1}^{n}\mathcal{T}_{k\text{div}%
}^{(p)}
\end{equation*}%
are the functionals $\mathcal{T}_{k}$ renormalized up to and including $n$
loops. By the locality of counterterms, each $\mathcal{T}_{k\text{div}%
}^{(p)} $ is local. Then, the functional $E_{n}=\sum_{k}\mathcal{T}_{nk}\hat{%
\delta}_{k}$ is renormalized up to and including $n$ loops, and satisfies 
\begin{equation}
\left\langle E_{n}\right\rangle =\sum_{k}\left\langle \mathcal{T}%
_{nk}\right\rangle \hat{\delta}_{k}=\mathcal{O}(\varepsilon )+\mathcal{O}%
(\hbar ^{n+1}),  \label{enr}
\end{equation}%
because each $\left\langle \mathcal{T}_{nk}\right\rangle $ is convergent up
to $\mathcal{O}(\hbar ^{n+1})$. Finally, the functional $E_{R}\equiv
E_{\infty }$ satisfies $\left\langle E_{R}\right\rangle =\mathcal{O}%
(\varepsilon )$.

In the procedure just outlined we have subtracted away all sorts of
contributions $\mathcal{T}_{k\text{div}}^{(p)}$, order by order. More
generally, we do not need to subtract those that, once multiplied by $\hat{%
\delta}_{k}$, give evanescent results. Indeed, collecting those evanescent
local parts inside a local functional $\Delta E$, anything we have said so
far for $E$ can be repeated for $\Delta E$. We reach the conclusion that $%
\left\langle E_{R}\right\rangle =\mathcal{O}(\varepsilon )$ even if we
\textquotedblleft forget\textquotedblright\ to subtract any evanescent local
parts.

Once we have renormalized $E$ so that $\left\langle E_{R}\right\rangle $ is
evanescent to all orders, we can apply the same procedure to the
bifunctional $Y=E\hspace{0.01in}B$, where $B$ is an arbitrary local
functional. The outcome is that we can find a $\mathcal{O}(\hbar )$ local
functional $F_{R}$, such that the local bifunctional $Y_{R}=E_{R}B_{R}+F_{R}$
is renormalized and the average $\left\langle Y_{R}\right\rangle _{\Gamma }$
is evanescent to all orders.

More precisely, we can iterate the renormalization of $Y$ as follows. Write $%
Y=EB=\sum_{k}\hat{\delta}_{k}\mathcal{U}_{k}$, where $\mathcal{U}_{k}=%
\mathcal{T}_{k}B$. Let $B_{n}$ denote the functional $B$ renormalized up to
and including $n$ loops. Inductively assume that the $n$-loop renormalized $%
\mathcal{U}_{k}$ have the form $\mathcal{U}_{nk}=\mathcal{T}_{nk}B_{n}+%
\mathcal{C}_{nk}$, where $\mathcal{C}_{nk}$ are local functionals. Define $%
Y_{n}=\sum_{k}\hat{\delta}_{k}\mathcal{U}_{nk}=E_{n}B_{n}+F_{n}$, where $%
F_{n}=\sum_{k}\hat{\delta}_{k}\mathcal{C}_{nk}$. Clearly, $\langle
Y_{n}\rangle _{\Gamma }=\mathcal{O}(\varepsilon )+\mathcal{O}(\hbar ^{n+1})$%
, because each $\left\langle \mathcal{U}_{nk}\right\rangle _{\Gamma }$ is
convergent up to $\mathcal{O}(\hbar ^{n+1})$. By the locality of
counterterms, the $(n+1)$-loop contributions $\mathcal{U}_{nk}^{(n+1)}$ to $%
\langle \mathcal{U}_{nk}\rangle _{\Gamma }$ are made of a local divergent
part $U_{nk\text{div}}^{(n+1)}$, plus a generically nonlocal convergent
part. Consequently, the $(n+1)$-loop contributions to $\langle Y_{n}\rangle
_{\Gamma }$ are the sum of a local divergent part, a local nonevanescent
part, plus a generically nonlocal evanescent part. If we define $\mathcal{U}%
_{n+1\hspace{0.01in}k}=\mathcal{T}_{n+1\hspace{0.01in}k}B_{n+1}+\mathcal{C}%
_{n+1\hspace{0.01in}k}$, where $B_{n+1}$ is the functional $B$ renormalized
up to and including $n+1$ loops, and $\mathcal{C}_{n+1\hspace{0.01in}k}=%
\mathcal{C}_{nk}-\mathcal{U}_{nk\text{div}}^{(n+1)}$, we see that $%
\left\langle \mathcal{U}_{n+1\hspace{0.01in}k}\right\rangle _{\Gamma }$ is
convergent up to $\mathcal{O}(\hbar ^{n+2})$, and so $\langle Y_{n+1}\rangle
_{\Gamma }=\mathcal{O}(\varepsilon )+\mathcal{O}(\hbar ^{n+2})$, where $%
Y_{n+1}=\sum_{k}\hat{\delta}_{k}\mathcal{U}_{n+1\hspace{0.01in}%
k}=E_{n+1}B_{n+1}+F_{n+1}$, and $F_{n+1}=\sum_{k}\hat{\delta}_{k}\mathcal{C}%
_{n+1\hspace{0.01in}k}$. The conclusion also holds if we \textquotedblleft
forget\textquotedblright\ to subtract any evanescent local parts of $E_{n+1}$
and/or $F_{n+1}$. The subtraction can be iterated in $n$ so that in the end $%
\left\langle Y_{R}\right\rangle _{\Gamma }$ is evanescent to all orders in $%
\hbar $, where $Y_{R}=Y_{\infty }$.

\renewcommand{\thesection}{C}

\section{Integrating equation (\protect\ref{teta})}

\renewcommand{\theequation}{\thesection.\arabic{equation}}

\setcounter{equation}{0}

In this appendix we integrate the equations (\ref{teta}) and (\ref{gy}).
First, we recall how to integrate the simpler equation%
\begin{equation}
\frac{\partial X}{\partial \theta }=(X,V),  \label{problo}
\end{equation}%
for the functional $X(\Phi ,K,\theta )$, given the functional $V(\Phi
,K,\theta )$. Expanding in powers of $\theta $, write%
\begin{equation*}
V(\Phi ,K,\theta )=\sum_{n=0}^{\infty }\theta ^{n}V_{n}(\Phi ,K).
\end{equation*}%
We want to show that there exists a canonical transformation $\Phi
,K\rightarrow \Phi ^{\prime },K^{\prime }$, with generating functional 
\begin{equation}
\mathcal{F}(\Phi ,K^{\prime },\theta )=\int \Phi ^{\alpha }K_{\alpha
}^{\prime }+\sum_{n=1}^{\infty }\theta ^{n}\mathcal{F}_{n}(\Phi ,K^{\prime
}),  \label{effa}
\end{equation}%
such that 
\begin{equation*}
X^{\prime }(\Phi ^{\prime },K^{\prime })\equiv X(\Phi (\Phi ^{\prime
},K^{\prime },\theta ),K(\Phi ^{\prime },K^{\prime },\theta ),\theta )
\end{equation*}%
is independent of $\theta $.

We can derive conditions on the unknown functionals $\mathcal{F}_{n}$ by
applying formula (\ref{bu}), which relates the functional $V$ of (\ref%
{problo}) to the canonical transformation $\mathcal{F}$. A sufficient
condition to have $\partial X^{\prime }/\partial \theta =0$ is $V=\mathcal{%
\tilde{F}}_{\theta }$, where $\mathcal{F}_{\theta }=\partial \mathcal{F}%
/\partial \theta $. In other words,%
\begin{equation*}
0=\sum_{n=0}^{\infty }\theta ^{n}\left[ V_{n}(\Phi ,K)-(n+1)\mathcal{F}%
_{n+1}(\Phi ,K^{\prime })\right] ,\qquad K_{\alpha }=K_{\alpha }^{\prime
}+\sum_{n=1}^{\infty }\theta ^{n}\frac{\delta \mathcal{F}_{n}(\Phi
,K^{\prime })}{\delta \Phi ^{\alpha }}.
\end{equation*}%
The first equation can be solved for $\mathcal{F}_{n+1}$ by working
recursively in $n$. It is sufficient to express each $V_{k}(\Phi ,K)$ as a
functional of $\Phi $ and $K^{\prime }$, by using the second equation, and
then set the coefficient of $\theta ^{n}$ to zero. This proves that the
desired canonical transformation (\ref{effa}) does exist. Clearly, $%
X^{\prime }(\Phi ^{\prime },K^{\prime })$ coincides with $X(\Phi ^{\prime
},K^{\prime },0)$. Therefore, expressing everything by means of fields and
sources without primes, we get%
\begin{equation*}
X(\Phi ,K,\theta )=X(\Phi ^{\prime }(\Phi ,K,\theta ),K^{\prime }(\Phi
,K,\theta ),0).
\end{equation*}

Now, assume that a functional $Y(\Phi ,K,\theta )$ satisfies%
\begin{equation}
\frac{\partial Y}{\partial \theta }=(Y,V)+G,  \label{yt}
\end{equation}%
where $V(\Phi ,K,\theta )$ and $G(\Phi ,K,\theta )$ are two other
functionals. Define a new functional $\tilde{G}$ and a map $\mathcal{L}%
_{\theta }:Z\rightarrow \mathcal{L}_{\theta }Z$, where $Z$ is a functional,
as

\begin{equation*}
\tilde{G}(\Phi ,K,\theta )=\int_{0}^{\theta }\mathrm{d}\bar{\theta}\hspace{%
0.01in}G(\Phi ,K,\bar{\theta}),\qquad \mathcal{L}_{\theta }Z(\Phi ,K,\theta
)=\int_{0}^{\theta }\mathrm{d}\bar{\theta}\hspace{0.01in}\left( Z_{\bar{%
\theta}},V_{\bar{\theta}}\right) ,
\end{equation*}%
where $Z_{\bar{\theta}}=Z(\Phi ,K,\bar{\theta})$ and $V_{\bar{\theta}%
}=V(\Phi ,K,\bar{\theta})$. Observe that%
\begin{equation*}
\frac{\partial }{\partial \theta }\mathcal{L}_{\theta }Z=(Z,V).
\end{equation*}%
Then, equation (\ref{yt}) turns into equation%
\begin{equation*}
\frac{\partial \tilde{Y}}{\partial \theta }=(\tilde{Y},V),\qquad \text{for }%
\tilde{Y}=Y-\sum_{n=0}^{\infty }\mathcal{L}_{\theta }^{n}\tilde{G}.
\end{equation*}%
Note that the terms $\mathcal{L}_{\theta }^{n}\tilde{G}$ are at least $%
\mathcal{O}(\theta ^{n+1})$. Using the result found above, the canonical
transformation $\Phi ,K\rightarrow \Phi ^{\prime },K^{\prime }$ given by
formula (\ref{effa}) is such that the transformed functional 
\begin{equation*}
\tilde{Y}^{\prime }(\Phi ^{\prime },K^{\prime })\equiv \tilde{Y}(\Phi (\Phi
^{\prime },K^{\prime },\theta ),K(\Phi ^{\prime },K^{\prime },\theta
),\theta )
\end{equation*}%
is $\theta $ independent. Finally, if $G=\mathcal{O}(u^{n})$ for some
expansion parameter $u$ (which is $\varepsilon $ or $\hbar $, when we apply
this theorem in subsection \ref{indproof}) and $V$ is regular in $u$, then
the canonical transformation $\Phi ,K\rightarrow \Phi ^{\prime },K^{\prime }$
is also regular in $u$, which implies%
\begin{equation*}
Y(\Phi (\Phi ^{\prime },K^{\prime },\theta ),K(\Phi ^{\prime },K^{\prime
},\theta ),\theta )=\tilde{Y}^{\prime }(\Phi ^{\prime },K^{\prime })+%
\mathcal{O}(u^{n}).
\end{equation*}%
Setting $\theta =0$, we get%
\begin{equation*}
Y(\Phi ^{\prime },K^{\prime },0)=\tilde{Y}^{\prime }(\Phi ^{\prime
},K^{\prime })+\mathcal{O}(u^{n}).
\end{equation*}%
Hence, expressing everything by means of fields and sources without primes, 
\begin{equation*}
Y(\Phi ,K,\theta )=Y(\Phi ^{\prime }(\Phi ,K,\theta ),K^{\prime }(\Phi
,K,\theta ),0)+\mathcal{O}(u^{n}).
\end{equation*}%
In other words, the functional $Y(\Phi ,K,\theta )$ still evolves by means
of a canonical transformation, but only up to $\mathcal{O}(u^{n})$.

In most applications, the functionals $V$ and $G$ of equation (\ref{yt}) may
intrinsically depend on $Y$. For example, this happens when $Y$ is some
renormalized action (or the $\Gamma $ functional associated with it) and $V$%
, $G$ are (the averages of) some renormalized local functionals, calculated
with that action. We can disentangle this difficulty by expanding each
functional in powers of $\hbar $ and proceeding inductively in this
expansion. Writing%
\begin{equation*}
Y=\sum_{n=0}^{\infty }\hbar ^{n}Y_{n},\qquad V=\sum_{n=0}^{\infty }\hbar
^{n}V_{n},\qquad G=\sum_{n=0}^{\infty }\hbar ^{n}G_{n},
\end{equation*}%
we obtain the equations%
\begin{equation}
\frac{\partial Y_{n}}{\partial \theta }-(Y_{n},V_{0})=%
\sum_{k=0}^{n-1}(Y_{k},V_{n-k})+G_{n},  \label{ytn}
\end{equation}%
which have the same form as (\ref{yt}). The contributions $V_{k}$ and $G_{k}$
to $V$ and $G$ with $k\leqslant n$ do not depend on $Y_{n}$. For $k=0$ this
is obvious. For $k>0$ it is sufficient to observe that the vertices $Y_{n}$
of order $\hbar ^{n}$ of the renormalized action $Y$ can only contribute to
the one-particle irreducible diagrams associated with $V$ and $G$ that have $%
n+1$ or more loops. Indeed, at least one additional loop must be closed to
connect a vertex $Y_{n}$ with the insertions provided by $V$ or $G$. When $Y$
is the $\Gamma $ functional and $V$, $G$ are averages of local functionals,
we can argue similarly.

Now, assume that we have solved the equations (\ref{ytn}) for $n<\bar{n}$,
and consider the equations (\ref{ytn}) for $n=\bar{n}$. The unknown is $Y_{%
\bar{n}}$, while $V_{k}$ and $G_{k}$ with $k\leqslant \bar{n}$ are
independent of it. Thus, equations (\ref{ytn}) can be solved with the method
explained above. We conclude that the procedure we have given to solve the
equations (\ref{yt}) is well defined.

\renewcommand{\thesection}{D}

\section{Standard model coupled to quantum gravity}

\renewcommand{\theequation}{\thesection.\arabic{equation}}

\setcounter{equation}{0}

In this appendix we report some reference formulas for the standard model
coupled to quantum gravity. The classical fields $\phi $ contain the
vielbein $e_{\bar{\mu}}^{\bar{a}}$, the Yang-Mills gauge fields $A_{\bar{\mu}%
}^{a}$ and the matter fields, where the indices $a,b,\ldots $ refer to the
Yang-Mills gauge group (within which we include the Abelian subgroup) and $%
\bar{a},\bar{b},\ldots $ refer to the Lorentz group. The classical action $%
S_{c}(\phi )$ is equal to the sum $S_{c\text{SM}}+\Delta S_{c}$, where 
\begin{equation*}
S_{c\text{SM}}=\int \sqrt{|g|}\left[ -\frac{1}{2\kappa ^{2}}(R+2\Lambda _{%
\text{c}})-\frac{1}{4}F_{\bar{\mu}\bar{\nu}}^{a}F^{a\bar{\mu}\bar{\nu}}+%
\mathcal{L}_{m}\right] 
\end{equation*}%
and $\Delta S_{c}$ collects the invariants generated by renormalization as
counterterms, multiplied by independent parameters. Here, $R$ is the Ricci
curvature, $g$ is the determinant of the metric tensor, $F_{\bar{\mu}\bar{\nu%
}}^{a}$ is the Yang-Mills field strength, $\mathcal{L}_{m}$ is the matter
Lagrangian coupled to gravity, $\Lambda _{\text{c}}$ is the cosmological
constant, and $\kappa ^{2}=8\pi G$, where $G$ is Newton's constant.

The functional $S_{K}$ of formula (\ref{sbard}) reads 
\begin{eqnarray*}
S_{K} &=&\int (C^{\bar{\rho}}\partial _{\bar{\rho}}A_{\bar{\mu}}^{a}+A_{\bar{%
\rho}}^{a}\partial _{\bar{\mu}}C^{\bar{\rho}}-\partial _{\bar{\mu}%
}C^{a}-gf^{abc}A_{\bar{\mu}}^{b}C^{c})K_{A}^{\bar{\mu}a}+\int \left( C^{\bar{%
\rho}}\partial _{\bar{\rho}}C^{a}+\frac{g}{2}f^{abc}C^{b}C^{c}\right)
K_{C}^{a} \\
&&+\int (C^{\bar{\rho}}\partial _{\bar{\rho}}e_{\bar{\mu}}^{\bar{a}}+e_{\bar{%
\rho}}^{\bar{a}}\partial _{\bar{\mu}}C^{\bar{\rho}}+C^{\bar{a}\bar{b}}e_{%
\bar{\mu}\bar{b}})K_{\bar{a}}^{\bar{\mu}}+\int C^{\bar{\rho}}(\partial _{%
\bar{\rho}}C^{\bar{\mu}})K_{\bar{\mu}}^{C}+\int (C^{\bar{a}\bar{c}}\eta _{%
\bar{c}\bar{d}}C^{\bar{d}\bar{b}}+C^{\bar{\rho}}\partial _{\bar{\rho}}C^{%
\bar{a}\bar{b}})K_{\bar{a}\bar{b}}^{C} \\
&&\!\!\!\!\!\!{+\int \left( C^{\bar{\rho}}\partial _{\bar{\rho}}\bar{\psi}%
_{L}-\frac{i}{4}\bar{\psi}_{L}\sigma ^{\bar{a}\bar{b}}C_{\bar{a}\bar{b}}+g%
\bar{\psi}_{L}T^{a}C^{a}\right) K_{\psi }+\int K_{\bar{\psi}}\left( C^{\bar{%
\rho}}\partial _{\bar{\rho}}\psi _{L}-\frac{i}{4}\sigma ^{\bar{a}\bar{b}}C_{%
\bar{a}\bar{b}}\psi _{L}+gT^{a}C^{a}\psi _{L}\right) } \\
&&+\int \left( C^{\bar{\rho}}(\partial _{\bar{\rho}}\varphi )+g\mathcal{T}%
^{a}C^{a}\varphi \right) K_{\varphi }-\int B^{a}K_{\bar{C}}^{a}-\int B_{\bar{%
\mu}}K_{\bar{C}}^{\bar{\mu}}-\int B_{\bar{a}\bar{b}}K_{\bar{C}}^{\bar{a}\bar{%
b}},
\end{eqnarray*}%
where $\psi _{L}$ are left-handed fermions, $\varphi $ are scalars, while $%
T^{a}$ and $\mathcal{T}^{a}$ are the anti-Hermitian matrices\ associated
with their representations. The triplets $C^{a}$-$\bar{C}^{a}$-$B^{a}$, $C^{%
\bar{a}\bar{b}}$-$\bar{C}_{\bar{a}\bar{b}}$-$B^{\bar{a}\bar{b}}$ and $C^{%
\bar{\mu}}$-$\bar{C}_{\bar{\mu}}$-$B_{\bar{\mu}}$ collect the ghosts, the
antighosts and the Lagrange multipliers of Yang-Mills symmetry, local
Lorentz symmetry and diffeomorphisms, respectively. It is easy to check that 
$(S_{K},S_{K})=0$ in arbitrary $D$ dimensions.

Finally, the gauge fermion of formula (\ref{psif}) reads 
\begin{eqnarray*}
\Psi (\Phi ) &=&\int \sqrt{|g|}\bar{C}^{a}\left( g^{\bar{\mu}\bar{\nu}%
}\partial _{\bar{\mu}}A_{\bar{\nu}}^{a}+\frac{\xi }{2}B^{a}\right) +\int e%
\bar{C}_{\bar{a}\bar{b}}\left( \frac{1}{\kappa }e^{\bar{\rho}\bar{a}}g^{\bar{%
\mu}\bar{\nu}}\partial _{\bar{\mu}}\partial _{\bar{\nu}}e_{\bar{\rho}}^{\bar{%
b}}+\frac{\xi _{L}}{2}B^{\bar{a}\bar{b}}+\frac{\xi _{L}^{\prime }}{2}g^{\bar{%
\mu}\bar{\nu}}\partial _{\bar{\mu}}\partial _{\bar{\nu}}B^{\bar{a}\bar{b}%
}\right) \\
&&{-\int }\sqrt{|g|}\bar{C}_{\bar{\mu}}\left( \frac{1}{\kappa }\partial _{%
\bar{\nu}}g^{\bar{\mu}\bar{\nu}}+\frac{\xi _{G}}{\kappa }g^{\bar{\mu}\bar{\nu%
}}g_{\bar{\rho}\bar{\sigma}}\partial _{\bar{\nu}}g^{\bar{\rho}\bar{\sigma}}-%
\frac{\xi _{G}^{\prime }}{2}g^{\bar{\mu}\bar{\nu}}B_{\bar{\nu}}\right) ,
\end{eqnarray*}%
where $\xi $, $\xi _{L}$, $\xi _{L}^{\prime }$, $\xi _{G}$ and $\xi
_{G}^{\prime }$ are gauge-fixing parameters.

\renewcommand{\thesection}{E}

\section{Comparison with manifestly nonanomalous theories}

\label{s6}

\renewcommand{\theequation}{\thesection.\arabic{equation}}

\setcounter{equation}{0}

We have mentioned that an unexpected consequence of our results is that in
AB nonanomalous theories the beta functions of the couplings can depend on
the gauge-fixing parameters. It is interesting to better understand why this
does not happen in manifestly nonanomalous theories.

We actually begin with nongauge theories, that is to say theories that have
no gauge symmetries. There the action $S(\Phi ,K)$ does not even depend on
the sources $K$ and the canonical transformations are just arbitrary changes
of field variables.

Denote the classical action by $S(\phi )$, the renormalized action by $%
S_{R}(\phi )$ and the renormalized $\Gamma $ functional by $\Gamma _{R}(\phi
)$. We assume that $S_{R}$ and $\Gamma _{R}$ are defined by subtracting away
the divergences just as they come, in the minimal subtraction scheme.

Consider a local, perturbative change of field variables%
\begin{equation}
\psi ^{i}(\phi ,\theta )=\phi ^{i}+\mathcal{O}(\theta )  \label{psix}
\end{equation}%
for the classical action $S$. Let $S_{\theta }(\phi ,\theta )$ denote the
transformed classical action,%
\begin{equation*}
S_{\theta }(\phi ,\theta )=S(\psi (\phi ,\theta )),
\end{equation*}%
which obviously satisfies%
\begin{equation*}
\frac{\partial S_{\theta }}{\partial \theta }=\int \Delta \phi ^{i}\hspace{%
0.01in}\frac{\delta _{l}S_{\theta }}{\delta \phi ^{i}},
\end{equation*}%
where%
\begin{equation}
\Delta \phi ^{i}=\int \frac{\delta \psi ^{j}}{\delta \theta }\frac{\delta
_{l}\phi ^{i}}{\delta \psi ^{j}}.  \label{compa}
\end{equation}%
Denote the renormalized $S_{\theta }$ by $S_{R\theta }$ and the $\Gamma $
functional associated with it by $\Gamma _{R\theta }$.

We want to show that the change of field variables (\ref{psix}) on $S$ is
mapped onto a renormalized change of field variables on $S_{R}$ and a
nonlocal, convergent change of field variables on $\Gamma _{R}$. This
property is encoded into the equations of gauge dependence, which now\ read%
\begin{equation}
\frac{\partial S_{R\theta }}{\partial \theta }=\int \Delta _{R}\phi ^{i}%
\hspace{0.01in}\frac{\delta _{l}S_{R\theta }}{\delta \phi ^{i}},\qquad \frac{%
\partial \Gamma _{R\theta }}{\partial \theta }=\int \left\langle \Delta \phi
_{R}^{i}\right\rangle \frac{\delta _{l}\Gamma _{R\theta }}{\delta \phi ^{i}},
\label{mno}
\end{equation}%
where $\Delta _{R}\phi ^{i}$ is the renormalized version of the composite
field (\ref{compa}). Equations (\ref{mno}) are just particular cases of
equation (\ref{problo}), and can be integrated with the method explained in
appendix C. So doing, it is straightforward to prove that the $\theta $
dependences of both $S_{R\theta }$ and $\Gamma _{R\theta }$ are encoded into
pure changes of field variables, with no redefinitions of parameters.

We point out that the first equations of formula (\ref{mno}) are highly
nonlinear in $S_{R\theta }$, because $\Delta _{R}\phi ^{i}$, being a
renormalized composite field, intrinsically depends on $S_{R\theta }$.
Nevertheless, with the inductive procedure explained in appendix C we can
disentangle this dependence. Similarly, the equations satisfied by $\Gamma
_{R\theta }$ contain the average $\left\langle \Delta _{R}\phi
^{i}\right\rangle $ on the right-hand side, which is also determined by $%
S_{R\theta }$. The procedure to integrate the equations of $\Gamma _{R\theta
}$ is basically the same as the one for $S_{R\theta }$ and is again given in
appendix C.

Formulas (\ref{mno}) can be proved by induction, using the minimal
subtraction scheme. Let $S_{n}=S_{\theta }+\mathcal{O}(\hbar )\times $poles
and $\Delta _{n}\phi ^{i}=\Delta \phi ^{i}+\mathcal{O}(\hbar )\times $poles\
denote the action and the composite field (\ref{compa}) renormalized up to
and including $n$ loops. Assume that 
\begin{equation}
\mathcal{R}_{n}\equiv \frac{\partial S_{n}}{\partial \theta }-\int \Delta
_{n}\phi ^{i}\hspace{0.01in}\frac{\delta _{l}S_{n}}{\delta \phi ^{i}}=%
\mathcal{O}(\hbar ^{n+1}).  \label{incud}
\end{equation}%
Clearly, this assumption is satisfied for $n=0$. Moreover, in the minimal
subtraction scheme $\mathcal{R}_{n}$ is made of pure poles.

Differentiating the $\Gamma $ functional $\Gamma _{n}$, associated with $%
S_{n}$, with respect to $\theta $, we get 
\begin{equation}
\frac{\partial \Gamma _{n}}{\partial \theta }=\left\langle \frac{\partial
S_{n}}{\partial \theta }\right\rangle _{n}=\int \mathrm{d}^{D}x\left\langle
\Delta _{n}\phi ^{i}(x)\frac{\delta _{l}S_{n}}{\delta \phi ^{i}(x)}%
\right\rangle _{n}+\langle \mathcal{R}_{n}\rangle _{n}.  \label{a1}
\end{equation}%
Now, 
\begin{equation*}
\frac{\delta _{l}S_{n}}{\delta \phi ^{i}(x)}\exp \left( iS_{n}+i\int \phi
^{j}J_{j}\right) =-J_{i}(x)-i\frac{\delta _{l}}{\delta \phi ^{i}(x)}\exp
\left( iS_{n}+i\int \phi ^{j}J_{j}\right) .
\end{equation*}%
Using this formula inside (\ref{a1}) we can drop the last term by
integrating by parts, because when the derivative $\delta _{l}/\delta \phi
^{i}(x)$ acts on $\Delta _{n}\phi ^{i}(x)$ it gives zero in dimensional
regularization. Finally, we obtain%
\begin{equation}
\frac{\partial \Gamma _{n}}{\partial \theta }=-\int \left\langle \Delta
_{n}\phi ^{i}\right\rangle _{n}J_{i}+\langle \mathcal{R}_{n}\rangle
_{n}=\int \left\langle \Delta _{n}\phi ^{i}\right\rangle _{n}\frac{\delta
_{l}\Gamma _{n}}{\delta \phi ^{i}}+\langle \mathcal{R}_{n}\rangle _{n}.
\label{gnx}
\end{equation}%
Since $S_{n}$ and $\Delta _{n}\phi ^{i}$ are renormalized up to and
including $n$ loops, the $(n+1)$-loop divergent parts $\Gamma _{n\text{%
\hspace{0.01in}div}}^{(n+1)}$ and $\Delta _{n\text{\hspace{0.01in}div}%
}^{(n+1)}\phi ^{i}$ of $\Gamma _{n}$ and $\left\langle \Delta _{n}\phi
^{i}\right\rangle _{n}$ are local. Moreover, the $\mathcal{O}(\hbar ^{n+1})$
divergent part of $\langle \mathcal{R}_{n}\rangle _{n}$ coincides with the $%
\mathcal{O}(\hbar ^{n+1})$ part of $\mathcal{R}_{n}$, because $\mathcal{R}%
_{n}$ starts from $\mathcal{O}(\hbar ^{n+1})$ and it is just made of poles.
Thus, taking the $\mathcal{O}(\hbar ^{n+1})$ divergent parts of formula (\ref%
{gnx}) we get%
\begin{equation}
\frac{\partial \Gamma _{n\text{\hspace{0.01in}div}}^{(n+1)}}{\partial \theta 
}=\int \Delta _{n\text{div}}^{(n+1)}\phi ^{i}\hspace{0.01in}\frac{\delta
_{l}S_{\theta }}{\delta \phi ^{i}}+\int \Delta \phi ^{i}\hspace{0.01in}\frac{%
\delta _{l}\Gamma _{n\text{\hspace{0.01in}div}}^{(n+1)}}{\delta \phi ^{i}}+%
\frac{\partial S_{n}}{\partial \theta }-\int \Delta _{n}\phi ^{i}\hspace{%
0.01in}\frac{\delta _{l}S_{n}}{\delta \phi ^{i}}+\mathcal{O}(\hbar ^{n+2}).
\label{div}
\end{equation}%
Subtracting the divergences just as they come, we define%
\begin{equation*}
S_{n+1}=S_{n}-\Gamma _{n\text{\hspace{0.01in}div}}^{(n+1)},\qquad \Delta
_{n+1}\phi ^{i}=\Delta _{n}\phi ^{i}-\Delta _{n\text{div}}^{(n+1)}\phi ^{i}.
\end{equation*}%
Clearly, the $\Gamma $ functional $\Gamma _{n+1}$ associated with $S_{n+1}$
is renormalized up to and including $n+1$ loops. Using (\ref{div}), we find%
\begin{equation*}
\mathcal{R}_{n+1}\equiv \frac{\partial S_{n+1}}{\partial \theta }-\int
\Delta _{n+1}\phi ^{i}\hspace{0.01in}\frac{\delta _{l}S_{n+1}}{\delta \phi
^{i}}=\mathcal{O}(\hbar ^{n+2}).
\end{equation*}

Thus, the inductive assumption (\ref{incud}) is promoted to the next order.
The equations (\ref{mno}) follow by taking $n=\infty $ in (\ref{incud}) and (%
\ref{gnx}).

We see that in theories with no gauge symmetries a change of field variables
on the classical action does not generate redefinitions of parameters in the
renormalized $\Gamma $ functional: the parameters $\theta $ introduced by
the field redefinition do not propagate into the beta functions of the
couplings. Moreover, we do not need to re-fine-tune the finite local
counterterms.

Another approach to these issues was given in refs. \cite{fieldcov,masterf},
where the changes of field variables were mapped from the classical action
to the renormalized action and the (renormalized) generating functionals $Z$%
, $W$ and $\Gamma $, as well as a more general type of $\Gamma $ functional,
called master functional. That approach also shows that a change of field
variables does not affect the beta functions of the couplings, in the
theories that have no gauge symmetries.

Similar properties hold in manifestly nonanomalous gauge theories, where the
equations%
\begin{equation}
\frac{\partial S_{R\theta }}{\partial \theta }=(S_{R\theta },\tilde{Q}%
_{R\theta }),\qquad \frac{\partial \Gamma _{R\theta }}{\partial \theta }%
=(\Gamma _{R\theta },\langle \tilde{Q}_{R\theta }\rangle )  \label{nonn}
\end{equation}%
hold and can be integrated \cite{back}. Again, the conclusion is that a
canonical transformation acting on the classical action is converted into a
renormalized canonical transformation acting on the renormalized action, and
a nonlocal, convergent canonical transformation acting on the renormalized $%
\Gamma $ functional, with no effect on the beta functions of the couplings.
Equations (\ref{mno}) can also be obtained by switching off\ the sources $K$
in formulas (\ref{nonn}).

What \textquotedblleft goes wrong\textquotedblright\ in AB\ nonanomalous
theories, is that \textquotedblleft small things\textquotedblright , that is
to say evanescent terms $\mathcal{O}(\mathcal{\varepsilon })$, are around
all the time, and can generate unexpected finite corrections by simplifying
some divergences. For this reason, they force us to re-fine-tune the
subtraction scheme at every, even minor, modification of the framework in
which we formulate the theory. Yet, we have shown in the paper that we can
put their effects under control and preserve the correct physical properties.

\end{document}